\documentclass[11pt]{article}
\usepackage{amssymb}
\usepackage{amsmath}
\usepackage[dvips]{graphicx}
\usepackage{slashed}
\usepackage{subfig}

\DeclareMathOperator{\tr}{tr}

\begin{document}

\baselineskip=18pt
\numberwithin{equation}{section}
\allowdisplaybreaks
\pagestyle{myheadings}
\thispagestyle{empty}
\pagestyle{plain}
\baselineskip=19pt

\def\eg{{\it e.g.}}
\newcommand{\nc}{\newcommand}
\nc{\rnc}{\renewcommand}
\rnc{\d}{\mathrm{d}}
\nc{\D}{\partial}
\nc{\K}{\kappa}
\nc{\bK}{\bar{\K}}
\nc{\bN}{\bar{N}}
\nc{\bq}{\bar{q}}
\nc{\bp}{\bar{p}}
\nc{\vbq}{\vec{\bar{q}}}
\nc{\g}{\gamma}
\nc{\lrarrow}{\leftrightarrow}
\nc{\rg}{\sqrt{g}}
\rnc{\[}{\begin{equation}}
\rnc{\]}{\end{equation}}
\nc{\bea}{\begin{eqnarray}}
\nc{\eea}{\end{eqnarray}}
\nc{\nn}{\nonumber}
\rnc{\(}{\left(}
\rnc{\)}{\right)}
\nc{\q}{\vec{q}}
\nc{\x}{\vec{x}}
\nc{\ep}{\epsilon}
\nc{\tto}{\rightarrow}
\rnc{\inf}{\infty}
\rnc{\Re}{\mathrm{Re}}
\rnc{\Im}{\mathrm{Im}}
\nc{\z}{\zeta}
\nc{\mA}{\mathcal{A}}
\nc{\mB}{\mathcal{B}}
\nc{\mC}{\mathcal{C}}
\nc{\mD}{\mathcal{D}}
\nc{\mE}{\mathcal{E}}
\nc{\mF}{\mathcal{F}}
\rnc{\H}{\mathcal{H}}
\rnc{\L}{\mathcal{L}}
\nc{\<}{\langle}
\rnc{\>}{\rangle}
\nc{\fnl}{f_{NL}}
\nc{\fnleq}{f_{NL}^{equil.}}
\nc{\fnlloc}{f_{NL}^{local}}
\nc{\vphi}{\varphi}
\nc{\Lie}{\pounds}
\nc{\half}{\frac{1}{2}}
\nc{\bOmega}{\bar{\Omega}}
\nc{\bLambda}{\bar{\Lambda}}
\nc{\dN}{\delta N}
\nc{\gYM}{g_{\mathrm{YM}}}
\nc{\geff}{g_{\mathrm{eff}}}
\nc{\bg}{\hat{\gamma}}
\nc{\Oi}{\Omega_{[2]}}
\nc{\Oii}{\Omega_{[3]}}
\nc{\Ei}{E_{[2]}}
\nc{\Eii}{E_{[3]}}
\nc{\bOi}{\bar{\Omega}_{[2]}}
\nc{\bOii}{\bar{\Omega}_{[3]}}
\nc{\bEi}{\bar{E}_{[2]}}
\nc{\bEii}{\bar{E}_{[3]}}
\rnc{\a}{\bar{a}}
\rnc{\b}{\bar{b}}
\rnc{\c}{\bar{c}}
\rnc{\O}{\mathcal{O}}
\nc{\blambda}{\bar{\lambda}}
\nc{\oa}{\stackrel{\leftrightarrow}}
\newcommand{\lla}{\langle \! \langle}
\newcommand{\rra}{\rangle \! \rangle}
\newcommand{\lwick}{:\!}
\newcommand{\rwick}{\!:}
\newcommand{\p}{\partial}
\nc{\wT}{\widetilde{T}}

\begin{titlepage}

\begin{center}

\hfill

\vskip 1 cm {\Large \bf  Holographic predictions for cosmological 3-point functions}

\vskip 1.25 cm {\bf Adam Bzowski${}^1$, Paul McFadden${}^2$ and Kostas Skenderis${}^3$}
\\ {\vskip 0.5cm \it\small
${}^1\,$KdV Institute for Mathematics, Science Park 904, 1090 GL Amsterdam, the Netherlands.\\
${}^2\,$Perimeter Institute for Theoretical Physics, Waterloo Ontario N2L 2Y5, Canada.\\
${}^3\,$KdV Institute for Mathematics, Institute for Theoretical Physics, Gravitation \& Astro-Particle Physics Amsterdam,
Science Park 904, 1090 GL Amsterdam, the Netherlands.}

{\vskip 0.2cm \small
E-mail: {\tt a.w.bzowski@uva.nl, pmcfadden@perimeterinstitute.ca, k.skenderis@uva.nl} }

\end{center}

\vskip 1 cm

\begin{abstract}
\baselineskip=16pt

We present the holographic predictions for cosmological 3-point correlators,
involving both scalar and tensor modes, for a universe which started in a non-geometric holographic phase.
Holographic formulae relate the cosmological 3-point functions
to stress tensor correlation functions of a holographically dual three-dimensional non-gravitational
QFT.
We  compute these correlators at 1-loop order for a theory containing massless scalars, fermions and gauge fields,
and present an extensive analysis of the constraints due to Ward identities showing that they uniquely
determine the correlators up to a few constants.
We define shapes for all cosmological bispectra and compare the holographic shapes to
the slow-roll ones, finding that some are distinguishable while others, perhaps surprisingly, are not.

\end{abstract}

\end{titlepage}

\tableofcontents

\section{Introduction}
In a recent series of papers \cite{McFadden:2009fg,McFadden:2010na,
McFadden:2010jw,McFadden:2010vh,Easther:2011wh,McFadden:2011kk} 
we put forward a holographic
framework for inflationary cosmology and discussed a novel class of
models describing a universe that started in a non-geometric phase,
which is described holographically via a large-$N$ three-dimensional
QFT. The power spectra and the scalar bispectrum were computed in
\cite{McFadden:2009fg,McFadden:2010na} and \cite{McFadden:2010jw},
respectively, and in this paper we complete this program by computing the
non-Gaussianities that involve tensors. Non-Gaussianities involving tensors
are not expected to
be measurable in near-future experiments. Nevertheless, they are still
interesting theoretically and their structure has been the topic of several
recent papers \cite{Maldacena:2011nz,Soda:2011am,Shiraishi:2011st,Gao:2011vs}.

The holographic model is specified by providing the dual QFT and the
holographic dictionary that relates QFT correlation functions to
cosmological observables. We worked out the holographic dictionary
for non-Gaussianities involving tensors in \cite{McFadden:2011kk},
and in this paper we compute the relevant QFT correlation functions.
The models we discuss are based on perturbative three-dimensional
QFTs that admit a large-$N$ limit and have a
generalised conformal structure \cite{Jevicki:1998ub,Kanitscheider:2008kd}.
An example of such a theory is $SU(N)$ Yang-Mills theory coupled to
massless  scalars and fermions, with all fields
transforming in the adjoint of $SU(N)$. The non-Gaussianities are
extracted from the
3-point function of the stress tensor of this theory.
The leading order 1-loop computation of this 3-point function
is independent of the interactions of the QFT, and thus our main
task is to compute this 3-point function for free QFTs.

Since all leading order results depend only on the free theory, let us
briefly discuss the case in which the holographic model is a free QFT.
In such a model, the spectrum is the exactly scale-invariant
Harrison-Zel'dovich spectrum and the bispectrum is given
exactly by the results reported here, {\it i.e.}, the leading order results
are the exact answer in the free theory. The shapes associated
with the bispectrum may thus be considered as
the analogue of the exact scale-invariant spectrum for
higher point functions. We have seen in
\cite{McFadden:2010vh} that the scalar bispectrum shape for this model
is indeed special: it is exactly equal to
the factorisable equilateral
shape\footnote{ The holographic model is also the only model
that yields exactly this shape.} originally introduced in
\cite{Creminelli:2005hu}. One may thus anticipate that shapes
associated with the other 3-point functions will also have
special properties. Possible shapes for the bispectrum involving
only tensors have been discussed recently in \cite{Maldacena:2011nz}
and here we will define and discuss shapes for the bispectrum involving
both tensors and scalars.

The computation of the 3-point function of the stress tensor
at 1-loop is a non-trivial task even for free QFTs. We discuss and
develop several methods for evaluating the relevant Feynman
diagrams. The 1-loop result is constrained by Ward identities and these
provide a very non-trivial check of the expression we obtained
by a direct computation.

As mentioned above, to 1-loop order
only the free part of the QFT enters. The QFT consists of gauge
fields, fermions, minimal and conformal scalars. Conformal scalars
and fermions are conformal field theories and their 3-point functions
are constrained by conformal Ward identities.
As is well known (from a position space analysis) \cite{Osborn:1993cr}, the 3-point function
of the stress tensor in $d=3$ is uniquely fixed by conformal invariance
to be a linear combination of two conformal invariants, and is thus parametrised by two constants. 
(We assume parity is preserved).
Our computation is done in momentum space and we thus provide
the most general such 3-point functions in momentum space, where the
two parameters are the number of conformal scalars and the number of fermions.
The same computation was also recently reported in \cite{Maldacena:2011nz}\footnote{Our results agree with the ones in v2 of
\cite{Maldacena:2011nz}. Relative to \cite{Maldacena:2011nz}, we also computed semi-local
terms.}.
We have explicitly verified
that our results satisfy the conformal Ward identities.

Let us now turn to minimal scalars and gauge fields. In three dimensions
vectors are dual to scalars, so one may expect
that gauge fields  contribute the same as minimal scalars at
1-loop order. We will indeed verify that this is the case. Note that beyond
1-loop the two are expected to contribute differently. Minimal scalars
differ from conformal scalars in the way they couple to gravity, which in
flat spacetime is reflected in their having a different stress tensor.
More precisely, the stress tensor $T^\phi_{ij}$ for a minimal scalar
may be decomposed into a part $\wT^\phi_{ij}$ corresponding to the stress tensor for a conformal scalar
plus an ``improvement term'':
\[
\label{minconfdecomp}
T_{ij}^\phi = \wT_{ij}^\phi - \frac{1}{8} \left( \delta_{ij} \partial^2 - \partial_i \partial_j \right) \phi^2.
\]
It follows that the 3-point function of $T_{ij}^\phi$ may be computed from the 3-point functions
involving $\wT_{ij}^\phi$ and the dimension one operator ${\cal O}_1= \phi^2$. In turn, these
3-point functions are uniquely determined
by conformal invariance, up to constants \cite{Osborn:1993cr}.  Thus, effectively all
3-point functions are determined by conformal 3-point functions at this order, even though the
underlying theory is not conformal.

In \cite{Maldacena:2011nz}, the 3-point functions for tensors were
computed in a de Sitter background.  The de Sitter isometries act as the
conformal group at late times, and the 3-point functions are then
constrained by conformal invariance to be specific linear combinations
of the 3-point functions of conformal scalars and fermions.
Note that the de Sitter
result is the leading order approximation for slow-roll inflation. In
general one expects that (broken) conformal invariance would constrain
cosmological correlators in asymptotically de-Sitter slow-roll
inflation, see also \cite{Antoniadis:2011ib, Creminelli:2011mw}.

Our holographic
results are for a very different universe, but we have seen that all relevant 3-point functions are
essentially determined by conformal 3-point functions. One may then wonder how our results compare with those
of slow-roll inflation. The 3-point functions involving only tensors are determined by
the 3-point functions of conformal scalars and fermions, and, as in the discussion of \cite{Maldacena:2011nz},
they agree exactly with slow-roll inflation if the field content of the dual QFT is appropriately chosen.
The other 3-point functions (involving both scalar and tensor perturbations) are different but, perhaps surprisingly, they are rather similar. To quantify the differences we define (and plot) shape functions for all correlators, generalising
the notion of shape functions for 3-point functions of only scalars or only tensors.

This paper is organised as follows. In Section \ref{sec:dual} 
we discuss the
dual QFT and in Section \ref{sec:dict} we present the holographic
dictionary. In Section \ref{sec:eva} we compute all relevant QFT
correlation function by direct evaluation of the relevant Feynman integrals,
and in Section \ref{sec:Wid} we explain their structure using Ward identities.
The holographic predictions for the cosmological observables are
presented in Section \ref{sec:HolPredictions} and these results are
compared with the slow-roll ones in Section \ref{sec:slow-roll}.
We discuss our results in Section \ref{sec:disc}. Several technical results
are presented in four appendices: in Appendix \ref{sec:hel} we
summarise our notation and conventions for the helicity tensors,
in Appendix \ref{methods_app} we present three different methods for
evaluating the
relevant diagrams, in Appendix \ref{App_ghosts} we show that
ghosts and gauge fixing terms do not contribute in
correlators of the stress tensor and in Appendix \ref{app_WI} we
present the conformal and diffeomorphicm Ward identities.

\section{Dual QFT} \label{sec:dual}

As a dual QFT we consider super-renormalisable theories that admit a large
$N$ limit and contain one
dimensionful coupling constant.
A prototype example\footnote{A different example would be to consider
$O(N)$ models, see \cite{Anninos:2011ui} for a related discussion.}
 is three-dimensional $SU(\bN)$
Yang-Mills theory\footnote{We use the unconventional
notation $SU(\bN)$ as we reserve $N$ for the analytically continued
value $\bar{N} = - i N$.} coupled to a number of massless scalars and massless
fermions,
all transforming in the adjoint of $SU(\bN)$.
Theories of this type are typical in AdS/CFT
where they appear as the worldvolume theories of D-branes.
A general such model that admits a large $\bN$ limit is
\[
\label{Lfree}
S = \frac{1}{g_{\mathrm{YM}}^2}\int \d^3 x\,
 \tr\left( \frac{1}{4} F^I_{ij}F^I_{ij} +  \frac{1}{2} (\D\phi^J)^2
+  \frac{1}{2} (\D\chi^K)^2
+ \bar{\psi}^L \slashed{\D} \psi^L + \mathrm{interactions}\, \right),
\]
where for all fields, $\varphi= \varphi^{a}T^a$, and
$\mathrm{tr} T^a T^b = \delta^{ab}$. We work with the Wick rotated
QFT (of signature $(+,+,+)$). This is mostly for convenience; we could equally well have
stated all results in Lorentzian signature. The analytic
continuation relevant for cosmology (which will appear in \eqref{an_cont} below) is a different
continuation: it acts on the magnitude of the momentum.
The gamma matrices satisfy
$\{\gamma_i,\gamma_j\}=-2 \delta_{ij}$.
We consider $\mathcal{N}_A$ gauge fields $A^I$
($I=1,\ \ldots,\ \mathcal{N}_A)$,
$\mathcal{N}_\phi$ minimal scalars $\phi^J$ ($J=1,\ \ldots,\ \mathcal{N}_\phi)$,
$\mathcal{N}_\chi$ conformal scalars $\chi^K$ ($K=1,\ \ldots,\ \mathcal{N}_\chi)$
 and $\mathcal{N}_\psi$ fermions $\psi^L$  ($L=1,\ \ldots,\  \mathcal{N}_\psi)$.
Note that $g_{\mathrm{YM}}^2$ has dimension one in three dimensions.
In general, the Lagrangian \eqref{Lfree} will also contain dimension-four interaction terms (see \cite{McFadden:2010na}).
We will leave these interactions unspecified, however, as they do not contribute to the leading order calculations we perform here.

In the next section we will present the holographic formulae
that relate cosmological 3-point functions to correlation functions
of the dual QFT.  Generally speaking, the terms
appearing in these formulae (see (\ref{holo_zzz})-(\ref{holo_ggg}))
are either 3-point
functions of the stress tensor, or else semi-local terms ({\it i.e.}, terms which are analytic in two of the three momentum).
The semi-local terms
involve either 2-point functions of the stress tensor or 
the $\Upsilon$ tensor defined by coupling the QFT to gravity,
differentiating the
stress tensor w.r.t.~to the background metric and then setting the
background metric to the flat metric, 
\[
\label{Upsilon_def}
 \Upsilon_{ijkl}(\x_1,\x_2) = \frac{\delta T_{ij}(\x_1)}{\delta g^{kl}(\x_2)}\Big|_{0} =
2\frac{\delta^2 S}{\delta g^{ij}(\x_1)\delta g^{kl}(\x_2)}\Big|_{0}+\half T_{ij}(\x_1)\delta_{kl}\delta(\x_1-\x_2).
\]
The 2-point function takes the following general form
\[
\label{ABdef}
 \<\!\<T_{ij}(\bq)T_{kl}(-\bq)\>\!\> = A(\bq)\Pi_{ijkl}+B(\bq)\pi_{ij}\pi_{kl},
\]
where the double bracket notation 
suppresses the delta function associated with momentum conservation, {\it i.e.},
\[
 \<T_{ij}(\vbq_1)T_{kl}(\vbq_2)\> = (2\pi)^3\delta(\vbq_1+\vbq_2)\<\!\<T_{ij}(\bq_1)T_{kl}(-\bq_1)\>\!\>,
\]
and the transverse and transverse traceless projection operators are respectively 
\[
\label{projection_operators}
 \pi_{ij} = \delta_{ij}-\frac{\bq_i\bq_j}{\bq^2}, \qquad \Pi_{ijkl}=\frac{1}{2}\big(\pi_{ik}\pi_{jl}+\pi_{il}\pi_{jk}-\pi_{ij}\pi_{kl}\big).
\]

\begin{figure}[t]
\center
\includegraphics[width=5cm]{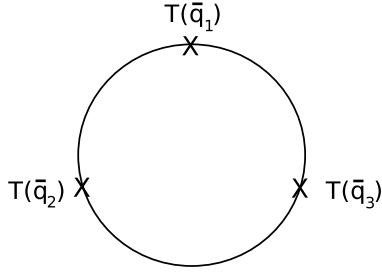}
\hspace{2pc}
\begin{minipage}[b]{14pc}
\caption{\label{1-loopfig}
1-loop contribution to the stress tensor 3-point function. 
\vspace{0.2cm}
\newline}
\end{minipage}
\end{figure}

The leading contribution to the 2- and 3-point function comes from
1-loop diagrams (see
Fig.~\ref{1-loopfig} for the 3-point function), which are of order $\bN^2$ and
involve only the
free part of the Lagrangian.  Interactions contribute to diagrams at
2-loop order and higher, but these are suppressed by factors of
$\geff^2$
relative to the 1-loop contribution and will be neglected here.
As discussed in \cite{McFadden:2009fg,McFadden:2010na}, $\geff^2$
is of the order of $n_s{-}1 \sim O(10^{-2})$. Indeed, fitting the
WMAP data to this model \cite{Easther:2011wh} leads to a small
value of
$\geff^2$ justifying the perturbative treatment.

For spatially flat cosmologies, the background metric seen by the dual QFT
is also flat.  The dual stress tensor is then given by
\[
 T_{ij} = \left.\frac{2}{\sqrt{g}} \frac{\delta S}{\delta g^{ij}}
\right|_{g_{ij}=\delta_{ij}}=
T^A_{ij}+T^\phi_{ij}+T^\psi_{ij}+T^\chi_{ij},
\]
where the contributions from the various fields (suppressing the interactions, as well as the ghost and gauge-fixing terms which we discuss in Appendix \ref{App_ghosts}) in \eqref{Lfree} are
\begin{align}
& T^A_{ij}=\frac{1}{g_{\mathrm{YM}}^2} \mathrm{tr} \big[F^I_{ik}F^I_{jk}-\delta_{ij}\frac{1}{4}F^I_{kl}F^I_{kl}\big], \label{Ta} \\
& T^\phi_{ij} =\frac{1}{g_{\mathrm{YM}}^2} \mathrm{tr} \big[\D_i\phi^J \D_j\phi^J-\delta_{ij}\frac{1}{2}(\D\phi^J)^2\big], \label{Tphi} \\
& T^\chi_{ij} = \frac{1}{g_{\mathrm{YM}}^2}\mathrm{tr} \big[\D_i\chi^K \D_j\chi^K-\frac{1}{8}\D_i\D_j(\chi^K)^2-\delta_{ij}\big(\frac{1}{2}(\D\chi^K)^2-\frac{1}{8}\D^2(\chi^K)^2\big)\big], \label{Tchi} \\
& T^\psi_{ij} = \frac{1}{g_{\mathrm{YM}}^2}\mathrm{tr} \big[\frac{1}{2}\bar{\psi}^L\g_{(i}{\oa{\D}}_{j)}\psi^L - \delta_{ij}\half\bar{\psi}^L\oa{\slashed{\D}}\psi^L \big], \label{Tpsi}
\end{align}
where $\oa{\D}_i \,=\, \stackrel{\rightarrow}{\D}_i -\stackrel{\leftarrow}{\D}_i$ and symmetrisation is performed with unit weight.
Note that the trace of the stress tensors for both conformally coupled scalars and for massless fermions vanish on shell.  This is a consequence of the Weyl invariance of the quadratic action for these fields (with the fields transforming non-trivially) when the action \eqref{Lfree} is appropriately coupled to gravity.

\section{Holographic formulae for cosmological 3-point functions} \label{sec:dict}

We review in this section the holographic formulae for cosmological 3-point functions derived in \cite{McFadden:2011kk}. These formulae relate the late-time
behaviour of the 3-point functions
of scalar  $\z$ and tensor $\bg_{ij}$
perturbations with correlation functions of the
stress tensor of the dual QFT. By late times we mean the end of the holographic epoch,
 which should be the beginning of hot big bang 
cosmology.\footnote{In particular, we will assume a smooth transition to hot big bang cosmology.
Developing a holographic theory of reheating is very interesting 
but will not be pursued here, see the comments at the end 
of Section 2 of \cite{Easther:2011wh} for a preliminary discussion.}
In subsequent sections we will use the hologragraphic
formulae to obtain the cosmological predictions for a universe
described holographically by a weakly coupled QFT.

More precisely, we consider 3-point functions involving
the curvature perturbation $\z$  on uniform energy density slices
and the transverse traceless tensor  $\bg_{ij}$
($\bg_{ii}=0$ and $\D_i\bg_{ij}=0$).
These variables are defined such that in comoving gauge, where the inflaton perturbation $\delta\vphi$ vanishes, the spatial part of the perturbed metric reads
\[ \label{pert-met}
g_{ij} = a^2 e^{2\z}[e^{\bg}]_{ij}=a^2 e^{2\z}(\delta_{ij}+\bg_{ij}+\frac{1}{2}\bg_{ik}\bg_{kj}).
\]
(For fully gauge-invariant expressions to quadratic order see Section 2.2 of \cite{McFadden:2011kk}.)
We find it useful to work in the helicity basis where
\[
\bg_{ij}(\q) = \bg^{(s)}(\q)\ep^{(s)}_{ij}(\q)
\]
and the helicity tensors $\ep^{(s)}_{ij}(\q)$ are summarised in
Appendix \ref{App_helicity}.

The holographic formulae for the in-in
3-point correlators are given by
\begin{align}
\label{holo_zzz}
&\<\!\<\z(q_1)\z(q_2)\z(q_3)\>\!\>  \nn\\[1ex]&\quad
 = -\frac{1}{256}\Big(\prod_i \Im [B(\bq_i)]\Big)^{-1}\times
\Im \Big[\<\!\<T(\bq_1)T(\bq_2)T(\bq_3)\>\!\> + 4\sum_i B(\bq_i) \nn\\[0ex]&\qquad\qquad\qquad\qquad\qquad\qquad\qquad\qquad
-2\Big( \<\!\<T(\bq_1)\Upsilon(\bq_2,\bq_3)\>\!\>+\mathrm{cyclic\,perms.}\Big)\Big], \\[2ex]
\label{holo_zzg}
&\<\!\<\z(q_1)\z(q_2)\bg^{(s_3)}(q_3)\>\!\>  \nn\\[1ex]& \quad
= -\frac{1}{32} \Big(\Im[B(\bq_1)]\Im[B(\bq_2)]\Im[A(\bq_3)]\Big)^{-1} \nn\\[1ex]& \qquad
\times \Im\Big[\<\!\<T(\bq_1)T(\bq_2)T^{(s_3)}(\bq_3)\>\!\> 
-2\big(\Theta_1^{(s_3)}(\bq_i)B(\bq_1)+\Theta_2^{(s_3)}(\bq_i)B(\bq_2)\big)  \nn\\[1ex]&\qquad\qquad\quad
-2\<\!\<\Upsilon(\bq_1,\bq_2)T^{(s_3)}(\bq_3)\>\!\>
-2\<\!\<T(\bq_1)\Upsilon^{(s_3)}(\bq_2,\bq_3)\>\!\>  
-2\<\!\<T(\bq_2)\Upsilon^{(s_3)}(\bq_1,\bq_3)\>\!\>
\Big],  \\[2ex]
\label{holo_zgg}
& \<\!\<\z(q_1)\bg^{(s_2)}(q_2)\bg^{(s_3)}(q_3)\>\!\> \nn\\[1ex]&\quad
= -\frac{1}{4}\Big(\Im[B(\bq_1)]\Im[A(\bq_2)]\Im[A(\bq_3)]\Big)^{-1} \nn\\[1ex]&\qquad
\times \Im\Big[\<\!\<T(\bq_1)T^{(s_2)}(\bq_2)T^{(s_3)}(\bq_3)\>\!\>
-\half \big(A(\bq_2)+A(\bq_3)\big)\theta^{(s_2s_3)}(\bq_i) 
-B(\bq_1) \Theta^{(s_2s_3)}(\bq_i)  \nn\\[1ex]&\qquad\quad
-2\<\!\<T(\bq_1)\Upsilon^{(s_2s_3)}(\bq_2,\bq_3)\>\!\> 
-2\<\!\<T^{(s_2)}(\bq_2)\Upsilon^{(s_3)}(\bq_1,\bq_3)\>\!\>
-2\<\!\<T^{(s_3)}(\bq_3)\Upsilon^{(s_2)}(\bq_1,\bq_2)\>\!\>
\Big], \\[2ex]
\label{holo_ggg}
&\<\!\<\bg^{(s_1)}(q_1)\bg^{(s_2)}(q_2)\bg^{(s_3)}(q_3)\>\!\> \nn\\[1ex]&\quad
= -\Big(\prod_i\Im[A(\bq_i)]\Big)^{-1}  
\times\Im \Big[2\<\!\<T^{(s_1)}(\bq_1)T^{(s_2)}(\bq_2)T^{(s_3)}(\bq_3)\>\!\> 
-\frac{1}{2}\Theta^{(s_1s_2s_3)}(\bq_i)\sum_i A(\bq_i) \nn\\[1ex]&\qquad \qquad\qquad\qquad \qquad\qquad\qquad
-4\Big(\<\!\<T^{(s_1)}(\bq_1)\Upsilon^{(s_2s_3)}(\bq_2,\bq_3)\>\!\> + \mathrm{cyclic\,perms.}\Big)
 \Big], 
\end{align}
where, as noted previously, the double bracket notation indicates correlators with the momentum-conserving delta function removed, {\it i.e.},
\begin{align}
\<\z(\q_1)\z(\q_2)\z(\q_3)\> = (2\pi)^3\delta(\sum\q_i)\<\!\<\z(q_1)\z(q_2)\z(q_3)\>\!\>,
\end{align}
and similarly for other correlators.
The coefficients $A(\bq_i)$ and $B(\bq_i)$
are related to 2-point function of the stress tensor (see (\ref{ABdef})),
while $T$ and $T^{(s)}$ are the trace and helicity-projected transverse traceless
part of $T_{ij}$, as defined in Appendix \ref{App_helicity}.  Similarly, $\Upsilon$, $\Upsilon^{(s)}$ and $\Upsilon^{(s_1 s_2)}$ are
the trace and helicity projections of the $\Upsilon$ tensor
(\ref{Upsilon_def}).  The theta functions, $\Theta_1^{(s)}$, $\Theta_2^{(s)}$,
$\Theta^{(s_1 s_2)}$, $\theta^{(s_1 s_2)}$ and $\Theta^{(s_1 s_2 s_3)}$,
represent specific contractions of the helicity tensors, and are described in Appendix \ref{App_helicity}.
The imaginary part in these formulae is taken after making the analytic continuation 
\[ \label{an_cont}
\bN = -iN, \qquad \bq_i=-iq_i.
\]
where $\bN$ is  the rank of the gauge group, see Section
\ref{sec:dual}, and $q=+\sqrt{\vec{q}\,^2}$ is the magnitude of the momentum.

The right-hand sides of (\ref{holo_zzz})-(\ref{holo_ggg})
were obtained by first deriving, using standard
gauge/gravity duality, the holographic 3-point functions of the stress
tensor along general holographic RG flows, either
asymptotically AdS or asymptotic to non-conformal brane
backgrounds\footnote{ The
non-conformal brane backgrounds are asymptotically AdS in the dual
frame \cite{Boonstra:1998mp}; 
the final formulae are however the same in both cases
\cite{McFadden:2010vh,McFadden:2011kk}.}
and then analytically continuing to the cosmological case.
These correlation functions are
in a flat background and are defined as usual (for example)
by the path integral formula
\[ \label{TQFT}
\< T_{i_1j_1}(\bq_1) 
\cdots
T_{i_nj_n}(\bq_n)\> = \int [d \varphi] T_{i_1j_1}(\bq_1)
\cdots T_{i_nj_n}(\bq_n) e^{-S_{QFT}[\varphi]}
\]
where $\varphi$ denotes collectively all fields of the boundary theory.
We work with the
Wick rotated QFT and correspondingly the bulk has Euclidean signature too.
This is convenient because bulk regularity in the interior
translates into the standard Bunch-Davies vacuum after analytic continuation
to cosmology \cite{McFadden:2009fg}. The holographic relations
(\ref{holo_zzz})-(\ref{holo_ggg}) may also be viewed as computing the
wavefunction of the universe extending 
the analysis of \cite{Maldacena:2002vr} (see also the recent \cite{Hertog:2011ky})
to a general class of FRW spacetimes.

Note that all terms appearing in the right-hand side 
numerators of (\ref{holo_zzz})-(\ref{holo_ggg}),
except for the 3-point functions,
are semi-local, {\it i.e.}, in position space two of the operators are
coincident (in momentum space this corresponds to the correlator
being non-analytic in only one of the three momenta). 
Due to the non-analytic powers of momenta appearing 
in the denominators of these formulae, however, 
semi-local terms in the numerator generate contributions to the bispectra
that are non-analytic in two momenta.
Hence, as discussed in \cite{McFadden:2010vh} in the case of the scalar bispectrum,
they may contribute, for example, to `local' type non-Gaussianity.
Note that these terms may be computed
unambiguously in perturbation theory (and we will do so in the
next section). In contrast, ultra-local terms,
{\it i.e.}, terms where all operators are coincident (or equivalently terms
analytic in all three momenta),
are in general scheme dependent since their value can be changed by
local finite counterterms.

One often defines the correlation function of the stress tensor by
first coupling the theory to a background metric $g_{ij}$,
differentiating w.r.t.~$g_{ij}$ and then setting  $g_{ij}=\delta_{ij}$.
One has to be careful, however, if one is to match with the
unambiguous expression in (\ref{TQFT}).
The 1-point function in the presence of the source $g_{ij}$ is defined by
\[ \label{defT}
\< T_{ij}(x) \>_g =
- \frac{2}{\sqrt{g(x)}} \frac{\delta}{\delta g^{ij}(x)}
\int [d \varphi] e^{-S_{QFT}[\varphi;g_{ij}]}
\]
Higher point functions are obtained by further functional differentiation
and then setting $g_{ij} = \delta_{ij}$. This procedure leads  to a
new insertion of the stress tensor when functional derivative
acts on $S_{QFT}[\varphi;g_{ij}]$, but also leads
to additional semi- and ulta-local terms. One source of such terms
are the factors of $1/\sqrt{g(x)}$ in (\ref{defT}) and for this reason some
authors (see for example \cite{Osborn:1993cr})
define the correlators without such factors, {\it i.e.},
\[ \label{osb}
< T_{i_1j_1}(x_1) \cdots T_{i_nj_n}(x_n)>  \equiv \left.
(-2)^n \frac{\delta}{\delta g^{i_1j_1}(x_1)} \cdots
\frac{\delta}{\delta g^{i_nj_n}(x_n)}
\int [d \varphi] e^{-S_{QFT}[\varphi;g_{ij}]} \right|_{g_{ij}=\delta_{ij}}
\]
Note however that these correlators differ from (\ref{TQFT}) (hence the different notation: $<T ...>$ instead of $\<T ... \>$) because
the stress tensor of the theory in a curved background also depends
on $g_{ij}$ and the functional differentiation leads to additional
insertions of
$\Upsilon_{ijkl}(\x_1,\x_2) = \delta T_{ij}(\x_1)/\delta g^{kl}(\x_2)\big|_{0}$.
A careful evaluation of all such semi-local terms is given in
Section 4.1 of \cite{McFadden:2011kk}. This is the origin of
most (but not all) semi-local terms in (\ref{holo_zzz})-(\ref{holo_ggg}).

To connect with holography let us now recall
that in gauge/gravity duality one identifies the fields parametrizing the
boundary conditions for the bulk fields with the sources of the dual
operators, and the (renormalised) on-shell gravitational action with the
generating functional of correlation functions
\cite{Gubser:1998bc,Witten:1998qj}. In particular, the boundary metric
$g_{(0)ij}$ in
Fefferman-Graham coordinates is identified with the source of stress
tensor \cite{deHaro:2000xn}.
One could also envision a holographic mapping where the source of the
stress tensor $g_{ij}$ is related to the boundary metric
$g_{(0)ij}$ via a local relation. Then, as was recently emphasized in
\cite{Maldacena:2011nz}, the two holographic maps would differ by (semi-)
local terms. Given any such relation one can straightforwardly work out the corresponding
holographic formulae, including all semi-local terms.
One can (at least partially) fix this potential ambiguity in the
holographic map by requiring that the (anomalous) Ward identities derived in
gravity and in QFT match. For example, the matching of the Weyl anomaly
in odd (bulk) dimensions \cite{Henningson:1998gx,Henningson:1998ey}
fixes $g_{ij}=g_{(0)ij}$. In even dimensions one would look to match the
semi-local terms in the dilatation Ward identity of higher point functions.
The formulae
(\ref{holo_zzz})-(\ref{holo_ggg}) we use here were derived using the standard
identification $g_{ij}=g_{(0)ij}$
and the definitions of $\zeta$ and $\hat{\gamma}_{ij}$ in (\ref{pert-met}) .

 \section{Evaluating the holographic formulae} \label{sec:eva}

In this section and throughout, we will work in Euclidean signature, and to leading order in $\geff^2$ and $1/\bar{N}$.
It will be useful to express our results in terms of the elementary symmetric polynomials of two and three variables, which we denote
\begin{align}
\label{sym_polys}
&\a_{123} = \bq_1+\bq_2+\bq_3, \qquad \b_{123} = \bq_1\bq_2+\bq_2\bq_3+\bq_3\bq_1, \qquad \c_{123}= \bq_1\bq_2\bq_3, \nn \\[1ex]
&\qquad\qquad\qquad\qquad \a_{12} = \bq_1+\bq_2, \qquad\qquad \b_{12} = \bq_1\bq_2,
\end{align}
with similar expressions for $\a_{23}$, $\b_{23}$, {\it etc}.
We also define
\begin{align}
\label{blambda_def}
 \blambda^2 &=(\bq_1+\bq_2+\bq_3)(-\bq_1+\bq_2+\bq_3)(\bq_1-\bq_2+\bq_3)(\bq_1+\bq_2-\bq_3) \nn\\
&= -\a_{123}(\a_{123}^3-4\a_{123}\b_{123}+8\c_{123}),
\end{align}
and the quantities
\[
\label{P_def}
 P_{ijkl} = 2\delta_{i(k}\delta_{l)j}-\delta_{ij}\delta_{kl}, \qquad
 \hat{P}_{ijkl} = \delta_{i(k}\delta_{l)j}-\delta_{ij}\delta_{kl}.
\]
Note that $\blambda$ is equal to 1/4 of the area of the triangle with
length sides equal to the momenta $\bq_1,\bq_2,\bq_3$ (Heron's formula).

\subsection{2-point functions}

In this subsection we recall the
contribution to the 2-point function \eqref{ABdef} from each of the individual fields
derived in \cite{McFadden:2009fg,McFadden:2010na}:
\begin{align}
\label{2ptfns}
A_\phi &= B_\phi=\frac{1}{256}\mathcal{N}_\phi \bN^2\bq^3, \qquad
A_\psi = \frac{1}{128}\mathcal{N}_\psi \bN^2 \bq^3, \qquad B_\psi = 0, \nn\\
A_A &= B_A=\frac{1}{256}\mathcal{N}_A \bN^2\bq^3, \qquad
A_\chi = \frac{1}{256}\mathcal{N}_\chi \bN^2\bq^3, \qquad B_\chi = 0,
\end{align}
and thus in total we have
\[
A = \frac{1}{256}\mathcal{N}_{(A)} \bN^2\bq^3, \qquad
B = \frac{1}{256}\mathcal{N}_{(B)} \bN^2\bq^3,
\]
where
\[
\label{NANBdef}
 \mathcal{N}_{(A)} =  \mathcal{N}_A + \mathcal{N}_\phi+\mathcal{N}_\chi+2\mathcal{N}_\psi, \qquad \mathcal{N}_{(B)} = \mathcal{N}_A + \mathcal{N}_\phi.
\]

\subsection{Contribution from minimal scalars}
\label{sec:min_scalars}

The 3-point function for minimal scalars is given by the integral
\begin{align}
\label{min_scalars_int}
& \<\!\<T^\phi_{ij}(\bq_1)T^\phi_{kl}(\bq_2)T^\phi_{mn}(\bq_3)\>\!\> \nn\\
&\qquad\qquad = \mathcal{N}_{\phi}\bN^2 P_{ijab}P_{klcd}P_{mnef}\int[\d \bq]\,\frac{\bq_a \bq_c (\bq-\bq_1)_b (\bq-\bq_1)_e (\bq+\bq_2)_d (\bq+\bq_2)_f}{\bq^2 (\bq-\bq_1)^2 (\bq+\bq_2)^2}.
\end{align}
Here, and throughout, we will make use of the shorthand notation
$
[\d \bq] = \d^3 \vec{\bq}/(2\pi)^{3}.
$
Evaluating this integral using the general methods discussed in Appendix \ref{methods_app}, we obtain
\begin{align}
\label{minscalars_3ptfns}
\<\!\<T_\phi(\bq_1)T_\phi(\bq_2)T_\phi^{(+)}(\bq_3)\>\!\>
&= -\frac{\mathcal{N}_\phi\bN^2\blambda^2}{1024\sqrt{2}\,\bq_3^2 \a_{123}^2}\,\big[3\a_{12}\bq_3^2 +2(3\a_{12}^2-4\b_{12})\bq_3+\a_{12}(3\a_{12}^2-4\b_{12}) \big], \nn\\[2ex]
\<\!\<T_\phi(\bq_1)T_\phi^{(+)}(\bq_2)T_\phi^{(+)}(\bq_3)\>\!\> &
 = -\frac{\mathcal{N}_\phi\bN^2}{8192\, \a_{123}^2 \b_{23}^2}\,(\bq_1-\a_{23})^2 \big[5\bq_1^7+20 \a_{23}\bq_1^6 +(29\a_{23}^2+6\b_{23})\bq_1^5
 \nn\\[1ex]&\quad
+ \a_{23}(17\a_{23}^2+21\b_{23})\bq_1^4
+3\a_{23}^2(\a_{23}^2+8\b_{23})\bq_1^3 +  2\a_{23}^3(\a_{23}^2+3\b_{23})\bq_1^2
\nn\\[1ex]
&\quad +(3\a_{23}^6-6\a_{23}^4\b_{23}-32\b_{23}^3)\bq_1+\a_{23}^5(\a_{23}^2-3\b_{23}) \big],  \nn\\[2ex]
\<\!\<T_\phi(\bq_1)T_\phi^{(+)}(\bq_2)T_\phi^{(-)}(\bq_3)\>\!\>
& = -\frac{\mathcal{N}_\phi\bN^2}{8192\, \b_{23}^2}\,(\bq_1^2-\a_{23}^2+4\b_{23})^2\big[5\bq_1^3-(\a_{23}^2+2\b_{23})\bq_1+\a_{23}(\a_{23}^2-3\b_{23})\big],
\nn\\[2ex]
\<\!\<T_\phi^{(+)}(\bq_1)T_\phi^{(+)}(\bq_2)T_\phi^{(+)}(\bq_3)\>\!\>
& =-\frac{\mathcal{N}_\phi\bN^2\blambda^2}{32768\sqrt{2}\, \a_{123}^4 \c_{123}^2}\,\big[3\a_{123}^9-7\a_{123}^7\b_{123}+5\a_{123}^6\c_{123}-64\c_{123}^3\big],
\nn\\[2ex]
\<\!\<T_\phi^{(+)}(\bq_1)T_\phi^{(+)}(\bq_2)T_\phi^{(-)}(\bq_3)\>\!\>
&=-\frac{\mathcal{N}_\phi\bN^2\blambda^2}{32768\sqrt{2}\,\a_{123}^2 \c_{123}^2}\,(\bq_3-\a_{12})^2\big[3\bq_3^5+4\a_{12}\bq_3^4+(\a_{12}^2-2\b_{12})\bq_3^3
\nn\\[1ex]&\quad
+\a_{12}(\a_{12}^2-3\b_{12})\bq_3^2
+4\a_{12}^2(\a_{12}^2-3\b_{12})\bq_3+\a_{12}^3(3\a_{12}^2-7\b_{12})\big].
\end{align}
All remaining 3-point functions for minimal scalars may be found from these via permutations and/or a parity transformation.
(The result for three insertions of the trace is given in \cite{McFadden:2010vh}.)

Turning now to evaluate the semi-local terms in the holographic formulae, for minimal scalars 
\[
\Upsilon^\phi_{ijkl}(\x_1,\x_2) = -\half (\delta_{ij}T^\phi_{kl}+P_{ijkl}T^\phi)\delta(\x_1-\x_2).
\]
We thus have
\[
\<\!\<T^\phi_{ij}(\bq_1)\Upsilon^\phi_{klmn}(\bq_2,\bq_3)\>\!\>=
-\half \delta_{kl}\<\!\<T^\phi_{ij}(\bq_1)T^\phi_{mn}(-\bq_1)\>\!\>-\half P_{klmn}\<\!\<T^\phi_{ij}(\bq_1)T^\phi(-\bq_1)\>\!\>,
\]
from which we may extract the helicity-projected components
\begin{align}
\<\!\<T_\phi(\bq_1)\Upsilon^{(s_3)}_\phi(\bq_2,\bq_3)\>\!\> &= -\frac{3}{2}B_\phi(\bq_1)\Theta_1^{(s_3)}(\bq_i), \nn\\[1ex]
\<\!\<T_\phi(\bq_1)\Upsilon^{(s_2 s_3)}_\phi(\bq_2,\bq_3)\>\!\> &= - B_\phi(\bq_1)\theta^{(s_2s_3)}(\bq_i), \nn\\[1ex]
\<\!\<T^{(s_1)}_\phi(\bq_1)\Upsilon_\phi(\bq_2,\bq_3)\>\!\> &= 0, \nn\\[1ex]
\<\!\<T^{(s_1)}_\phi(\bq_1)\Upsilon^{(s_3)}_\phi(\bq_2,\bq_3)\>\!\> &= -\frac{3}{8}A_\phi(\bq_1)\theta^{(s_1 s_3)}(\bq_i), \nn\\[1ex]
\<\!\<T^{(s_1)}_\phi(\bq_1)\Upsilon^{(s_2 s_3)}_\phi(\bq_2,\bq_3)\>\!\> &= 0.
\end{align}

\subsection{Contribution from fermions}

In momentum space
\[
 T^\psi_{ij}(\vbq_1) = \frac{1}{g_{\mathrm{YM}}^2}\mathrm{tr}\Big[\frac{i}{2}\hat{P}_{ijab}\int [\d\bq] (\bq_1-2\bq)_a :\!\bar{\psi}^L(\vbq)\g_{b}\psi^L(\vbq_1-\vbq)\!:\Big],
\]
from which it follows that the 3-point function is given by the integral

\begin{align}
& \<\!\<T^\psi_{ij}(\bq_1)T^\psi_{kl}(\bq_2)T^\psi_{mn}(\bq_3)\>\!\> \nn\\ &
= \frac{1}{4}\mathcal{N}_{\psi}\bN^2\hat{P}_{ijab}\hat{P}_{klcd}\hat{P}_{mnef}
\Gamma_{ubvfwd}
\int[\d \bq]\,\frac{\bq_u (\bq{-}\bq_1)_v (\bq{+}\bq_2)_w (2\bq{-}\bq_1)_a(2\bq{+}\bq_2)_c(2\bq{-}\bq_1{+}\bq_2)_e }{\bq^2 (\bq-\bq_1)^2 (\bq+\bq_2)^2},
\end{align}
where
\begin{align}
 \Gamma_{ubvfwd} &= \mathrm{tr}(\g_u\g_b\g_v\g_f\g_w\g_d) \nn\\ &= -2\delta_{ub}P_{vwdf}+2\delta_{uv}P_{bwdf}-2\delta_{uf}P_{bwvd}+2\delta_{uw}P_{bfvd}
-2\delta_{ud}P_{bfvw},
\end{align}
recalling that $\{\gamma_i,\gamma_j\}=-2 \delta_{ij}$.

Evaluating the integral explicitly, we find\footnote{\label{qual} The result for the momentum dependence of the
$(+++)$ and $(++-)$ components agrees with the result
reported in \cite{Maldacena:2011nz}, but the overall normalisation
(after taking into account the difference in conventions, see Appendix
\ref{App_helicity}) still differs from ours by a factor of four. Actually all correlators
in \cite{Maldacena:2011nz}, including the 2-point functions, differ from
ours by the same overall factor of four.}
\begin{align}
\label{fermion_3ptfns}
 \<\!\<T_\psi(\bq_1)T_\psi(\bq_2)T_\psi^{(+)}(\bq_3)\>\!\> &=  0,
\nn\\[2ex]
\<\!\<T_\psi(\bq_1)T_\psi^{(+)}(\bq_2)T_\psi^{(+)}(\bq_3)\>\!\> &= -\frac{\mathcal{N}_\psi\bN^2}{2048\, \b_{23}^2}\,\a_{23}(\a_{23}^2-3\b_{23})(\bq_1^2-\a_{23}^2)^2,
\nn\\[2ex]
\<\!\<T_\psi(\bq_1)T_\psi^{(+)}(\bq_2)T_\psi^{(-)}(\bq_3)\>\!\> &= -\frac{\mathcal{N}_\psi\bN^2}{2048\, \b_{23}^2}\,\a_{23}(\a_{23}^2-3\b_{23})(\bq_1^2-\a_{23}^2+4\b_{23})^2,
\nn\\[2ex]
\<\!\<T_\psi^{(+)}(\bq_1)T_\psi^{(+)}(\bq_2)T_\psi^{(+)}(\bq_3)\>\!\>
&=-\frac{\mathcal{N}_\psi\bN^2\blambda^2}{8192\sqrt{2}\,\a_{123}^4 \c_{123}^2}\,\big[\a_{123}^9-2\a_{123}^7\b_{123}+\a_{123}^6\c_{123}+32\c_{123}^3\big],
\nn\\[2ex]
\<\!\<T_\psi^{(+)}(\bq_1)T_\psi^{(+)}(\bq_2)T_\psi^{(-)}(\bq_3)\>\!\> &
= -\frac{\mathcal{N}_\psi\bN^2\blambda^2}{8192\sqrt{2}\,\a_{123}^2 \c_{123}^2}\,(\bq_3-\a_{12})^2\big[\bq_3^5 + \a_{12}\bq_3^4-\b_{12}\bq_3^3
\nn\\[1ex]&\qquad\qquad\qquad
+\a_{12}^2(\a_{12}^2-3\b_{12})\bq_3
 +\a_{12}^3(\a_{12}^2-2\b_{12}) \big].&
\end{align}

The correlators with only one trace may be written in the condensed form
\[
\label{proto_Ward1}
\<\!\<T_\psi(\bq_1)T_\psi^{(s_2)}(\bq_2)T_\psi^{(s_3)}(\bq_3)\>\!\>  =-\frac{1}{2}\big(A_\psi(\bq_2)+A_\psi(\bq_3)\big)\theta^{(s_2 s_3)}(\bq_i).
\]
This result is in fact fully determined by the dilatation Ward identity (accounting for its semi-local nature), as
we discuss in the next section. 

To compute the semi-local terms appearing in the holographic formulae, we find by explicit calculation that the operator
\[
\label{Upexppsi}
\Upsilon^{\psi}_{ijkl}(\x_1,\x_2) = C^{(\mathcal{M})}_{ijklmn}\mathcal{M}_{mn}(\vec{x}_1)\delta(\vec{x}_1-\vec{x}_2) +C^{(\mathcal{J})}_{ijklm}\mathcal{J}_{m}(\vec{x}_1)\D_n\delta(\vec{x}_1-\vec{x}_2),
\]
where partial derivatives are taken with respect to $\vec{x}_1$, and the local operators
\[
\mathcal{M}_{mn} =  \frac{1}{g_{\mathrm{YM}}^2}\mathrm{tr}\big[\frac{1}{2}\bar{\psi}^L\g_{m}{\oa{\D}}_n\psi^L\big], \qquad
\mathcal{J}_m = \frac{1}{g_{\mathrm{YM}}^2}\mathrm{tr}\big[\frac{1}{4}\bar{\psi}^L\g_m\psi^L\big],
\]
are associated with the coefficients
\begin{align}
C^{(\mathcal{M})}_{ijklmn}&=\delta_{i(k}\delta_{l)j}\delta_{mn}-\half\delta_{ij}\delta_{m(k}\delta_{l)n}-\half\delta_{m(k}\delta_{l)(i}\delta_{j)n},\nn\\[1ex]
C^{(\mathcal{J})}_{ijklm} &= \delta_{i(k}\delta_{l)j}\delta_{mn}
+\delta_{ij}\delta_{m(k}\delta_{l)n}
 -\delta_{ij}\delta_{kl}\delta_{mn}
-\delta_{m(k}\delta_{l)(i}\delta_{j)n}.
\end{align}
As might be anticipated from their respective conformal dimensions (see Section \ref{sec:Wid}),
\[
\<T^\psi_{kl}\mathcal{M}^{\,}_{mn}\> = \<T^\psi_{kl}T^\psi_{mn}\>,\qquad \<T^\psi_{kl}\mathcal{J}^{\,}_m\>=0,
\]
from which it follows that
\[
\label{TUppsi}
 \<\!\<T^\psi_{ij}(\bq_1)\Upsilon^\psi_{klmn}(\bq_2,\bq_3)\>\!\> = C^{(\mathcal{M})}_{klmnab} \<\!\<T^\psi_{ij}(\bq_1)T^\psi_{ab}(-\bq_1)\>\!\>.
\]
Projecting into the helicity basis, the components appearing in the holographic formulae are
\begin{align}
\label{contact_fermions}
\<\!\<T_\psi(\bq_1)\Upsilon^{(s_3)}_\psi(\bq_2,\bq_3)\>\!\> &= 0, \nn\\[1ex]
\<\!\<T_\psi(\bq_1)\Upsilon^{(s_2 s_3)}_\psi(\bq_2,\bq_3)\>\!\> &= 0, \nn\\[1ex]
\<\!\<T^{(s_1)}_\psi(\bq_1)\Upsilon_\psi(\bq_2,\bq_3)\>\!\> &= 0, \nn\\[1ex]
\<\!\<T^{(s_1)}_\psi(\bq_1)\Upsilon^{(s_3)}_\psi(\bq_2,\bq_3)\>\!\> &=  -\half A_\psi(\bq_1)\theta^{(s_1s_3)}(\bq_i), \nn\\[1ex]
\<\!\<T^{(s_1)}_\psi(\bq_1)\Upsilon^{(s_2 s_3)}_\psi(\bq_2,\bq_3)\>\!\> &= -\frac{1}{16}A_\psi(\bq_1)\Theta^{(s_1s_2s_3)}(\bq_i).
\end{align}

\subsection{Contribution from conformal scalars} \label{conf_sc}

As discussed in the introduction, the stress tensor $T^\phi_{ij}$ for minimal scalars may be decomposed as $T^\phi_{ij}=\wT^\phi_{ij}+\mathcal{C}_{ij}$, where $\wT^\phi_{ij}$ is the stress tensor for conformal scalars
and $\mathcal{C}_{ij}$ is an improvement term.  For fields in the adjoint representation, and in momentum space,
the improvement term takes the form
\[
 \mathcal{C}_{ij}(\vbq_1)= \frac{1}{g_\mathrm{YM}^2}\mathrm{tr}\Big[ \,\frac{1}{8}\,\bq_1^2 \pi_{ij}(\bq_1)\int[\d\bq]:\!\phi^J(\vbq)\phi^J(\vbq_1-\vbq)\!:\Big].
\]
Due to the presence of the projection operator $\pi_{ij}$, it follows that $T_\phi^{(s)}(\vbq) = \wT_\phi^{(s)}(\vbq)$ and hence
the conformal scalar 3-point function involving three helicities is equal to that for minimal scalars.
Similarly, the correlator
\[
\<\!\<\wT_\phi(\bq_1)\wT^{(s_2)}_\phi(\bq_2)\wT^{(s_3)}_\phi(\bq_3)\>\!\> =
\<\!\< T_\phi(\bq_1)T^{(s_2)}_\phi(\bq_2) T^{(s_3)}_\phi(\bq_3)\>\!\> -
\<\!\<\mathcal{C}(\bq_1)T^{(s_2)}_\phi(\bq_2)T^{(s_3)}_\phi(\bq_3)\>\!\>,
\]
where the latter term may be evaluated from the integral
\[
\<\!\<\mathcal{C}(\bq_1)T^\phi_{kl}(\bq_2) T^\phi_{mn}(\bq_3)\>\!\> = \half \mathcal{N}_\phi\bN^2 \bq_1^2 P_{klab}P_{mncd}\int[\d\bq]\frac{\bq_a(\bq+\bq_2)_b(\bq-\bq_1)_c(\bq+\bq_2)_d}{\bq^2(\bq-\bq_1)^2(\bq+\bq_2)^2}.
\]
Thus, to evaluate the conformal scalar 3-point function involving two helicities,  only this integral needs to be computed 
since we already have the result for minimal scalars.
Finally, evaluating the trace $\wT_\phi(\vbq)$ directly, it is straightforward to show that the conformal scalar 3-point function involving only one helicity vanishes.

In light of these considerations, the 3-point functions for the conformal scalar field $\chi$ are\footnote{The result for momentum dependence of the $(+++)$ and $(++-)$ components agrees with the result
reported in v2 of \cite{Maldacena:2011nz}, but the overall normalisation
(after taking into account the difference in conventions, see Appendix
\ref{App_helicity}) still differs from ours by a factor of four,
see also footnote \ref{qual}.}
\begin{align}
\label{confscalars_3ptfns}
\<\!\<T_\chi(\bq_1)T_\chi(\bq_2)T_\chi^{(s_3)}(\bq_3)\>\!\> &= 0, \nn\\[1ex]
\<\!\<T_\chi(\bq_1)T_\chi^{(s_2)}(\bq_2)T_\chi^{(s_3)}(\bq_3)\>\!\> &=
\frac{1}{4}\frac{\mathcal{N}_\chi}{\mathcal{N}_\psi}\,  \<\!\<T_\psi(\bq_1)T_\psi^{(s_2)}(\bq_2)T_\psi^{(s_3)}(\bq_3)\>\!\>, \nn\\[1ex]
\<\!\<T_\chi^{(s_1)}(\bq_1)T_\chi^{(s_2)}(\bq_2)T_\chi^{(s_3)}(\bq_3)\>\!\> &=
\frac{\mathcal{N}_\chi}{\mathcal{N}_\phi} \,\<\!\<T_\phi^{(s_1)}(\bq_1)T_\phi^{(s_2)}(\bq_2)T_\phi^{(s_3)}(\bq_3)\>\!\>.
\end{align}

Turning now to the semi-local terms in the holographic formulae, by direct calculation 
\begin{align}
\label{Upexpchi}
\Upsilon^\chi_{ijkl}(\x_1,\x_2) &= -\half\big(\delta_{ij}T^\chi_{kl}+P_{ijkl}T^\chi\big)\delta(\x_1-\x_2) \nn\\&\quad
+\frac{1}{16}\Big[C^{(1)}_{ijklmn}\delta(\x_1-\x_2)\D_m\D_n + C^{(2)}_{ijklmn}(\D_m\delta(\x_1-\x_2))\D_n \nn\\&\qquad\qquad +C^{(3)}_{ijklmn}(\D_m\D_n\delta(\x_1-\x_2))\Big] \O^\chi(\x_1), 
\end{align}
where partial derivatives are again taken with respect to $\x_1$, the dimension one operator
\[
\O^\chi = \frac{1}{g_{\mathrm{YM}}^2}\tr [(\chi^K)^2],
\]
and the prefactors are
\begin{align}
C^{(1)}_{ijklmn} &= \delta_{ij}\delta_{k(m}\delta_{n)l}+2\delta_{i(k}\delta_{l)j}\delta_{mn}-\delta_{ij}\delta_{kl}\delta_{mn},\nn\\
C^{(2)}_{ijklmn} &=2\delta_{ij}\delta_{k(m}\delta_{n)l}+ \delta_{i(k}\delta_{l)j}\delta_{mn}-\delta_{ij}\delta_{kl}\delta_{mn}
-2\delta_{m(i}\delta_{j)(k}\delta_{l)n},\nn\\
C^{(3)}_{ijklmn} &= \delta_{ij}\delta_{k(m}\delta_{n)l}+ \delta_{i(k}\delta_{l)j}\delta_{mn}-\delta_{ij}\delta_{kl}\delta_{mn}
-2\delta_{m(i}\delta_{j)(k}\delta_{l)n}+\delta_{kl}\delta_{i(m}\delta_{n)j}.
\end{align}
The precise form of these prefactors is not important, however, since
\[
\<T^\chi_{ij} \O^\chi \> = 0
\]
due to the differing conformal dimension of the two operators, hence
\[
\label{TUpchi}
\<\!\<T^\chi_{ij}(\bq_1)\Upsilon^\chi_{klmn}(\bq_2,\bq_3)\>\!\> = -\half \delta_{kl}\<\!\<T^\chi_{ij}(\bq_1)T^\chi_{mn}(-\bq_1)\>\!\>.
\]
The helicity-projected components appearing in the holographic formulae are then
\begin{align}
\label{contact_confscalars}
\<\!\<T_\chi(\bq_1)\Upsilon^{(s_3)}_\chi(\bq_2,\bq_3)\>\!\> &= 0, \nn\\[1ex]
\<\!\<T_\chi(\bq_1)\Upsilon^{(s_2 s_3)}_\chi(\bq_2,\bq_3)\>\!\> &= 0, \nn\\[1ex]
\<\!\<T^{(s_1)}_\chi(\bq_1)\Upsilon_\chi(\bq_2,\bq_3)\>\!\> &= 0, \nn\\[1ex]
\<\!\<T^{(s_1)}_\chi(\bq_1)\Upsilon^{(s_3)}_\chi(\bq_2,\bq_3)\>\!\> &=  -\frac{3}{8} A_\chi(\bq_1)\theta^{(s_1s_3)}(\bq_i), \nn\\[1ex]
\<\!\<T^{(s_1)}_\chi(\bq_1)\Upsilon^{(s_2 s_3)}_\chi(\bq_2,\bq_3)\>\!\> &= 0.
\end{align}

\subsection{Contribution from gauge fields}

The stress tensor for the gauge fields is given by (\ref{Ta}) plus the
contribution due to the ghosts and gauge-fixing terms. The latter
contributions are BRST exact however, and thus they should not contribute
to the 3-point function. Indeed, we show in Appendix \ref{App_ghosts} that
their contribution cancels.

Introducing the Hodge-dual field strength $G_i^{I}$, where
\begin{equation}
F^{I}_{ij} = \epsilon_{ijk} G_k^{I}, \qquad \qquad T_{ij}^A = \frac{1}{g^2_{\mathrm{YM}}}P_{ijab} \tr G_a^I G_b^I,
\end{equation}
and evaluating its propagator, one finds
the contribution from gauge fields to the 3-point function is given by
\[
\<\!\<T^A_{ij}(\bq_1)T^A_{kl}(\bq_2)T^A_{mn}(\bq_3)\>\!\> = -\mathcal{N}_A \bN^2 P_{ijab}P_{klcd}P_{mnef}\int [\d\bq]\pi_{ac}(\bq)\pi_{be}(\bq-\bq_1)\pi_{df}(\bq+\bq_2).
\]
Upon closer examination, this integral may equivalently be expressed in terms of the 2- and 3-point functions for minimal scalars,

\begin{align}
\label{gauge3ptfn}
\frac{\mathcal{N}_\phi}{\mathcal{N}_A}
\<\!\<T^A_{ij}(\bq_1)T^A_{kl}(\bq_2)T^A_{mn}(\bq_3)\>\!\> &=
\<\!\<T^\phi_{ij}(\bq_1)T^\phi_{kl}(\bq_2)T^\phi_{mn}(\bq_3)\>\!\> - Q_{klmnab}\<\!\<T^\phi_{ij}(\bq_1)T^\phi_{ab}(-\bq_1)\>\!\> \nn\\
&\quad - Q_{mnijab}\<\!\<T^\phi_{kl}(\bq_2)T^\phi_{ab}(-\bq_2)\>\!\>- Q_{ijklab}\<\!\<T^\phi_{mn}(\bq_3)T^\phi_{ab}(-\bq_3)\>\!\>,
\end{align}
where
\[
\label{Odef}
 Q_{ijklmn} = P_{ijac}P_{klbc}\hat{P}_{abmn}.
\]
This result is a consequence of the fact that $G_i$ may be identified with the operator $\partial_i \phi$, where $\phi$ is a massless scalar field. The appearance of the various semi-local terms in \eqref{gauge3ptfn} then reflects the fact that $\partial_i G_i$ vanishes identically, while $\partial^2 \phi$ vanishes on-shell only.

Turning now to evaluate the semi-local terms appearing in the holographic formulae, a short calculation reveals the operator
\begin{align}
 \Upsilon^A_{ijkl}(\x_1,\x_2) &
= -\half \Big[\delta_{ij}T^A_{kl}+P_{ijkl}T^A  + Q_{ijklmn}T^A_{mn}\Big]\delta(\x_1-\x_2).
\end{align}
Making use of the fact that the 2-point functions for gauge fields and for minimal scalars coincide (see \eqref{2ptfns}),
it then follows that
\[
\frac{\mathcal{N}_\phi}{\mathcal{N}_A} \<\!\<T^A_{ij}(\bq_1)\Upsilon^A_{klmn}(\bq_2,\bq_3)\>\!\> = \<\!\<T^\phi_{ij}(\bq_1)\Upsilon^\phi_{klmn}(\bq_2,\bq_3)\>\!\> -\half Q_{klmnab}\<\!\<T^\phi_{ij}(\bq_1)T^\phi_{ab}(-\bq_1)\>\!\>.
\]
Thus, from \eqref{gauge3ptfn},
\begin{align}
 & \frac{\mathcal{N}_\phi}{\mathcal{N}_A}
\Big[ \<\!\<T^A_{ij}(\bq_1)T^A_{kl}(\bq_2)T^A_{mn}(\bq_3)\>\!\>
- 2\Big(\<\!\<T^A_{ij}(\bq_1)\Upsilon^A_{klmn}(\bq_2,\bq_3)\>\!\> + \mathrm{cyclic\,\,perms}\Big)\Big] \nn\\[1ex]&\qquad
=  \<\!\<T^\phi_{ij}(\bq_1)T^\phi_{kl}(\bq_2)T^\phi_{mn}(\bq_3)\>\!\>
- 2\Big(\<\!\<T^\phi_{ij}(\bq_1)\Upsilon^\phi_{klmn}(\bq_2,\bq_3)\>\!\> + \mathrm{cyclic\,\,perms}\Big).
\end{align}
This particular combination, suitably projected, appears in the numerator of all the holographic formulae for cosmological 3-point functions (namely \eqref{holo_zzg}, \eqref{holo_zgg} and \eqref{holo_ggg}).  Thus, since the 2-point functions for gauge fields and minimal scalars also coincide, we see that gauge fields and minimal scalars necessarily make {\it identical} contributions to all cosmological 3-point functions.
Since scalars and vectors are dual in three dimensions, this result is perhaps not unexpected, and indeed similar behaviour was noted in \cite{McFadden:2010vh} for the case of the scalar bispectrum.

\section{Ward identities} \label{sec:Wid}

In the previous section we computed all relevant 3-point functions and semi-local terms
by direct computation of 1-loop Feynman diagrams. In this section
we elucidate the structure of these correlators by ascertaining the
extent to which they are determined by Ward identities.

\subsection{Minimal scalars from conformal scalars}
\label{sec:phifromchi}

As noted previously, the stress tensor for minimal scalars may be decomposed as
\[
T^\phi_{ij} = \wT^\phi_{ij} -\frac{1}{8}(\delta_{ij}\p^2-\p_i\p_j) \O_1, \qquad \O_1 =\frac{1}{g_{\mathrm{YM}}^2}\tr[ (\phi^J)^2],
\]
where $\wT^\phi_{ij}$ is the stress tensor for a conformal scalar field and $\O_1$ is a dimension one scalar operator.
The 3-point functions of $T^\phi_{ij}$ may thus be expressed in terms of 3-point functions of the conformal fields $\wT^\phi_{ij}$ and $\O_1$.
Specifically, we find
\begin{align}
\label{phifromchi}
\lla T^{(s_1)}_\phi(\bq_1) T^{(s_2)}_\phi(\bq_2) T^{(s_3)}_\phi (\bq_3)\rra
&=  \lla \wT^{(s_1)}_\phi(\bq_1) \wT^{(s_2)}_\phi(\bq_2)\wT^{(s_3)}_\phi(\bq_3) \rra\, , \nn\\[1ex]
\lla T_\phi(\bar{q}_1) T^{(s_2)}_\phi(\bar{q}_2) T^{(s_3)}_\phi(\bar{q}_3) \rra & =  \lla \wT_\phi(\bar{q}_1) \wT^{(s_2)}_\phi(\bar{q}_2) \wT^{(s_3)}_\phi(\bar{q}_3) \rra + \frac{\bar{q}_1^2}{4} \lla {\cal O}_1(\bar{q}_1) \wT^{(s_2)}_\phi(\bar{q}_2) \wT^{(s_3)}_\phi(\bar{q}_3) \rra\, , \nn\\[1ex]
\lla T_\phi(\bar{q}_1) T_\phi(\bar{q}_2) T^{(s_3)}_\phi(\bar{q}_3) \rra & =  \lla \wT_\phi(\bar{q}_1) \wT_\phi(\bar{q}_2) \wT^{(s_3)}_\phi(\bar{q}_3) \rra + \frac{\bar{q}_1^2}{4} \lla {\cal O}_1(\bar{q}_1) \wT_\phi(\bar{q}_2) \wT^{(s_3)}_\phi(\bar{q}_3)
\rra \nonumber \\
& \quad+ \: \frac{\bar{q}_2^2}{4}
\lla \wT_\phi(\bar{q}_1) {\cal O}_1(\bar{q}_2) \wT^{(s_3)}_\phi(\bar{q}_3) \rra
+ \frac{\bar{q}_1^2 \bar{q}_2^2}{16}
\lla {\cal O}_1(\bar{q}_1) {\cal O}_1(\bar{q}_2) \wT^{(s_3)}_\phi(\bar{q}_3) \rra ,\nonumber \\[1ex]
\lla T_\phi(\bar{q}_1) T_\phi(\bar{q}_2) T_\phi(\bar{q}_3) \rra & =  \lla \wT_\phi(\bar{q}_1) \wT_\phi(\bar{q}_2) \wT_\phi(\bar{q}_3) \rra + \left[ \frac{\bar{q}_1^2}{4} \lla {\cal O}_1(\bar{q}_1) \wT_\phi(\bar{q}_2) \wT_\phi(\bar{q}_3) \rra + 2 \text{ perm. } \right] \nonumber  \\
&\quad + \: \left[ \frac{\bar{q}_1^2 \bar{q}_2^2}{16}
\lla {\cal O}_1(\bar{q}_1) {\cal O}_1(\bar{q}_2) \wT_\phi(\bar{q}_3) \rra + 2 \text{ perm. } \right]
\nonumber \\&\quad
 + \: \frac{\bar{q}_1^2 \bar{q}_2^2 \bar{q}_3^2}{64}
\lla {\cal O}_1(\bar{q}_1) {\cal O}_1(\bar{q}_2) {\cal O}_1(\bar{q}_3) \rra.
\end{align}
Recalling that gauge fields contribute the same as minimal scalars, the computation of general
3-point functions thus reduces to computing a set of 2- and 3-point functions in a CFT.
(Note that free fermions are also a CFT.)
These correlators are in turn (almost) uniquely determined by Ward identities, as we now show.

\subsection{Trace Ward identity}

In light of the above, we are interested in correlation functions of the stress tensor and
of a scalar operator $\mathcal{O}_\Delta$ of dimension $\Delta=1$ in a three-dimensional CFT. We will be more general however
and discuss the case of any $d$ and
$\Delta$, provided only that $\Delta\neq d$. 
The trace Ward identity in the
presence of a source $\phi_0$ for  $\mathcal{O}_\Delta$ reads
\begin{equation} \label{e:wardT}
\langle T(x) \rangle_s = (\Delta - d) \phi_0 \langle \mathcal{O}_\Delta(x) \rangle_s.
\end{equation}
This Ward identity implies that $n$-point functions involving an insertion
of the trace of the stress tensor are given by semi-local terms
involving $(n-1)$-point functions. These relations can be obtained
by functionally differentiating (\ref{e:wardT}) w.r.t.~the sources
$(n-1)$ times
and then setting them to zero.
Noting that all 1-point functions vanish, for the 2-point functions we find
\[
\label{2ptWds}
\<T(x)\O_\Delta(y)\>=0,\qquad \<T(x)T_{kl}(y)\>=0,
\]
and for the 3-point functions,
\begin{align}
\langle T(x) \mathcal{O}_\Delta(y) \mathcal{O}_\Delta(z) \rangle & =  \langle \frac{\delta T(x)}{\delta \phi_0(y)} \mathcal{O}_\Delta(z) \rangle + \langle \frac{\delta T(x)}{\delta \phi_0(z)} \mathcal{O}_\Delta(y) \rangle
\nonumber \\[0ex]
&\quad + \: (d - \Delta) \left[ \delta(x - y) \langle \mathcal{O}_\Delta(x) \mathcal{O}_\Delta(z) \rangle + \delta(x - z) \langle \mathcal{O}_\Delta(x) \mathcal{O}_\Delta(y) \rangle \right],
\label{TOO} \\[2ex]
\langle T(x) T_{kl}(y) \mathcal{O}_\Delta(z) \rangle & =
2 \langle \frac{\delta T(x)}{\delta g^{kl}(y)}
\mathcal{O}_\Delta(z) \rangle + \langle \frac{\delta T(x)}{\delta \phi_0(z)} T_{kl}(y) \rangle + \langle T(x) \frac{\delta T_{kl}(y)}{\delta \phi_0(z)} \rangle
\nonumber \\[0ex]
&\qquad
+ (d - \Delta) \delta(x - z) \langle T_{kl}(y) \mathcal{O}_\Delta(x) \rangle,
\label{TTO} \\[2ex]
\langle T(x) T_{kl}(y) T_{mn}(z) \rangle & =
 2 \langle \frac{\delta T(x)}{\delta g^{kl}(y)} T_{mn}(z) \rangle + 2 \langle \frac{\delta T(x)}{\delta g^{mn}(z)} T_{kl}(y) \rangle + 2 \langle T(x) \frac{\delta T_{kl}(y)}{\delta g^{mn}(z)} \rangle.  
\label{TTT}
\end{align}
As the stress tensor in the presence of sources has $\phi_0$ dependence
\begin{equation}
T_{ij}[g,\phi_0] = T_{ij}[g,\phi_0=0] - g_{ij} \phi_0 \mathcal{O}_\Delta,
\end{equation}
we may in addition identify
\begin{equation}
\frac{\delta T(x)}{\delta \phi_0(y)} = - d \delta(x - y) \mathcal{O}_\Delta(x).
\end{equation}
Then, since the CFT correlator $\langle T_{ij} \mathcal{O}_\Delta \rangle$ vanishes for any operator $\mathcal{O}_\Delta$ with dimension different to $d$, equations \eqref{TOO} and \eqref{TTO} reduce to
\begin{eqnarray}
\langle T(x) \mathcal{O}_\Delta(y) \mathcal{O}_\Delta(z) \rangle & = & - \Delta \left[ \delta(x - y) \langle \mathcal{O}_\Delta(x) \mathcal{O}_\Delta(z) \rangle + \delta(x - z) \langle \mathcal{O}_\Delta(x) \mathcal{O}_\Delta(y) \rangle \right], \label{WI:TOO} \\[1ex]
\langle T(x) T_{kl}(y) \mathcal{O}_\Delta(z) \rangle & = &
 2 \langle \frac{\delta T(x)}{\delta g^{kl}(y)} \mathcal{O}_\Delta(z) \rangle.
\label{WI:TTO}
\end{eqnarray}
Finally, it is convenient to express \eqref{TTT}, \eqref{WI:TOO} and \eqref{WI:TTO} in momentum space in terms of the $\Upsilon$ tensor defined in \eqref{Upsilon_def}.
Projecting into the helicity basis, we obtain the complete set of trace Ward identities
\begin{align}
\<\!\<T(\bq_1)\mathcal{O}_\Delta(\bq_2)\mathcal{O}_\Delta(\bq_3)\>\!\> &= -\Delta \big[\<\!\<\O_\Delta(\bq_2)\O_\Delta(-\bq_2)\>\!\>+\<\!\<\O_\Delta(\bq_3)\O_\Delta(-\bq_3)\>\!\>\big],\nn\\[1ex]
\<\!\<T(\bq_1)T(\bq_2)\O_\Delta(\bq_3)\>\!\> &= 2\<\!\<\Upsilon(\bq_1,\bq_2)\O_\Delta(\bq_3)\>\!\>,\nn\\[1ex]
\<\!\<T(\bq_1)T^{(s_2)}(\bq_2)\O_\Delta(\bq_3)\>\!\> &= 2\<\!\<\Upsilon^{(s_2)}(\bq_1,\bq_2)\O_\Delta(\bq_3)\>\!\>, \nn\\[1ex]
\<\!\<T(\bq_1)T(\bq_2)T(\bq_3)\>\!\> &= 2\big[\<\!\<T(\bq_1)\Upsilon(\bq_2,\bq_3)\>\!\>+\<\!\<T(\bq_2)\Upsilon(\bq_3,\bq_1)\>\!\>+
\<\!\<T(\bq_3)\Upsilon(\bq_1,\bq_2)\>\!\>\big], \nn\\[1ex]
 \<\!\<T(\bq_1)T(\bq_2)T^{(s_3)}(\bq_3)\>\!\>
&= 2\big[\<\!\<T(\bq_1)\Upsilon^{(s_3)}(\bq_2,\bq_3)\>\!\>
+ \<\!\<T(\bq_2)\Upsilon^{(s_3)}(\bq_1,\bq_3)\>\!\> \nn\\&\quad
+ \<\!\<\Upsilon(\bq_1,\bq_2)T^{(s_3)}(\bq_3)\>\!\>\big], \nn\\[0ex]
 \<\!\<T(\bq_1)T^{(s_2)}(\bq_2)T^{(s_3)}(\bq_3)\>\!\>
&= \frac{1}{2}\big(A(\bq_2)+A(\bq_3)\big)\theta^{(s_2 s_3)}(\bq_i)
+ 2\big[\<\!\<T(\bq_1)\Upsilon^{(s_2 s_3)}(\bq_2,\bq_3)\>\!\> \nn\\&\quad
+\<\!\<T^{(s_2)}(\bq_2)\Upsilon^{(s_3)}(\bq_1,\bq_3)\>\!\>
+\<\!\<T^{(s_3)}(\bq_3)\Upsilon^{(s_2)}(\bq_1,\bq_2)\>\!\>\big],
\label{traceWards}
\end{align}
where $A(\bq)$ is the transverse traceless piece of the stress tensor 2-point function defined in \eqref{ABdef} (for conformal fields the trace piece $B(\bq)$ vanishes as a consequence of \eqref{2ptWds}).

Comparing with our holographic formulae \eqref{holo_zzz}, \eqref{holo_zzg} and \eqref{holo_zgg}, we immediately see that conformal fields make no contribution to the numerators of these formulae, as found earlier by explicit calculation.
An important consequence of this, as we will see in Section \ref{sec:HolPredictions}, is that the $\z\z\z$, $\z\z\bg$ and $\z\bg\bg$ cosmological shape functions are forced to be independent of the field content of the dual QFT.

Further insight may be distilled from the trace Ward identities \eqref{traceWards} by replacing the semi-local contact terms on the r.h.s.~with 2-point functions of $T_{ij}$ and $\mathcal{O}_\Delta$.
On general grounds, the $\Upsilon$ tensor has an expansion
in terms of local operators of dimension less than or equal to $d$,
and for fermions and conformal scalars we computed this explicitly in \eqref{Upexppsi} and \eqref{Upexpchi}.
Then, as we found in the analysis leading to \eqref{TUppsi} and \eqref{TUpchi},
only operators of dimension $d$ contribute to the correlator $\<T_{ij}\Upsilon_{klmn}\>$, permitting it to be expressed in terms of $\<T_{ij}T_{kl}\>$.
Substituting into \eqref{traceWards} our previous results \eqref{contact_fermions} and \eqref{contact_confscalars} for the semi-local terms $\<T_{ij}\Upsilon_{klmn}\>$, we obtain
\begin{align}
\label{TTTresults}
\<\!\<T_\chi(\bq_1)T_\chi(\bq_2)T_\chi^{(s_3)}(\bq_3)\>\!\> &=
\<\!\<T_\psi(\bq_1)T_\psi (\bq_2)T_\psi^{(s_3)}(\bq_3)\>\!\> = 0, \nn\\
\<\!\<T_\chi(\bq_1)T_\chi^{(s_2)}(\bq_2)T_\chi^{(s_3)}(\bq_3)\>\!\>  &=-\frac{1}{4}\big(A_\chi(\bq_2)+A_\chi(\bq_3)\big)\theta^{(s_2 s_3)}(\bq_i), \nn\\
\<\!\<T_\psi(\bq_1)T_\psi^{(s_2)}(\bq_2)T_\psi^{(s_3)}(\bq_3)\>\!\> & =-\frac{1}{2}\big(A_\psi(\bq_2)+A_\psi(\bq_3)\big)\theta^{(s_2 s_3)}(\bq_i).
\end{align}
Thus, all our earlier results in \eqref{fermion_3ptfns} involving the trace $T_\psi$ are in fact a consequence of the Ward identities (noting also \eqref{proto_Ward1}), and similarly for  all our results in \eqref{confscalars_3ptfns} involving $T_\chi$.
For the latter, note that $A_\chi(\bq) = (\mathcal{N}_\chi/2\mathcal{N}_\psi)A_\psi(\bq)$ from \eqref{2ptfns}, hence from \eqref{TTTresults} we have
\[
\<\!\<T_\chi(\bq_1)T_\chi^{(s_2)}(\bq_2)T_\chi^{(s_3)}(\bq_3)\>\!\> =
\frac{\mathcal{N}_\chi}{4\mathcal{N}_\psi}
\<\!\<T_\psi(\bq_1)T_\psi^{(s_2)}(\bq_2)T_\psi^{(s_3)}(\bq_3)\>\!\>.
\]
As well as confirming earlier calculations,
these formulae additionally serve as a check of the overall sign in our 3-point function integrals.

To check the results of our 3-point function calculations for minimal scalars using \eqref{phifromchi},
we must also evaluate the semi-local terms on the r.h.s.~of \eqref{traceWards} involving the correlator 
$\<\mathcal{O}_1\widetilde{\Upsilon}^\phi_{ijkl}\>$, where $\widetilde{\Upsilon}^\phi_{ijkl}$ denotes the $\Upsilon$ tensor for conformal scalars.
The expansion for this latter quantity may be read off from \eqref{Upexpchi} (replacing $\chi$ with $\phi$).
The correlator
$\<\mathcal{O}_1\widetilde{\Upsilon}^\phi_{ijkl}\>$ receives contributions only from terms of dimension one in this expansion,
and so we find
\begin{align}
\<\!\<\widetilde{\Upsilon}_\phi(\bq_1,\bq_2)\mathcal{O}_1(\bq_3)\>\!\> &= \frac{1}{16}(\a_{12}^2-2\b_{12}-\bq_3^2)\<\!\<\mathcal{O}_1(\bq_3)\mathcal{O}_1(-\bq_3)\>\!\>,
\nn\\
\<\!\<\widetilde{\Upsilon}^{(s_2)}_\phi(\bq_1,\bq_2)\mathcal{O}_1(\bq_3)\>\!\> & 
= \frac{3}{32}\bq_3^2 \Theta_3^{(s_2)}(\bq_i)\<\!\<\O_1(\bq_3)\O_1(-\bq_3)\>\!\>.
\end{align}
Substituting these expressions into \eqref{traceWards}, we obtain
\begin{align}
\label{yetmoreresults}
\<\!\<\wT_\phi(\bq_1)\mathcal{O}_1(\bq_2)\mathcal{O}_1(\bq_3)\>\!\> &= -\<\!\<\O_1(\bq_2)\O_1(-\bq_2)\>\!\>-\<\!\<\O_1(\bq_3)\O_1(-\bq_3)\>\!\>, \nn\\[1ex]
\<\!\<\wT_\phi(\bq_1)\wT_\phi(\bq_2)\O_1(\bq_3)\>\!\> &= \frac{1}{8}(\a_{12}^2-2\b_{12}-\bq_3^2)\<\!\<\mathcal{O}_1(\bq_3)\mathcal{O}_1(-\bq_3)\>\!\>,
 \nn\\[1ex]
\<\!\<\wT_\phi(\bq_1)\wT_\phi^{(s_2)}(\bq_2)\O_1(\bq_3)\>\!\> &
=\frac{3}{16}\bq_3^2 \Theta_3^{(s_2)}(\bq_i)\<\!\<\O_1(\bq_3)\O_1(-\bq_3)\>\!\>, \\[1ex]
\<\!\<\wT_\phi(\bq_1)\wT_\phi (\bq_2)\wT_\phi(\bq_3)\>\!\> &=0,
\end{align}
where for the latter equation we used \eqref{TUpchi}.
The trace Ward identities thus supply all terms appearing on the r.h.s.~of \eqref{phifromchi}
that involve the trace $\wT_\phi$.

\subsection{Conformal Ward identities}

In the previous subsection we showed how the trace Ward identities
determine the 3-point functions involving the trace of the stress
tensor in terms of 2-point functions. Thus to determine all correlation
functions,  it remains to obtain
\[ \label{conf_corr}
\langle T^{(s_1)} T^{(s_2)} T^{(s_3)} \rangle, \qquad
\langle {\cal O}_1 \wT^{(s_2)}_\phi \wT^{(s_3)}_\phi \rangle, \qquad
\langle {\cal O}_1 {\cal O}_1 \wT^{(s_3)}_\phi \rangle, \qquad
\langle {\cal O}_1 {\cal O}_1 {\cal O}_1 \rangle,
\qquad \< {\cal O}_1 {\cal O}_1 \rangle.
\]
These may be directly computed using the methods described in Appendix \ref{methods_app}.  For conformal scalars and fermions, the result for $\< T^{(s_1)} T^{(s_2)} T^{(s_3)}\>$ is
given in \eqref{fermion_3ptfns} and \eqref{confscalars_3ptfns}. The remaining correlators are found to be (we suppress a common overall factor of $\bar{N}^2\mathcal{N}_\phi$ )
\begin{align}
\label{O1corrs}
\lla {\cal O}_1(\bar{q}) {\cal O}_1(-\bar{q}) \rra &= \frac{1}{4 \bar{q}}, \nn\\[1ex]
\lla {\cal O}_1(\bar{q}_1) {\cal O}_1(\bar{q}_2) {\cal O}_1(\bar{q}_3) \rra & =  \frac{1}{\c_{123}}, \nn\\[1ex]
\lla \wT_\phi^{(s_1)}(\bar{q}_1) {\cal O}_1(\bar{q}_2) {\cal O}_1(\bar{q}_3) \rra & =  \frac{\bar{\lambda}^2}{16 \sqrt{2}}  \frac{(2\bar{q}_1 + \bar{a}_{23})}{\bar{a}_{123}^2  \b_{23} \bar{q}_1^2}, \nn\\[1ex]
\lla \wT_\phi^{(+)}(\bar{q}_1) \wT_\phi^{(-)}(\bar{q}_2) {\cal O}_1(\bar{q}_3) \rra & =  \frac{1}{2048 \: \b_{12}^2 \bar{q}_3}\,(\bq_3^2-\a_{12}^2+4\b_{12})^2 (\a_{12}^2+2\b_{12} - 5 \bar{q}_3^2), \nn\\[1ex]
\lla \wT_\phi^{(+)}(\bar{q}_1) \wT_\phi^{(+)}(\bar{q}_2) \O_1(\bar{q}_3) \rra & =  \frac{(\bar{a}_{12} - \bar{q}_3)^2}{2048 \:  \bar{a}_{123}^2\b_{12}^2 \bar{q}_3} \big[ -5 \bar{q}_3^6 - 20 \bar{a}_{12} \bar{q}_3^5 -  (29 \bar{a}_{12}^2 + 6 \bar{b}_{12}) \bar{q}_3^4 \nn \\
&\qquad \qquad - \: 8 \bar{a}_{12} (2 \bar{a}_{12}^2 + 3  \bar{b}_{12} ) \bar{q}_3^3 + \bar{a}_{12}^2(\bar{a}_{12}^2 - 36  \bar{b}_{12})\bar{q}_3^2  \nonumber \\
& \qquad\qquad  + \: 4 \bar{a}_{12}^3 (\bar{a}_{12}^2 - 6  \bar{b}_{12})\bar{q}_3 + \bar{a}_{12}^6 - 6 \bar{a}_{12}^4 \bar{b}_{12} + 32 \bar{b}_{12}^3 \big].
\end{align}
Subsituting these expressions into \eqref{phifromchi}, along with those in \eqref{yetmoreresults},  we recover all the results for minimal scalars listed in \eqref{minscalars_3ptfns} that involve the trace $T_\phi$.
(One may additionally check we recover the result for $\<T_\phi T_\phi T_\phi\>$ in equation (101) of \cite{McFadden:2010vh}.)

The correlation functions \eqref{O1corrs} are almost uniquely determined by
conformal Ward identities. This was discussed in position space in
\cite{Osborn:1993cr}, and more recently in momentum space in
\cite{Maldacena:2011nz} (see also \cite{Giombi:2011rz}). 
More precisely, the 3-point function
of the stress tensor is unique, up to two constants
(which is our case may be taken to be the number of conformal
scalars and the number of free fermions), and all remaining
3-point functions are unique up to an overall constant.
Thus, a non-trivial check of the correlators listed above is to
verify that they indeed satisfy the special
conformal Ward identities. (The scale Ward identity is satisfied by
inspection.)
These identities take the form of differential equations that the correlators must satisfy,
and are listed explicitly in Appendix \ref{app_CWI}
(they may be obtained by Fourier transforming the position space Ward identities
whose derivation is discussed, for example, in \cite{DiFrancesco:1997nk}).
We have checked that our conformal correlators satisfy these Ward identities.

\section{Holographic predictions for cosmological 3-point functions}
\label{sec:HolPredictions}

Having computed all relevant QFT quantities we can now evaluate the
holographic formulae. It is instructive to first use the trace Ward
identities and the relation of minimal scalars to conformal scalars
in order to express the cosmological 3-point functions in terms of
CFT correlations functions. This yields
\begin{align}
\label{results_bonanza}
\lla \zeta (q_1)\zeta(q_2) \zeta (q_3) \rra
& =  - \frac{2^4}{N^4 \mathcal{N}_{(B)}^2 \prod_{i=1}^3 \left( q_i^4 \lla {\cal O}_1(q_i) {\cal O}_1(-q_i) \rra\right)}  \nonumber \\
& \quad\quad \times \left[ q_1^2q_2^2q_3^2 \lla {\cal O}_1(q_1) {\cal O}_1(q_2) {\cal O}_1(q_3) \rra
\right. \nonumber \\& \qquad \quad\quad \left.
+ \left( 2 q_1^2 (q_1^2 - q_2^2 - q_3^2)
\lla {\cal O}_1(q_1) {\cal O}_1(-q_1) \rra + 2\ {\rm perm.} \right) \right], \nonumber \\[1ex]
\lla \zeta (q_1)\zeta(q_2) \hat{\gamma}^{(s_3)} (q_3) \rra
& =  - \frac{2^{11}}{N^4 \mathcal{N}_{(A)} \mathcal{N}_{(B)} q_3^3 \prod_{i=1}^2 \left( q_i^4 \lla {\cal O}_1(q_i) {\cal O}_1(-q_i) \rra\right)} \nonumber \\
& \quad\quad \times \left[ q_1^2 q_2^2 \lla \mathcal{O}_1(q_1) \mathcal{O}_1(q_2) \wT_\phi^{(s_3)}(q_3) \rra \right. \nonumber \\
& \qquad \quad\quad \left. + \left(\: q_1^4 \Theta_1^{(s_3)}(q_i) \lla \mathcal{O}_1(q_1) \mathcal{O}_1(-q_1) \rra + ( q_1 \leftrightarrow q_2 )\right) \right], \nonumber \\[1ex]
\lla \zeta (q_1)\hat{\gamma}^{(s_2)}(q_2) \hat{\gamma}^{(s_3)} (q_3) \rra
& =  - \frac{2^{14}}{N^4 \mathcal{N}_{(A)}^2 q_2^3 q_3^3 q_1^4 \lla \mathcal{O}_1(q_1) \mathcal{O}_1(-q_1) \rra} \nonumber \\
& \quad\quad \times \left[ 16 q_1^2 \lla \mathcal{O}_1(q_1) \wT_\phi^{(s_2)}(q_2) \wT_\phi^{(s_3)}(q_3) \rra \right. \nonumber \\
& \left. \qquad \quad\quad + \: q_1^4 \left( 2 \theta^{(s_2 s_3)}(q_i) - \Theta^{(s_2 s_3)}(q_i) \right) \lla \mathcal{O}_1(q_1) \mathcal{O}_1(-q_1) \rra \right],
\end{align}
where $\mathcal{N}_{(A)}$ and $\mathcal{N}_{(B)}$ are defined in
\eqref{NANBdef} and we have made the dependence on the number of fields explicit by considering the
${\cal O}_1$ and $\wT^{(s)}_\phi$ correlators to be those of a single field.
(The analytic continuation \eqref{an_cont} has also been implicitly performed;
the correlators appearing above are therefore those in \eqref{O1corrs} with $\bq_i$ replaced by $q_i$.)
As discussed in the previous section, the trace Ward identities imply that
the numerators of the holographic formulae \eqref{holo_zzz}-\eqref{holo_zgg} for the above
correlators receive no contribution from
conformal fields and are therefore proportional to $\mathcal{N}_{(B)}$, the number of non-conformal fields.
The dependence of these correlators on the field content is then simply given by an overall factor, amounting to $\mathcal{N}_{(B)}$ divided by the corresponding factors in the denominators of the holographic formulae.

For the $\bg \bg \bg$ correlator, we find
\begin{align}
\label{ggg_conf}
&\lla \hat{\gamma}^{(s_1)}(q_1)\hat{\gamma}^{(s_2)}(q_2) \hat{\gamma}^{(s_3)} (q_3) \rra  \nonumber \\
& \qquad = - \frac{2^{24}}{N^4 \mathcal{N}_{(A)}^3 q_1^3 q_2^3 q_3^3} \left[ \mathcal{N}_\psi \left( 2 \lla T_{\psi}^{(s_1)} T_{\psi}^{(s_2)} T_{\psi}^{(s_3)} \rra - \frac{\Theta^{(s_1 s_2 s_3)}(q_i)}{512} \sum_{i=1}^3 q_i^3 \right) \right. \nonumber \\
& \qquad \qquad \left. + \: ( \mathcal{N}_\phi + \mathcal{N}_\chi + \mathcal{N}_A ) \left( 2 \lla T_{\chi}^{(s_1)} T_{\chi}^{(s_2)} T_{\chi}^{(s_3)} \rra - \frac{\Theta^{(s_1 s_2 s_3)}(q_i)}{512} \sum_{i=1}^3 q_i^3 \right) \right],
\end{align}
considering again the correlators to be those of a single field so as to make the dependence on the number of fields explicit.
It then turns out that
\begin{equation}  \label{++-}
2 \lla T_{\chi}^{(+)} T_{\chi}^{(+)} T_{\chi}^{(-)} \rra = \lla T_{\psi}^{(+)} T_{\psi}^{(+)} T_{\psi}^{(-)} \rra + \frac{\Theta^{(++-)}(q_i)}{1024} \sum_{i=1}^3 q_i^3,
\end{equation}
{\it i.e.}, these correlators differ only by the helicity projection of a semi-local term. Thus, while the
3-point function of the stress tensor at separated points
in general depends on two constants in $d=3$ \cite{Osborn:1993cr},
only one combination survives the $(++-)$ helicity projection.  (This was also shown in \cite{Maldacena:2011nz} using the conformal Ward identities.)  The specific form of the semi-local term in (\ref{++-}) is then such that the $\bg^{(+)} \bg^{(+)} \bg^{(-)}$ correlation function depends on the field content through an overall multiplicative constant only. On the other hand,
\begin{equation}
2 \lla T_{\chi}^{(+)} T_{\chi}^{(+)} T_{\chi}^{(+)} \rra = \lla T_{\psi}^{(+)} T_{\psi}^{(+)} T_{\psi}^{(+)} \rra + \frac{\Theta^{(+++)}(q_i)}{1024} \sum_{i=1}^3 q_i^3 + \frac{\lambda^2}{128 \sqrt{2}} \frac{c_{123}}{a_{123}^4},
\end{equation}
and so these correlators differ by the helicity projections of both a semi-local and a non-local term.  The non-local term reflects the fact that both solutions for the 3-point function of the stress tensor at separated points survive the $(+++)$ helicity projection, and leads in turn to a more complicated dependence on the QFT field content in the $\bg^{(+)}\bg^{(+)}\bg^{(+)}$ correlator.

Returning to \eqref{results_bonanza} and substituting in our results for the remaining correlators,
we first recover the result derived in \cite{McFadden:2010vh},
\begin{align}
\<\!\<\z(q_1)\z(q_2)\z(q_3)\>\!\>
&= \frac{512}{\mathcal{N}_{(B)}^2 N^4} \Big(\prod_i q_i^{-3}\Big)\big({-}2q_1q_2q_3-\sum_i q_i^3+(q_1q_2^2+5\,\mathrm{perms})\big) \nn\\ & = \frac{512}{\mathcal{N}_{(B)}^2 N^4}\frac{\lambda^2}{a_{123}c_{123}^3},
\end{align}
showing that the scalar bispectrum exactly coincides with the equilateral template.
In the first line, note that all the terms but the one proportional to $q_1 q_2 q_3$ originate
from semi-local terms in the numerator of the holographic formula \eqref{holo_zzz}. Without their contribution we would not have been able to distinguish the equilateral shape from others involving a similar factor of $q_1 q_2 q_3$ in the numerator (for example, the orthogonal shape \cite{Senatore:2009gt}, for which the corresponding numerator is $-8q_1q_2q_3 -3\sum_iq_i^3+3(q_1 q_2^2 + 5\,\mathrm{perms})$).
In fact, due to the factor of $\prod_i q_i^{-3}$ coming from the product of 2-point functions in the denominator of the holographic formula, the semi-local term $\sum_i q_i^3$ in the numerator generates a contribution to the bispectrum of exactly the `local' type.
It is therefore essential to include the contribution of all semi-local terms in the holographic formulae, as we have been careful to do.

For the remaining correlators, we find
\begin{align}
\label{main_hol_results}
\<\!\<\z(q_1)\z(q_2)\bg^{(+)}(q_3)\>\!\> & = \frac{2048}{\sqrt{2}N^4 \mathcal{N}_{(A)}\mathcal{N}_{(B)}} \frac{\lambda^2}{a_{123}^2 c_{123}^3 q_3^2} \Big[(a_{123}^3-a_{123}b_{123}-c_{123})-a_{123}q_3^2\Big],\nn\\[2ex]
\<\!\<\z(q_1)\bg^{(+)}(q_2)\bg^{(+)}(q_3)\>\!\> &= -\frac{512}{N^4\mathcal{N}_{(A)}^2  b_{23}^5 q_1^2}(q_1^2-a_{23}^2)^2
\Big[(q_1^2-a_{23}^2+2b_{23})+\frac{32 b_{23}^3}{a_{123}^4}\Big],
\nn\\[2ex]
\<\!\<\z(q_1)\bg^{(+)}(q_2)\bg^{(-)}(q_3)\>\!\> &= -\frac{512}{N^4\mathcal{N}_{(A)}^2 b_{23}^5 q_1^2}(q_1^2-a_{23}^2+4b_{23})^2(q_1^2-a_{23}^2+2b_{23}),
\nn\\[2ex]
 \<\!\<\bg^{(+)}(q_1)\bg^{(+)}(q_2)\bg^{(+)}(q_3)\>\!\>
&=\frac{1024}{\sqrt{2}N^4 \mathcal{N}_{(A)}^2}\frac{\lambda^2 a_{123}^2}{c_{123}^5} \Big[ (a_{123}^3-a_{123}b_{123}-c_{123})-\Big(1-4\frac{\mathcal{N}_\psi}{\mathcal{N}_{(A)}}\Big)\frac{64 c_{123}^3}{a_{123}^6}\Big], \nn\\[2ex]
 \<\!\<\bg^{(+)}(q_1)\bg^{(+)}(q_2)\bg^{(-)}(q_3)\>\!\> &= \frac{1024}{\sqrt{2}N^4 \mathcal{N}_{(A)}^2}\frac{\lambda^2}{a_{123}^2c_{123}^5}(q_3-a_{12})^4(a_{123}^3-a_{123}b_{123}-c_{123}).
\end{align}

We would now like to define corresponding shape functions, {\it i.e.}, we wish
to write these correlators as bispectra:
a product of power spectra times a
shape function. To do so,
we first define the dimensionless 2-point amplitudes
\[
\label{2ptampdef}
\mathcal{A}(\z\z)= q^3 \<\!\<\z(q)\z(-q)\>\!\> = \frac{32}{N^2 \mathcal{N}_{(B)}},\qquad \mathcal{A}(\bg\bg) = q^3\<\!\<\bg^{(+)}(q)\bg^{(+)}(-q)\>\!\> = \frac{256}{N^2 \mathcal{N}_{(A)}},
\]
and similarly the dimensionless 3-point amplitudes, {\it e.g.},
\[
 \mathcal{A}(\z\z\bg^{(+)})=  q_1^2 q_2^2 q_3^2\, \<\!\<\z(q_1)\z(q_2)\bg^{(+)}(q_3)\>\!\>,
\]
with analogous expressions for the other correlators.
Physically, these quantitites parametrise the contribution per logarithmic interval of wavenumbers to the corresponding position-space expectation values with all insertions at the same point, {\it e.g.},
\[
\<\zeta^2(\vec{x})\> = \frac{1}{2\pi^2}\int (\d \ln q) \mathcal{A}(\z\z), \qquad
\<\zeta^2(\vec{x})\bg^{(+)}(\vec{x})\> = \frac{1}{8\pi^4}\int \(\prod_i\d\ln q_i\) \mathcal{A}(\z\z\bg^{(+)}),
\]
where the latter integral ranges over all possible triangle side lengths in momentum space.
(For reference, the usual logarithmic power spectrum is simply $\Delta_\z^2 = (1/2\pi^2)\mathcal{A}(\z\z)$.)
The dimensionless 3-point amplitudes may now be naturally re-expressed as a product of dimensionless 2-point amplitudes and a purely momentum-dependent shape function:
\begin{align}
\label{shapes_def}
  \mathcal{A}(\z\z\bg^{(s_3)})  &= \mathcal{A}(\z\z)\mathcal{A}(\bg\bg) \mathcal{S}(\z\z\bg^{(s_3)}), \qquad
 \mathcal{A}(\z\bg^{(s_2)}\bg^{(s_3)})  = \mathcal{A}^2(\bg\bg) \mathcal{S}(\z\bg^{(s_2)}\bg^{(s_3)}), \nn\\[1ex]&\qquad\quad
 \mathcal{A}(\bg^{(s_1)}\bg^{(s_2)}\bg^{(s)})  = \mathcal{A}^2(\bg\bg) \mathcal{S}(\bg^{(s_1)}\bg^{(s_2)}\bg^{(s)}).
\end{align}
Explicitly, the shape functions are given by
\begin{align}
\label{hol_shapes}
\mathcal{S}(\z\z\bg^{(+)}) &=
 \frac{1}{4\sqrt{2}} \frac{\lambda^2}{a_{123}^2 c_{123} q_3^2} \Big[(a_{123}^3-a_{123}b_{123}-c_{123})-a_{123}q_3^2\Big], \nn\\[2ex]
\mathcal{S}(\z\bg^{(+)}\bg^{(+)}) &=  -\frac{1}{128 b_{23}^3}(q_1^2-a_{23}^2)^2
\Big[(q_1^2-a_{23}^2+2b_{23})+\frac{32 b_{23}^3}{a_{123}^4}\Big], \nn\\[2ex]
\mathcal{S}(\z\bg^{(+)}\bg^{(-)}) &=
-\frac{1}{128 b_{23}^3}(q_1^2-a_{23}^2+4b_{23})^2(q_1^2-a_{23}^2+2b_{23}), \nn\\[2ex]
\mathcal{S}(\bg^{(+)}\bg^{(+)}\bg^{(+)}) &=\frac{1}{64\sqrt{2}}\frac{\lambda^2 a_{123}^2}{c_{123}^3} \Big[ (a_{123}^3-a_{123}b_{123}-c_{123})-\Big(1-4\frac{\mathcal{N}_\psi}{\mathcal{N}_{(A)}}\Big)\frac{64 c_{123}^3}{a_{123}^6}\Big], \nn\\[2ex]
\mathcal{S}(\bg^{(+)}\bg^{(+)}\bg^{(-)}) &= \frac{1}{64\sqrt{2}}\frac{\lambda^2}{a_{123}^2c_{123}^3}(q_3-a_{12})^4(a_{123}^3-a_{123}b_{123}-c_{123}).
\end{align}
Thus, with the sole exception of $\mathcal{S}(\bg^{(+)}\bg^{(+)}\bg^{(+)})$, all the shape functions defined in this manner are independent of the field content of the dual QFT.  
(Indeed, this was our motivation in selecting the factors of $\mathcal{A}(\z\z)$ and $\mathcal{A}(\bg\bg)$ appearing in \eqref{shapes_def}.)
From our previous discussion, we see that for the shape functions involving one or more factors of $\z$ this property is a consequence of the trace Ward identities, which limit the field content-dependence of the corresponding bispectra to a single overall factor.
The independence of $\mathcal{S}(\bg^{(+)}\bg^{(+)}\bg^{(-)})$ from the QFT field content arises similarly from the fact that the corresponding bispectrum depends on the field content via an overall factor only.  As we saw above, this latter property relies on both the conformal Ward identities and the precise form of the semi-local terms appearing in the holographic formula.

\begin{figure}[p]
\centering
\subfloat[$\mathcal{S}(\bg^{(+)}\bg^{(+)}\bg^{(-)})$]{\label{fig:Sppm}\includegraphics[width=0.30\textwidth]{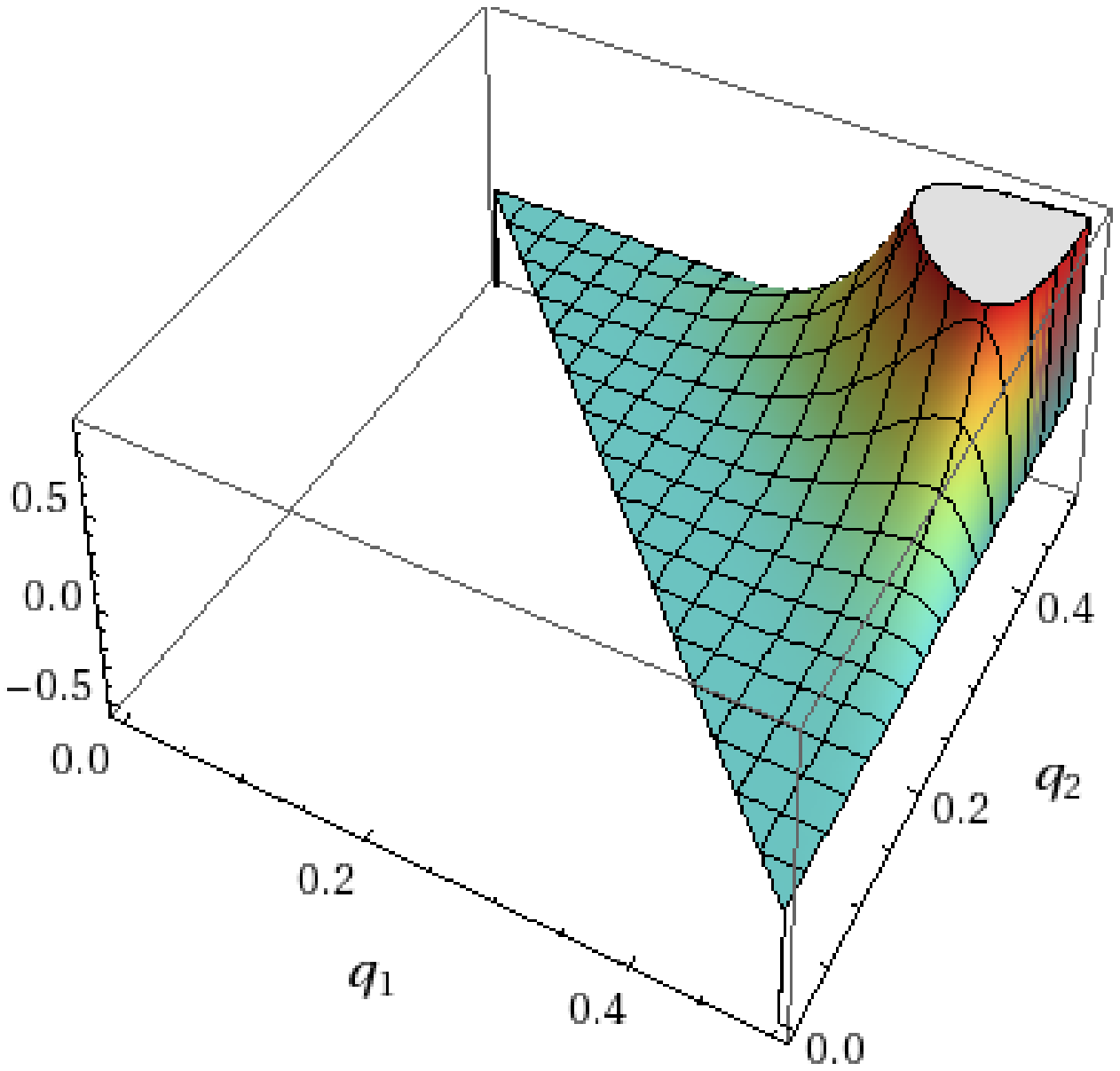}}
\\
\subfloat[$\mathcal{S}(\bg^{(+)}\bg^{(+)}\bg^{(+)})$]{\label{fig:Holppp}\includegraphics[width=0.293\textwidth]{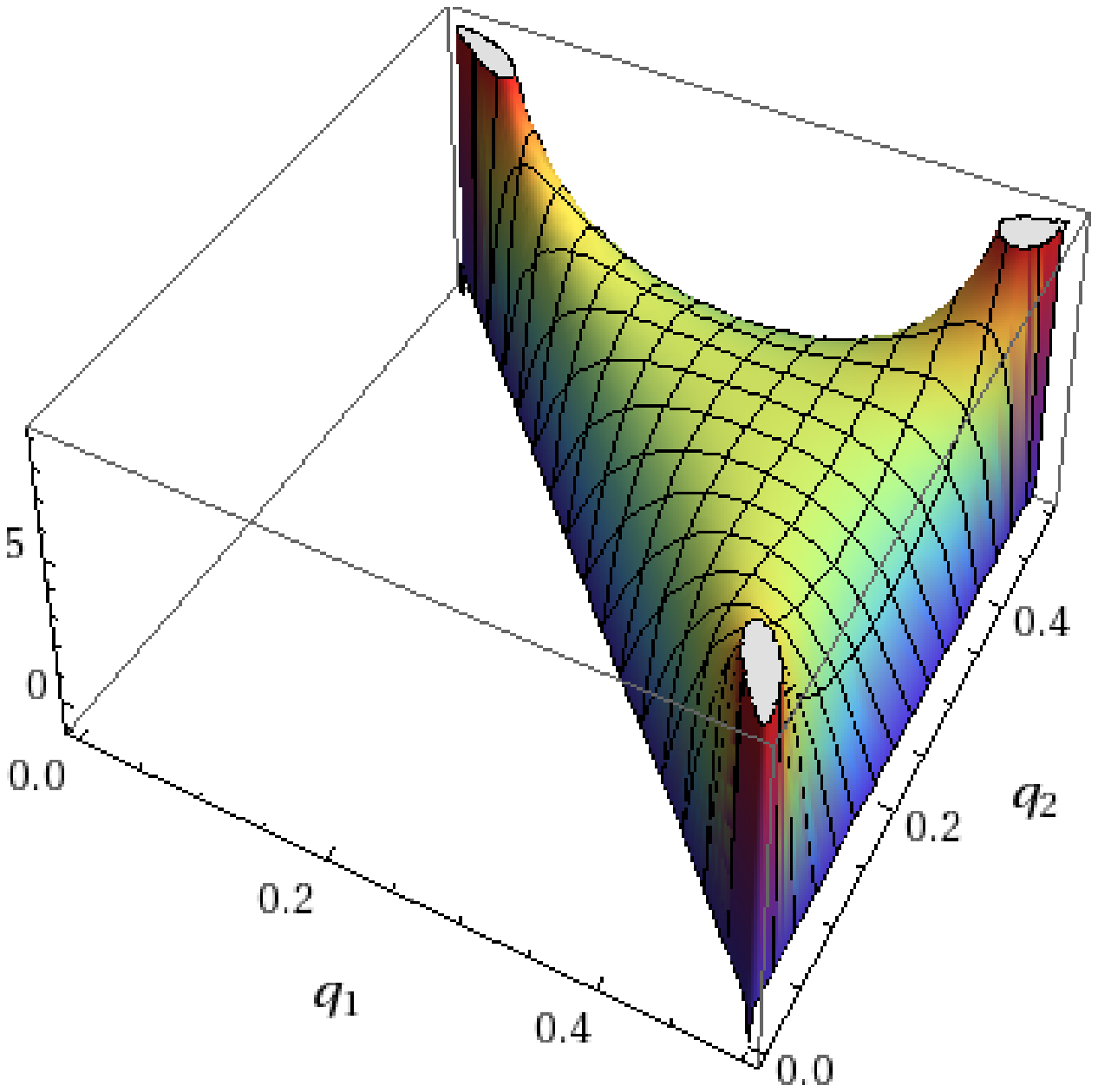}}
\hspace{0.01\textwidth}
\subfloat[$\mathcal{S}_{SR}(\bg^{(+)}\bg^{(+)}\bg^{(+)})$]{\label{fig:SRppp}\includegraphics[width=0.293\textwidth]{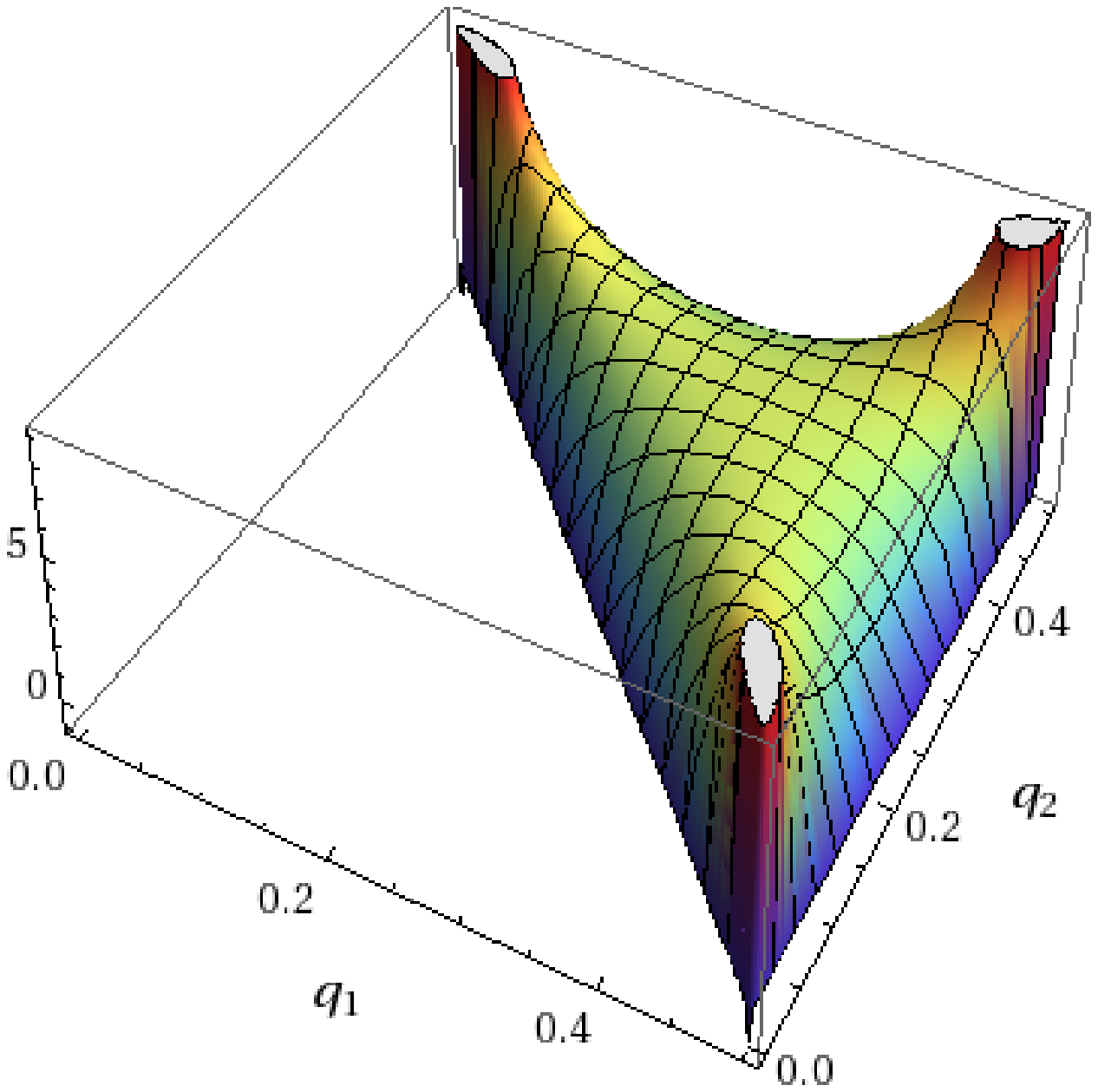}}
\hspace{0.01\textwidth}
\subfloat[$\Delta \mathcal{S}(\bg^{(+)}\bg^{(+)}\bg^{(+)})$]{\label{fig:Deltappp}\includegraphics[width=0.30\textwidth]{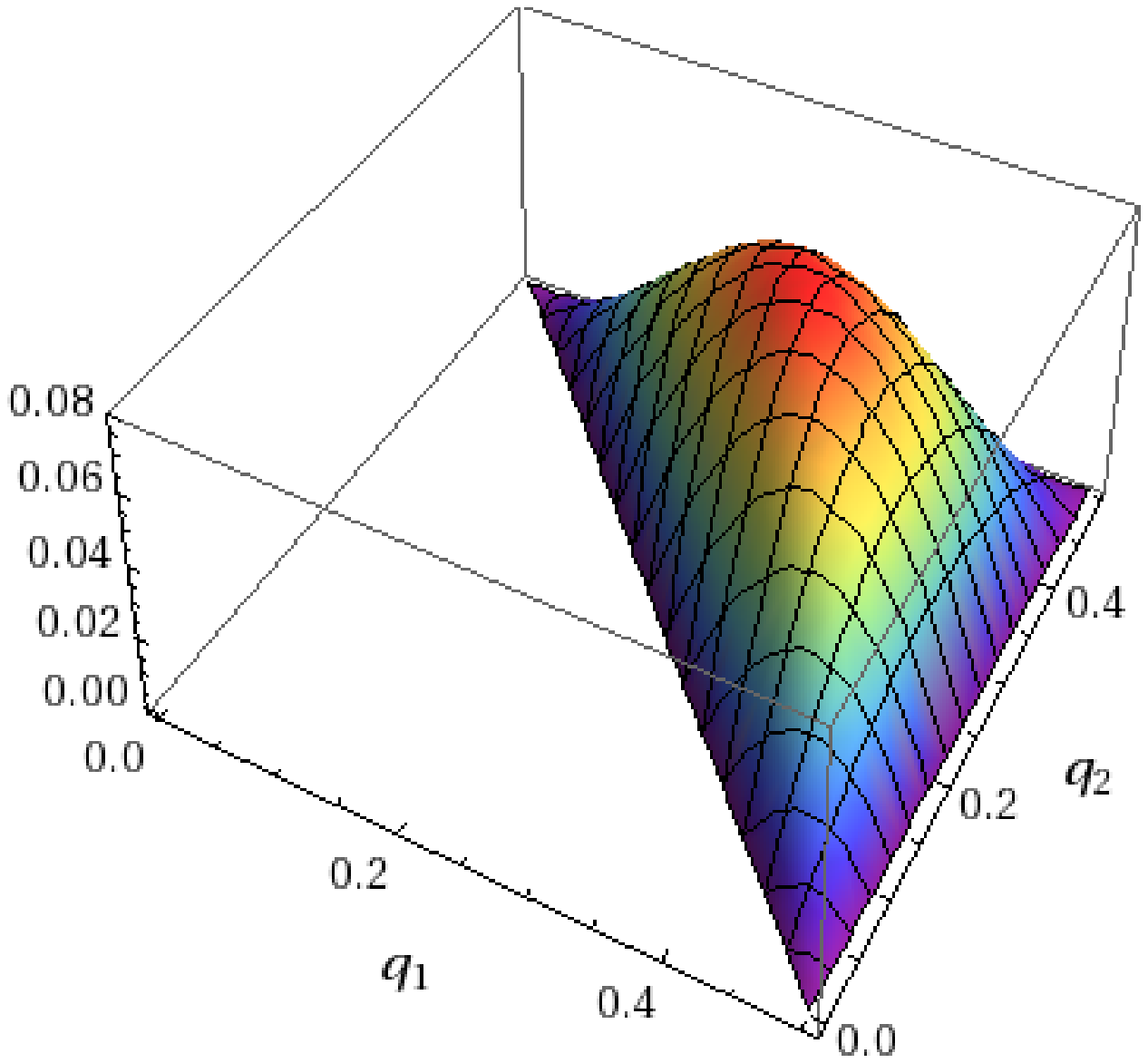}}
\\
\subfloat[$\mathcal{S}(\z\z\bg^{(+)})$]{\label{fig:Holzzp}\includegraphics[width=0.291\textwidth]{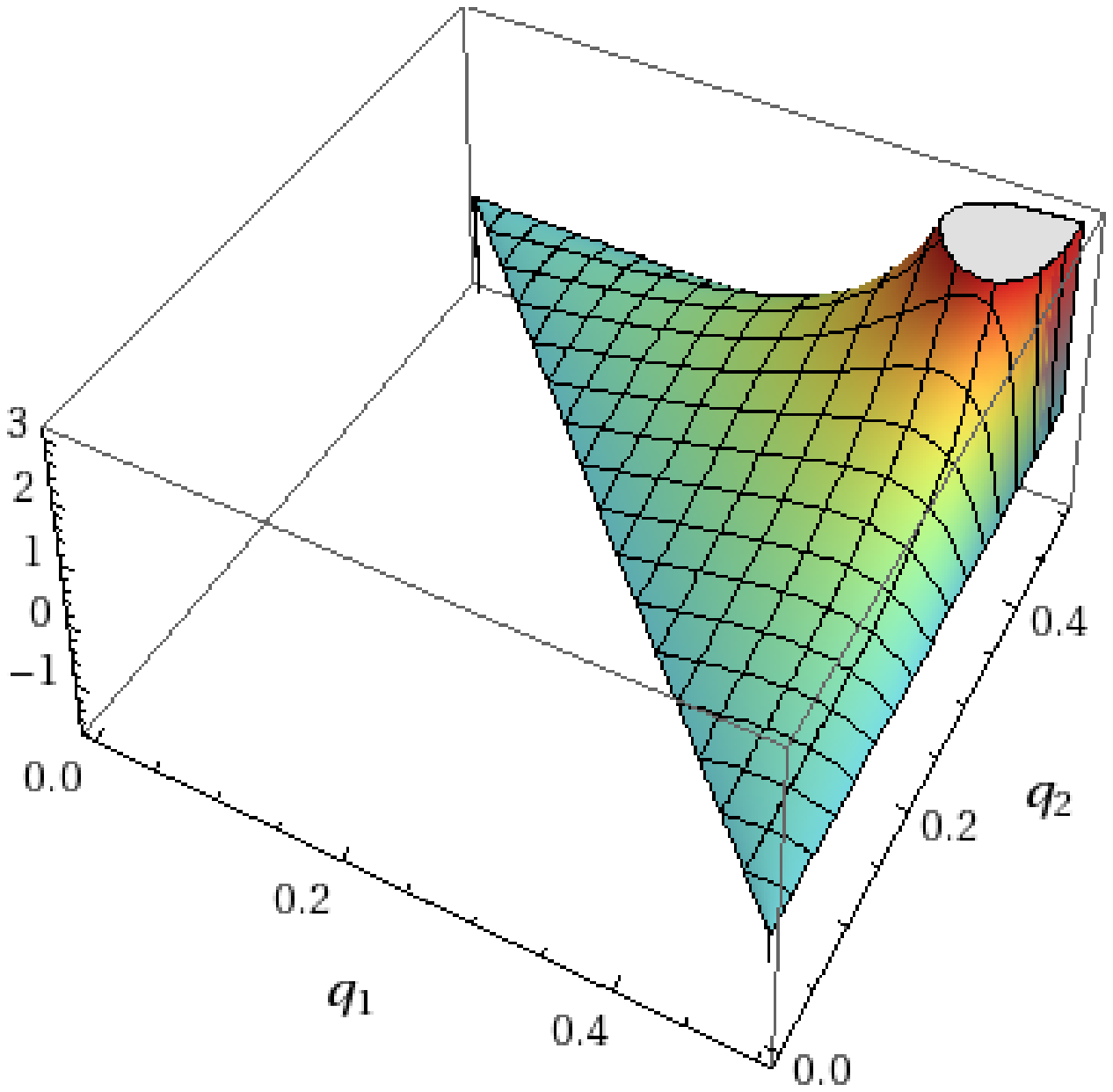}}
\hspace{0.01\textwidth}
\subfloat[$\mathcal{S}_{SR}(\z\z\bg^{(+)})$]{\label{fig:SRzzp}\includegraphics[width=0.291\textwidth]{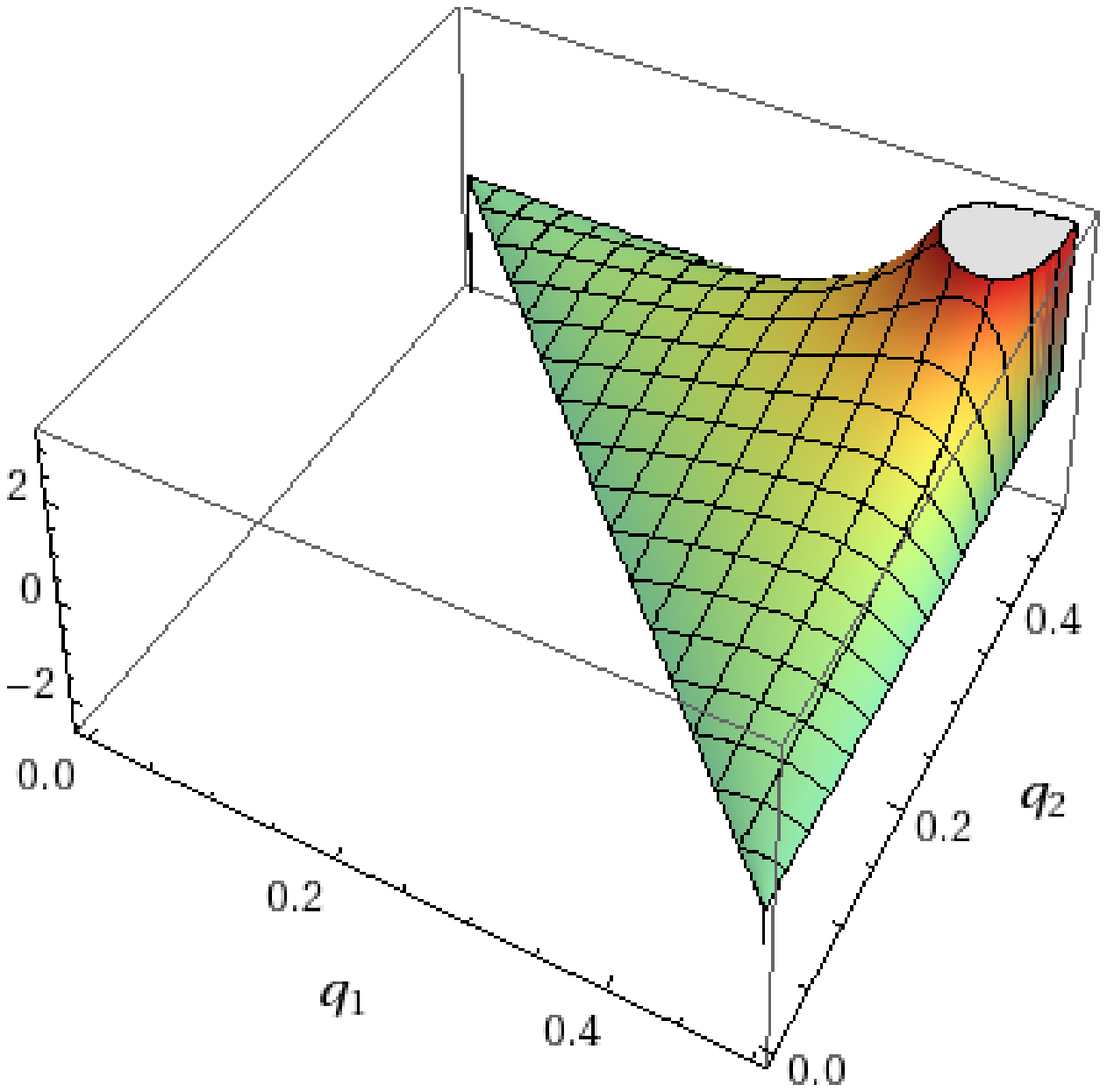}}
\hspace{0.01\textwidth}
\subfloat[$\Delta \mathcal{S}(\z\z\bg^{(+)})$]{\label{fig:Deltazzp}\includegraphics[width=0.30\textwidth]{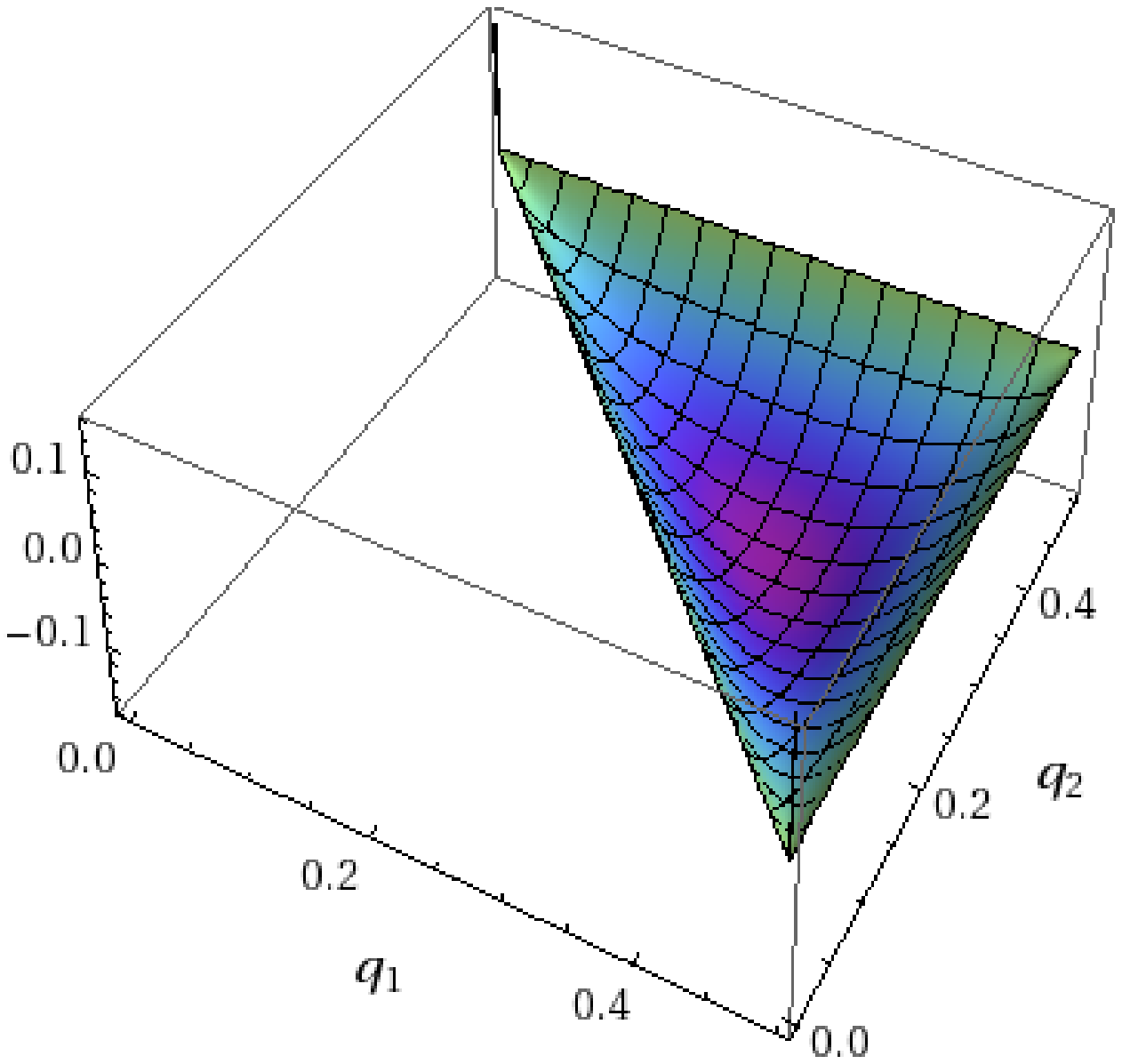}}
\caption{\label{Shapes1} Isoperimetric plots displaying the holographic and slow-roll shape functions, as well as the difference between them ({\it e.g.}, $\Delta \mathcal{S}(\bg^{(+)}\bg^{(+)}\bg^{(+)})=\mathcal{S}(\bg^{(+)}\bg^{(+)}\bg^{(+)})-\mathcal{S}_{SR}(\bg^{(+)}\bg^{(+)}\bg^{(+)})$).
The invariance of the shape functions under a rescaling $q_i\rightarrow \lambda q_i$ of all momenta has been exploited to set $q_1+q_2+q_3=1$, constraining the allowed momentum values to those displayed.   Each plot is symmetric under interchange of the appropriate momenta as expected.
Note that  $\mathcal{S}(\bg^{(+)}\bg^{(+)}\bg^{(-)})$ (shown in plot (a)) coincides for the holographic and slow-roll models.
In plots (\ref{fig:Holppp}) and (\ref{fig:Deltappp}) we have set $\mathcal{N}_\psi=\mathcal{N}_{(A)}$ to maximise $\Delta \mathcal{S}(\bg^{(+)}\bg^{(+)}\bg^{(+)})$ for illustrative purposes.
}
\end{figure}

\begin{figure}[p]
\centering
\subfloat[$\mathcal{S}(\z\bg^{(+)}\bg^{(+)})$]{\label{fig:Holzpp}\includegraphics[width=0.30\textwidth]{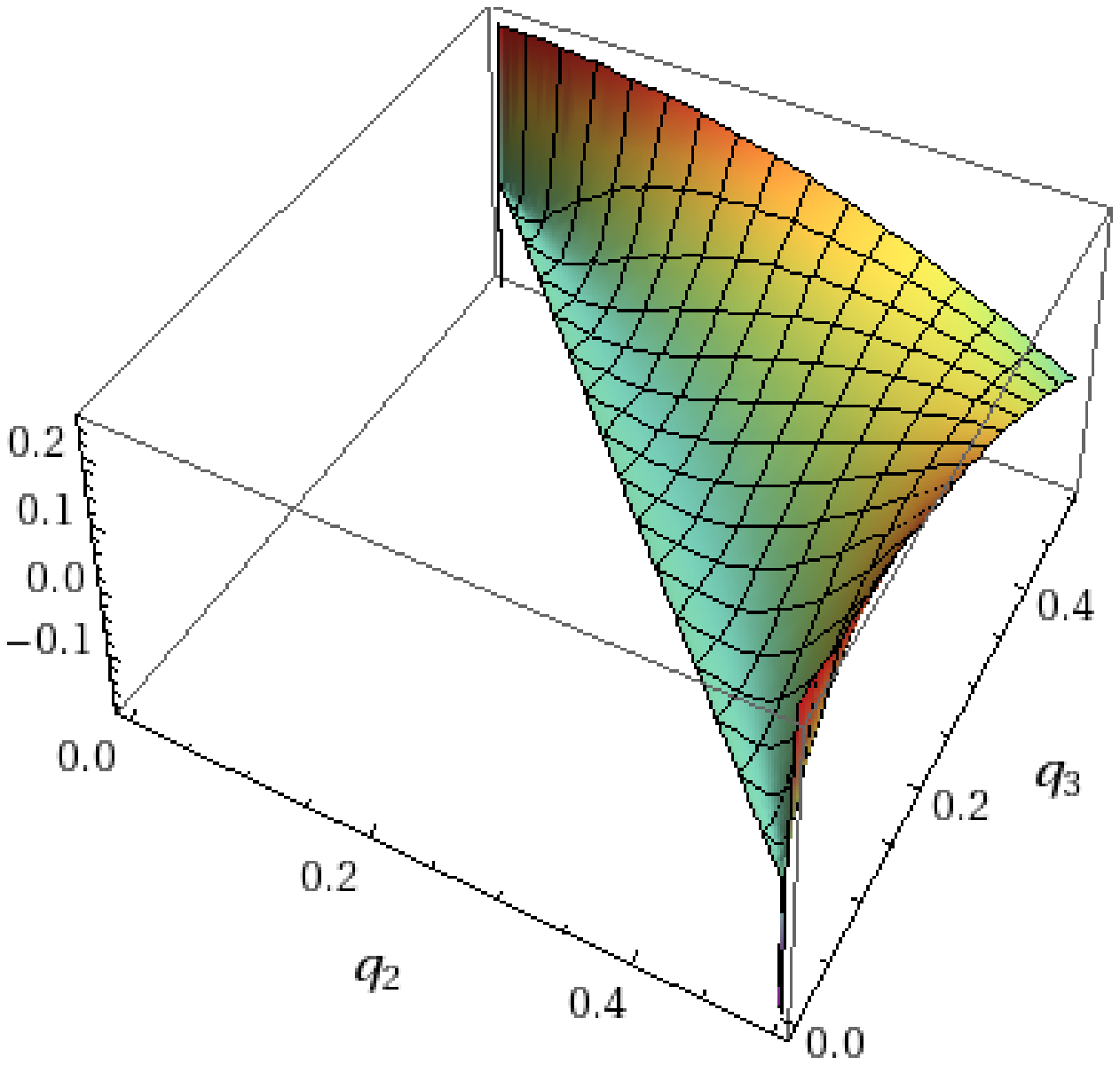}}
\hspace{0.01\textwidth}
\subfloat[$\mathcal{S}_{SR}(\z\bg^{(+)}\bg^{(+)})$]{\label{fig:SRzpp}\includegraphics[width=0.3\textwidth]{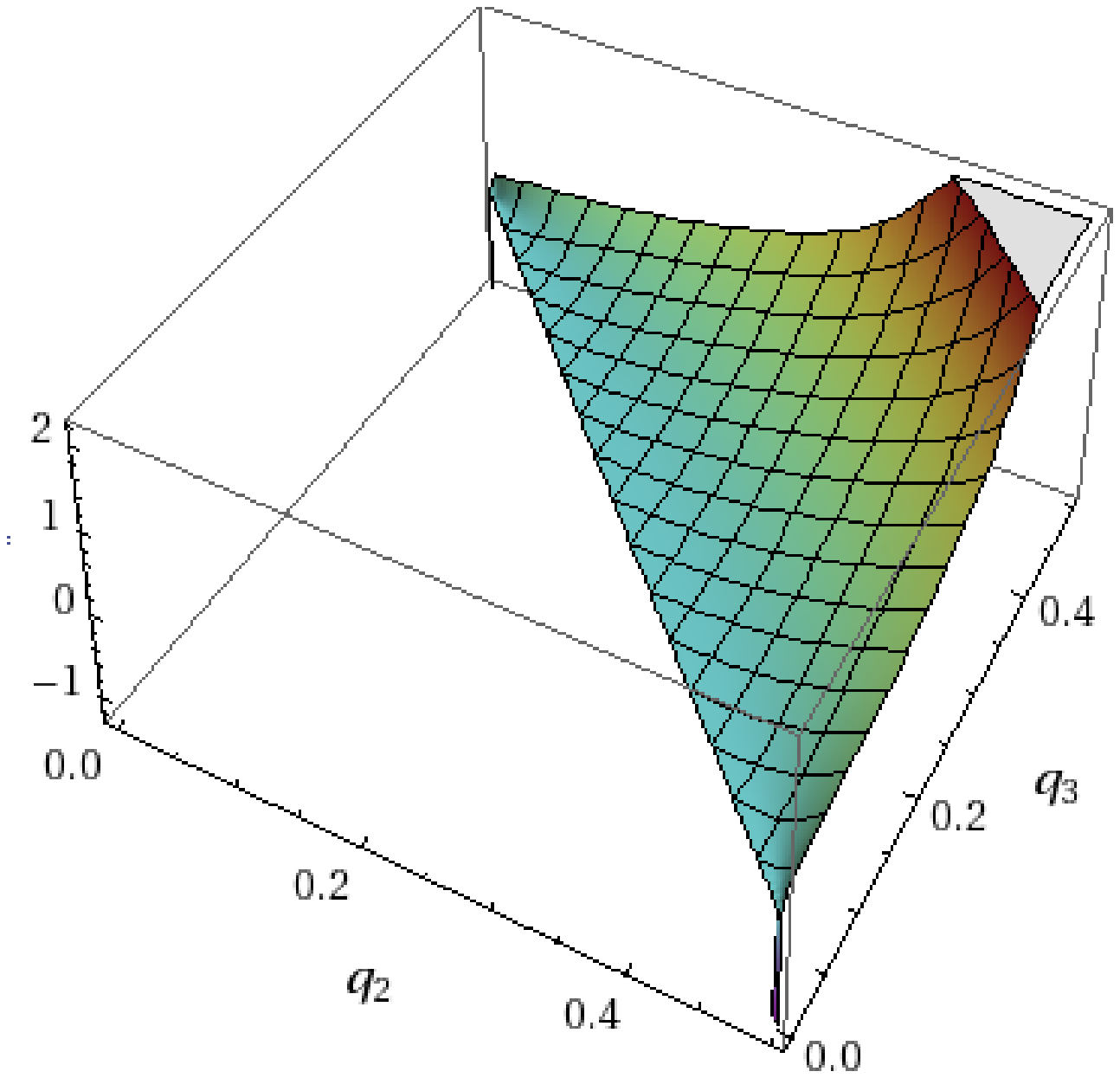}}
\hspace{0.01\textwidth}
\subfloat[$\Delta \mathcal{S}(\z\bg^{(+)}\bg^{(+)})$]{\label{fig:Deltazpp}\includegraphics[width=0.3\textwidth]{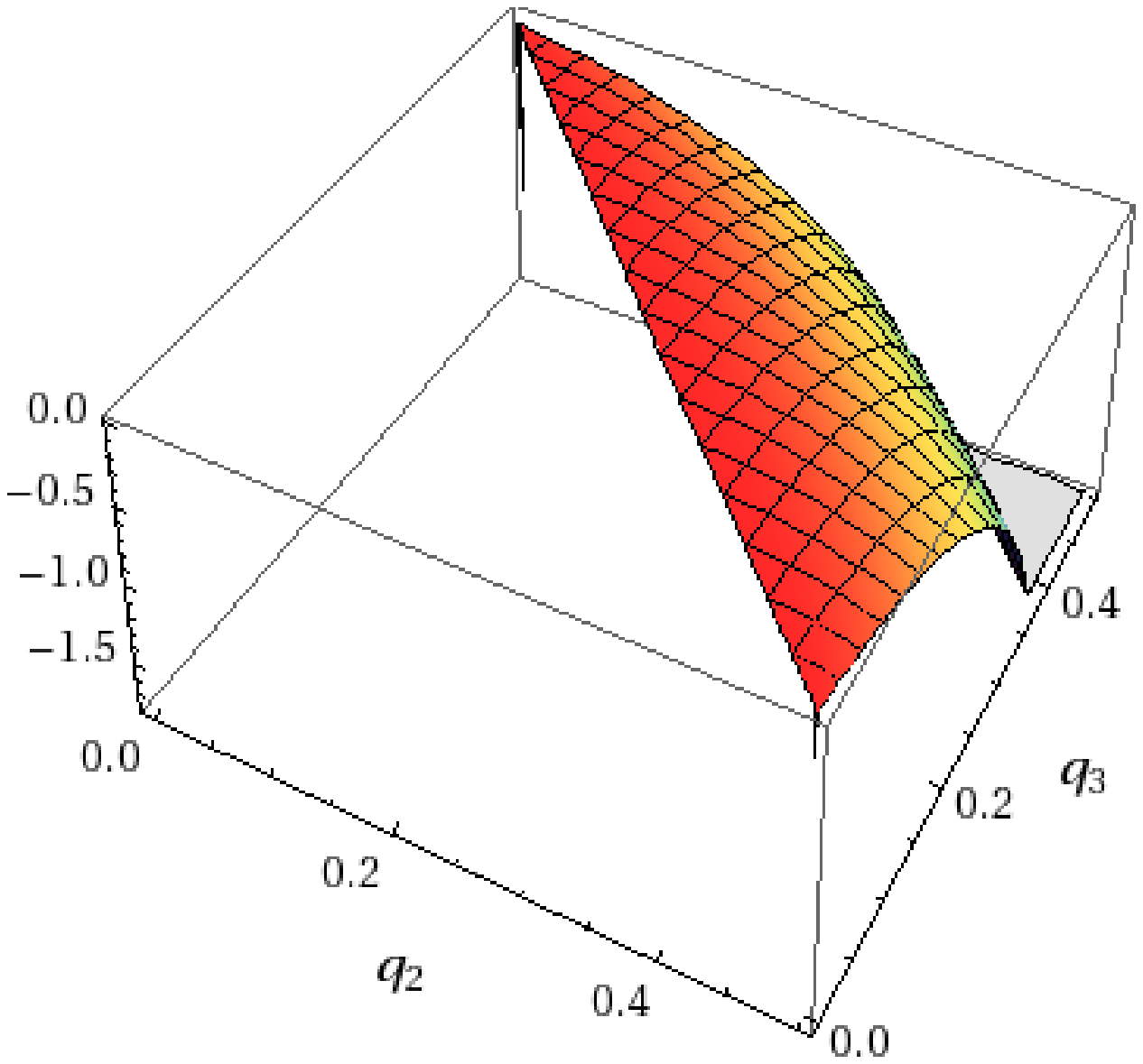}}
\\
\subfloat[$\mathcal{S}(\z\bg^{(+)}\bg^{(-)})$]{\label{fig:Holzpm}\includegraphics[width=0.3\textwidth]{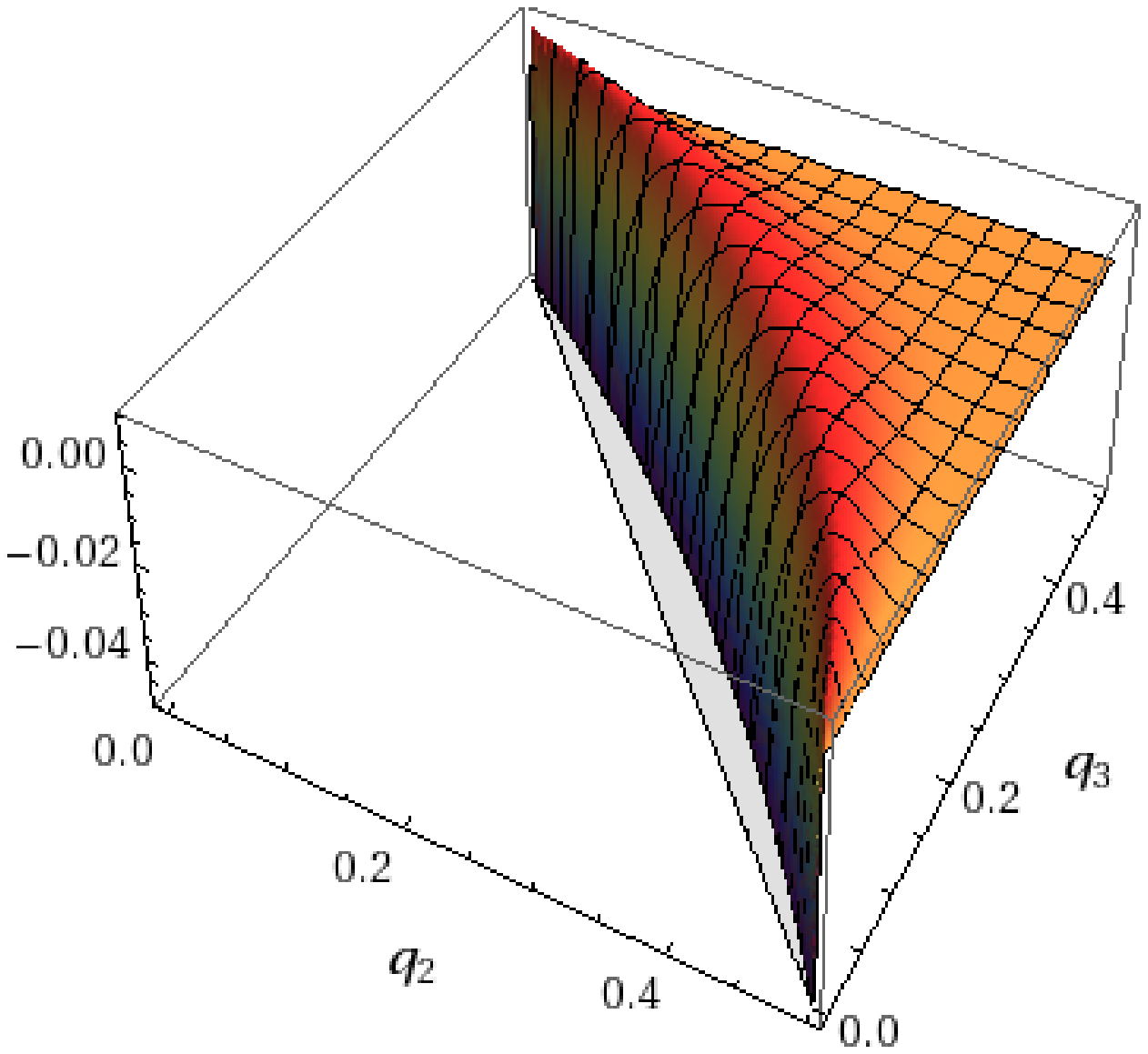}}
\hspace{0.01\textwidth}
\subfloat[$\mathcal{S}_{SR}(\z\bg^{(+)}\bg^{(-)})$]{\label{fig:SRzpm}\includegraphics[width=0.3\textwidth]{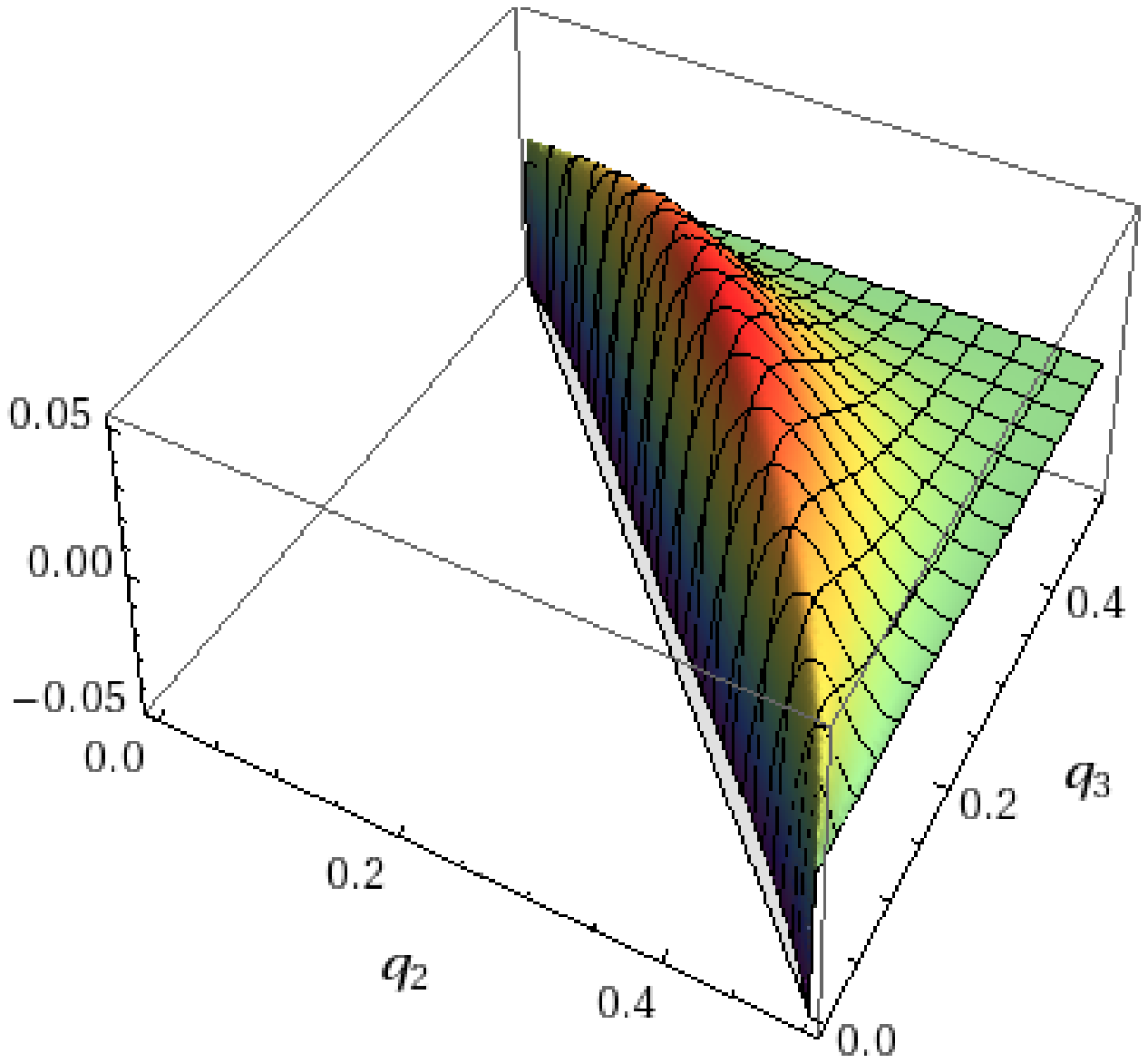}}
\hspace{0.01\textwidth}
\subfloat[$\Delta \mathcal{S}(\z\bg^{(+)}\bg^{(-)})$]{\label{fig:Deltazpm}\includegraphics[width=0.3\textwidth]{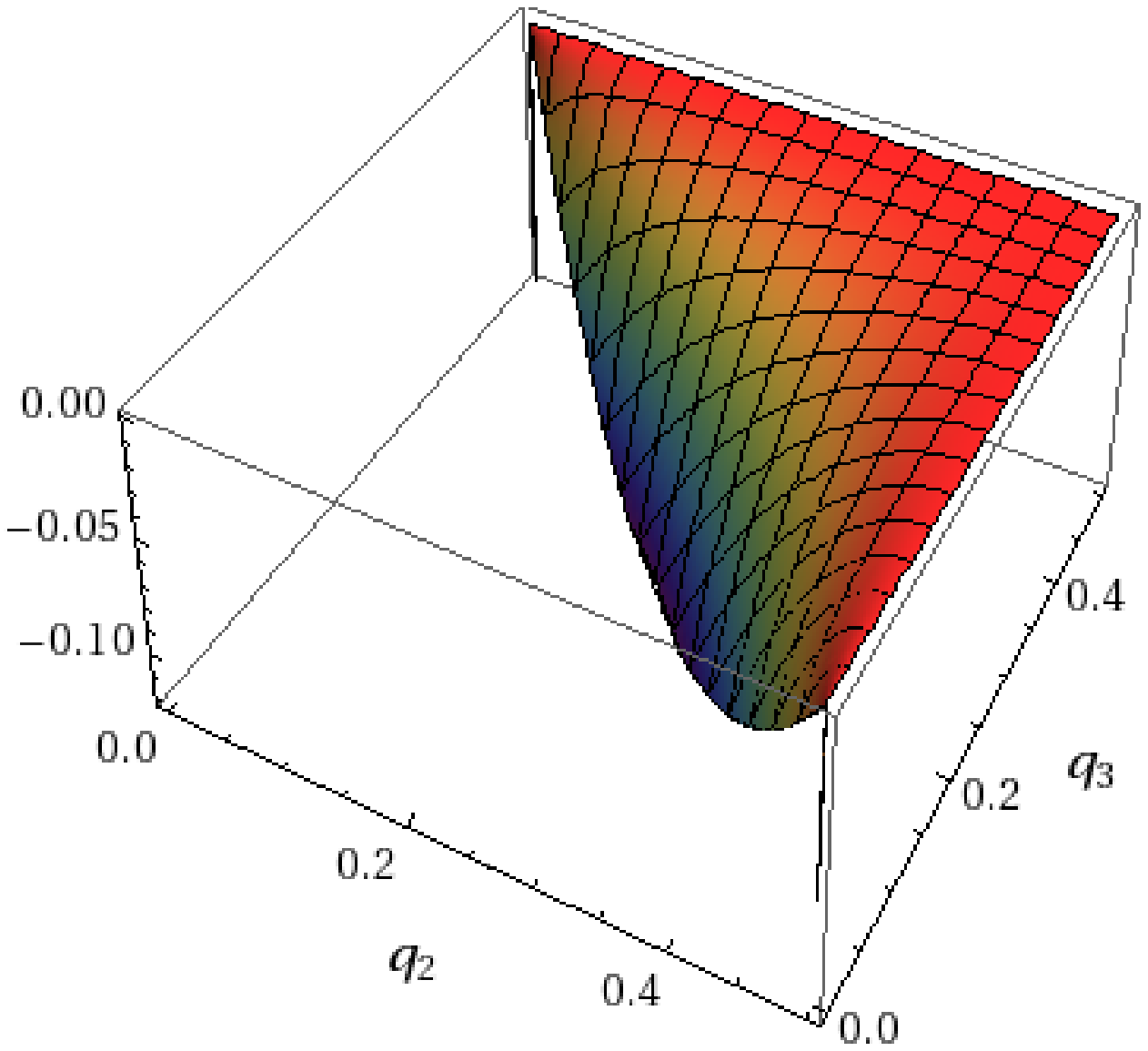}}
\caption{\label{Shapes2} Isoperimetric plots for holographic and slow-roll shape functions continued.  In plots (\ref{fig:Holzpm}) and (\ref{fig:SRzpm}) note that both shape functions are actually finite along the line $q_3=1/2-q_2$ ({\it i.e.}, $q_1=1/2$); we have simply restricted the plot range to exhibit the overall shape more clearly.}
\end{figure}

We have plotted the holographic shape functions in Figs.~\ref{Shapes1} and \ref{Shapes2} (along with their counterparts for slow-roll inflation which we discuss in the next section).
In these figures we have adopted the expedient of scaling all momenta such that $q_1+q_2+q_3=1$ (note that the shape functions are invariant under a constant rescaling of all momenta).  By the usual triangle inequalities, the allowed range for any two momenta, say $q_1$ and $q_2$, is then $0\le q_1 \le 1/2$ and $1/2-q_1\le q_2 \le 1/2$ as displayed.  In each case, we have chosen to plot the two momenta under whose interchange the shape function is symmetric.

Note that the usual plotting convention adopted for the scalar bispectrum $\mathcal{S}(\z\z\z)$ (namely, ordering the momenta $q_1\ge q_2\ge q_3$ and then scaling $q_1$ to unity, with the triangle inequality then constraining $q_2 \ge 1-q_3$) is not applicable to the correlators considered here, since in each case (with the sole exception of $\mathcal{S}(\bg^{(+)}\bg^{(+)}\bg^{(+)})$) one of the three momenta is distinguished and so the required ordering of momenta cannot be accomplished without loss of generality.  Without this initial ordering step, rescaling one of the momenta to unity then fails to yield an upper bound on the magnitude of the remaining momenta, resulting in a plot with unbounded area.  This problem is neatly sidestepped by constraining the total perimeter of the triangle to be unity, instead of the length of one the sides.


\section{Comparison with slow-roll results} \label{sec:slow-roll}

Slow-roll inflation predicts the correlators of three gravitons are
\begin{align}
\label{SRgggbispec}
 \<\!\<\bg^{(+)}(q_1)\bg^{(+)}(q_2)\bg^{(+)}(q_3)\>\!\>_{SR} &= \frac{\K^4H_*^4}{64 \sqrt{2}}\,\frac{\lambda^2a_{123}^2}{c_{123}^5}(a_{123}^3-a_{123}b_{123}-c_{123}), \nn \\[1ex]
 \<\!\<\bg^{(+)}(q_1)\bg^{(+)}(q_2)\bg^{(-)}(q_3)\>\!\>_{SR} &=
\frac{\K^4H_*^4}{64\sqrt{2}}\,\frac{\lambda^2}{a_{123}^2c_{123}^5}(q_3-a_{12})^4(a_{123}^3-a_{123}b_{123}-c_{123}).
\end{align}
We may recover these results exactly from our holographic model by setting
\[
\label{condition}
2\mathcal{N}_\psi = \mathcal{N}_\phi+\mathcal{N}_A+\mathcal{N}_\chi, \qquad \frac{1}{256}N^2 \mathcal{N}_{(A)} = \frac{1}{\K^2 H_*^2}.
\]
In particular, the latter relation is also consistent with matching the amplitude of the graviton 2-point function of slow-roll inflation and the holographic model.
The first relation is consistent with that found in v2 of \cite{Maldacena:2011nz} for the special case where $\mathcal{N}_A=\mathcal{N}_\phi=0$.
(Note however that our careful treatment of the semi-local terms in the holographic formulae enables us to correctly recover the {\it entire} slow-roll bispectrum \eqref{SRgggbispec}.)
For general QFT field content (for which the first relation in \eqref{condition} is not satisfied), the $\bg^{(+)}\bg^{(+)}\bg^{(+)}$ and $\bg^{(+)}\bg^{(+)}\bg^{(-)}$ holographic bispectra in \eqref{main_hol_results} coincide precisely with the corresponding bispectra derived in \cite{Maldacena:2011nz} for slow-roll inflation in which one includes an additional term in the action proportional to the Weyl tensor cubed.
Relative to \cite{Maldacena:2011nz}, our QFT additionally contains non-conformal fields, and our treatment of the semi-local terms enables us to recover the cosmological result exactly.

The remaining slow-roll results are
\begin{align}
\<\!\<\z(q_1)\z(q_2)\bg^{(+)}(q_3)\>\!\>_{SR} & = \frac{\K^4 H_*^4}{16\sqrt{2}\ep_*} \frac{\lambda^2}{a_{123}^2 c_{123}^3 q_3^2} \big[a_{123}^3-a_{123}b_{123}-c_{123}\big],\nn\\[2ex]
\<\!\<\z(q_1)\bg^{(+)}(q_2)\bg^{(+)}(q_3)\>\!\>_{SR} &= -\frac{\K^4H_*^4}{128 b_{23}^5 q_1^2}(q_1^2-a_{23}^2)^2
\Big[(q_1^2-a_{23}^2+2b_{23})-\frac{8 b_{23}^2}{q_1 a_{123}}\Big],
\nn\\[2ex]
\<\!\<\z(q_1)\bg^{(+)}(q_2)\bg^{(-)}(q_3)\>\!\>_{SR} &= -\frac{\K^4H_*^4}{128\, b_{23}^5 q_1^2}(q_1^2-a_{23}^2+4b_{23})^2\big[(q_1^2-a_{23}^2+2b_{23})-\frac{8 b_{23}^2}{q_1 a_{123}}\big].
\end{align}
While these differ from the predictions of the holographic model, interestingly the difference is only in the last term.

Evaluating the shape functions, for slow-roll inflation the 2-point amplitudes defined analogously to \eqref{2ptampdef} are
\[
\mathcal{A}_{SR}(\z\z) = \frac{\K^2H_*^2}{4\ep_*},\qquad \mathcal{A}_{SR}(\bg\bg) = \K^2 H_*^2.
\]
The slow-roll shape functions then differ from their holographic counterparts by at most a single term:
\begin{align}
\mathcal{S}_{SR}(\z\z\bg^{(+)}) &= \mathcal{S}(\z\z\bg^{(+)})+\frac{1}{4\sqrt{2}}\frac{\lambda^2}{a_{123}c_{123}}, \nn\\[2ex]
\mathcal{S}_{SR}(\z\bg^{(+)}\bg^{(+)}) &= \mathcal{S}(\z\bg^{(+)}\bg^{(+)})+\frac{1}{16 a_{123}c_{123}}(q_1^2-a_{23}^2)^2\Big(1+\frac{4c_{123}}{a_{123}^3}\Big), \nn\\[2ex]
\mathcal{S}_{SR}(\z\bg^{(+)}\bg^{(-)}) &= \mathcal{S}(\z\bg^{(+)}\bg^{(-)})+\frac{1}{16 a_{123} c_{123}}(q_1^2-a_{23}^2+4b_{23})^2, \nn\\[2ex]
\mathcal{S}_{SR}(\bg^{(+)}\bg^{(+)}\bg^{(+)}) &=\mathcal{S}(\bg^{(+)}\bg^{(+)}\bg^{(+)})+\frac{\lambda^2}{\sqrt{2}\,a_{123}^4}\Big(1-\frac{4\mathcal{N}_\psi}{\mathcal{N}_{(A)}}\Big), \nn\\[2ex]
\mathcal{S}_{SR}(\bg^{(+)}\bg^{(+)}\bg^{(-)}) &= \mathcal{S}(\bg^{(+)}\bg^{(+)}\bg^{(-)}).
\end{align}
The holographic and slow-roll shape functions, as well as the difference terms in the expressions above, are plotted in Figs.~\ref{Shapes1} and \ref{Shapes2}.
From these figures it is apparent that the holographic and slow-roll shape functions share the same broad qualititative features in all cases except for $\z\bg^{(+)}\bg^{(+)}$: here, $\mathcal{S}_{SR}(\z\bg^{(+)}\bg^{(+)})$ has a simple pole as the momentum $q_1$ associated with $\zeta$ vanishes, whereas the corresponding holographic shape function has a zero.

At a more quantitative level, a rough indication of the distinguishability of the holographic and slow-roll shape functions may be obtained by evaluating the cosine orthogonality measure proposed in \cite{Fergusson:2008ra} (following earlier work in \cite{Babich:2004gb}),
\[
\label{orth}
C(\mathcal{S},\mathcal{S}') =\frac{F(\mathcal{S},\mathcal{S}')}{\sqrt{F(\mathcal{S},\mathcal{S})F(\mathcal{S}',\mathcal{S}')}},
\]
where the weighted inner product
\[
\label{inner1}
F(\mathcal{S},\mathcal{S}') = \int\d q_1\d q_2 \d q_3 \frac{1}{a_{123}}\mathcal{S}(q_1,q_2,q_3) \mathcal{S}'(q_1,q_2,q_3).
\]
Writing $q_1=\alpha \hat{q}_1$, $q_2=\alpha \hat{q_2}$ and $q_3=\alpha(1-\hat{q_1}-\hat{q_2})$, the integral over $\alpha$ in the inner product factors out, since all shape functions we consider here are scale-invariant, {\it i.e.}, independent of $\alpha$.  This overall factor may then be discarded since its contribution to the cosine measure $C(\mathcal{S},\mathcal{S}')$ cancels between numerator and denominator. We may thus replace \eqref{inner1} with the two-dimensional integral
\[
\label{inner2}
F(\mathcal{S},\mathcal{S}') = \int\d \hat{q}_1\d \hat{q}_2 \mathcal{S}(\hat{q}_1,\hat{q}_2,1-\hat{q}_1-\hat{q}_2) \mathcal{S}'(\hat{q}_1,\hat{q}_2,1-\hat{q}_1-\hat{q}_2),
\]
where the shape functions here are precisely the isoperimetric shape functions plotted in Figs.~\ref{Shapes1} and \ref{Shapes2}.

Naively, one might expect the domain of integration would be $0<\hat{q}_1<1/2$ and $1/2-\hat{q}_1<\hat{q}_2<1/2$.
Since however several of the shape functions have poles when one or more of the triangle sides are taken to zero, as we see from Figs.~\ref{Shapes1} and \ref{Shapes2}, one must further restrict the domain of integration in order to obtain finite inner products.  The physical justification for this procedure is that any real observation is only sensitive to momenta in some range $q_{\mathrm{min}}< q_i < q_{\mathrm{max}}$.  We will therefore restrict all rescaled momenta $\hat{q}_i>\ep$, where the cutoff $\ep = q_{\mathrm{min}}/2q_{\mathrm{max}}\sim 5\times 10^{-4}$.  The domain of integration $0<\hat{q}_1<1/2$ and $1/2-\hat{q}_1<\hat{q}_2<1/2$ is thus further restricted by the conditions $\hat{q}_1>\ep$, $\hat{q}_2>\ep$, and $1-\hat{q}_2-\hat{q}_3>\ep$.  For shape functions with poles at the corners, the orthogonality measure \eqref{orth} will depend on the cutoff $\ep$, reflecting the fact that our ability to resolve the shape functions concerned depends on how sensitive we are to the corners of the distribution.

Having thus carefully defined the orthogonality measure, one may now numerically evaluate the orthogonality measure between each holographic shape function and its slow-roll counterpart.  Rounding to two decimal places,
\begin{align}
&C(\bg^{(+)}\bg^{(+)}\bg^{(+)}) \approx 1.00, \quad C(\z\bg^{(+)}\bg^{(+)}) = 0.33, \quad 
C(\z\bg^{(+)}\bg^{(-)})=0.67, \quad C(\z\z\bg^{(+)}) \approx 1.00. \nn
\end{align}
Values close to unity indicate nearly indistinguishable shape functions, while smaller values correspond to shape functions that are more orthogonal.  (For comparison, the overlap between the standard local and equilateral shape functions evaluates to $C=0.34$ with our cutoff prescription.)
Overall, these values confirm one's impression by eye from Figs.~\ref{Shapes1} and \ref{Shapes2}; namely, that the holographic and slow-roll shape functions are nearly indistiguishable for the cases $\bg^{(+)}\bg^{(+)}\bg^{(+)}$ and $\z\z\bg^{(+)}$, while in the case $\z\bg^{(+)}\bg^{(+)}$ the two shape functions may be distinguished by the presence or absence of a pole as the momentum $q_1$ associated with $\z$ vanishes.

\section{Discussion} \label{sec:disc}

In this paper we computed the complete set of bispectra
(and defined and extracted the corresponding shapes\footnote{ To our
knowledge, shapes other than those relevant for purely scalar or purely tensor
bispectra have not been discussed before.})
for a class of holographic models of the very early universe 
based on perturbative QFT.
 The leading 1-loop result actually depends
only on the free part of the QFT, so in particular
our results are also the complete
answer when the dual QFT is free.
The field content of the dual theory includes gauge fields,
massless fermions, massless minimal and conformal scalars and thus the
parameters that can appear in the results are the number of species
for each type of field.  The bispectra could, a priori, depend on these in
a complicated way, but it turns out that we get instead (nearly)
universal results that are independent of all details of the dual QFT,
within the class of the theories we consider. Thus, 
these models make clean and precise predictions.

One can trace this universality to the specific form of the holographic
map, the fact that to leading order the QFT is free,
symmetry considerations and properties of $d=3$ theories.
Let us explain this.
Firstly, in three dimensions, vectors are dual to scalars so
one may anticipate that the contribution due to gauge fields (at
1-loop order) is equal to that of the contribution due to mininal
scalars, and we indeed find this to be the case. Taking this into
account, the answer could then depend on three parameters, the number
of conformal scalars, ${\cal N}_\chi$, the number of fermions, ${\cal
  N}_\psi$ and the total number of gauge fields plus minimal scalars,
${\cal N}_{(B)}$.  The trace Ward identity of the dual QFT and the
specific form of the holographic formulae then imply
that, in all correlators involving at least one factor of $\zeta$, the
field content appears only as a multiplicative factor and is such that the
corresponding shape functions are completely independent of the
field content.

Let us now turn to correlators involving
only tensors: these are effectively determined by the 3-point function
of the stress tensor of a CFT. In three dimensions, this 3-point function is
parametrised by two constants, which in our case are related to the
field content. Indeed, the shape corresponding to three positive
helicity gravitons does depend on the field content, but
surprisingly the shape for two positive and one negative helicity
graviton is independent of the field content. This can be explained in part
by the fact that the $\<T^{(+)} T^{(+)} T^{(-)}\>$ correlator
(at separated points) is actually
uniquely fixed by conformal invariance up to a single constant.
We emphasize however that this by itself
is not sufficient to explain the independence of the corresponding
shape function from the
field content, as the specific form of the semi-local terms (both in the
holographic map and in $\<T^{(+)} T^{(+)} T^{(-)}\>$) is crucial
for this to happen.

Our calculations carefully 
include all such semi-local contributions. 
In the holographic formulae for the bispectra,
these contributions appear as 
 terms in the numerator that are non-analytic in only one of the three momenta.
Since the denominator of the holographic formulae is however non-analytic in all three momenta, 
the net contribution  of these semi-local terms to the bispectra is in fact non-analytic in {\it two} of the three momenta.
Semi-local terms in the holographic formulae may thus contribute, for example, to `local'-type non-Gaussianity behaving as 
$1/q_1^3 q_2^{3} + \mathrm{perms}$.  
Contributions of this nature therefore play a crucial role in allowing different cosmological shapes to be distinguished.

To get a feeling for our results we also computed the corresponding
slow-roll results and compared them with the holographic results. 
Firstly, comparing the power spectra one obtains a relation between the
parameters  $N^2$, ${\cal N}_{(A)}$ and ${\cal N}_{(B)}$ of the QFT
and the parameters  $\kappa^2$, $H_*^2$ and $\epsilon_*$ of the slow-roll model.
Comparing the 3-point functions, we find that the
$\bg^{(+)}\bg^{(+)}\bg^{(-)}$ correlators agree exactly,   
while the
$\bg^{(+)}\bg^{(+)}\bg^{(+)}$ correlators can be made to agree 
if one imposes that that the field content satisfies the relation
$2\mathcal{N}_\psi = \mathcal{N}_\phi+\mathcal{N}_A+\mathcal{N}_\chi$.
As explained in \cite{Maldacena:2011nz}, these slow-roll correlators are
constrained by the late-time de Sitter isometries to satisfy
conformal Ward identities, and thus at separated points they should be
expressible in terms of the 3-point  functions of conformal
scalars and free fermions. Indeed, the linear combination found in v2 of \cite{Maldacena:2011nz}
is the same as the one we find (setting
$\mathcal{N}_\phi=\mathcal{N}_A=0$ in our relation). 
By taking into account the contribution from semi-local terms, however, we are further able to correctly recover every individual
term appearing in the graviton bispectra.

There is no apparent reason for the remaining slow-roll and holographic correlators to agree.
Nevertheless we find rather similar results. To quantify the difference
we used the cosine orthogonality measure of \cite{Fergusson:2008ra} 
to obtain a first indication of
the distinguishability of the corresponding shapes. We find that
the shapes for $\z \z \bg^{(s)}$ are nearly indistinguishable, while for
 $\z \bg^{(+)} \bg^{(+)}$, the two shapes may be distinguished (as a consequence of differing behaviour
 in the squeezed limit where the momentum associated with the $\z$ goes to zero), 
with the case of $\z \bg^{(+)} \bg^{(-)}$ lying in between.

All in all, we have a rather complete understanding of this class
of models and their phenomenology. There are still a few things
to be understood better: what constrains the semi-local contributions to
the tensor correlators, and why are the holographic results apparently
close to slow-roll ones?
One can presumably also understand the squeezed limit of the correlators
using Ward identities. On a whole, however, the structure of these
models is reasonably firmly understood. It would be interesting to arrive at a
similar level of understanding for the class holographic models
that are based on deformations of conformal field theories.

\section*{Acknowledgments}
We thank Raphael Flauger for discussion.
This work is part of the research program of the Stichting voor
Fundamenteel Onderzoek der Materie (FOM), which is financially
supported by the  Nederlandse Organisatie voor Wetenschappelijk
Onderzoek (NWO). KS and AB acknowledge support via an NWO
Vici grant, and PM via an NWO Veni grant.
Research at the Perimeter Institute is supported by the Government of Canada
through Industry Canada and by the Province of Ontario through the Ministry of
Research \& Innovation.

\appendix

\section{Helicity tensors} \label{sec:hel} 

\label{App_helicity}

This appendix summarises our notation and conventions for 
helicity tensors and their contractions.
To facilitate the comparison of our results with those of \cite{Maldacena:2011nz},
we also briefly review the spinor helicity formalism of this latter work.

We use helicity tensors $\ep^{(s)}_{ij}(\vec{\bq})$ satisfying the standard identities
\[
\label{PiTTdecomp}
\Pi_{ijkl}(\bq) = \half\ep^{(s)}_{ij}(\vbq)\ep^{(s)}_{kl}(-\vbq), \qquad
 \ep^{(s)}_{ij}(\vbq)\ep^{(s')}_{ij}(-\vbq) = 2\delta^{ss'}.
\]
where helicities $s_i$ take values $\pm 1$ and our conventions for are those of \cite{WeinbergBook} (see p.~233)).
We may go from a tensor basis to a helicity basis
by contracting with $\ep^{(s)}_{ij}(\vec{\bq})$. Explicitly, the
trace and helicity components of the stress tensor $T_{ij}$ and the $\Upsilon_{ijkl}$
tensor are defined by
\begin{align}
&T(\bq) = \delta_{ij} T_{ij}(\bq), \qquad
T^{(s)}(\bq)= \half\ep^{(s)}_{ij}(-\vbq) T_{ij}(\bq), \qquad \Upsilon(\bq_1,\bq_2) =\delta_{ij}\delta_{kl} \Upsilon_{ijkl}(\bq_1,\bq_2),\\
& \Upsilon^{(s)}(\bq_1,\bq_2) =\half \delta_{ij}\ep^{(s)}_{kl}(-\vbq_2) \Upsilon_{ijkl}(\bq_1,\bq_2), \quad
\Upsilon^{(s_1 s_2)}(\bq_1,\bq_2) =\frac{1}{4} \ep^{(s_1)}_{ij}(-\vbq_1) \ep^{(s_2)}_{kl}(-\vbq_2) \Upsilon_{ijkl}(\bq_1,\bq_2). \nonumber
\end{align}
The various contractions of helicity tensors appearing in the main text are
\begin{align}
\label{Theta_def}
&\Theta_1^{(s_3)}(\bq_i) = \pi_{ij}(\bq_1)\ep_{ij}^{(s_3)}(-\vbq_3), \qquad
&& \Theta_2^{(s_3)}(\bq_i) = \pi_{ij}(\bq_2)\ep_{ij}^{(s_3)}(-\vbq_3), \nn\\
& \Theta^{(s_2s_3)}(\bq_i) = \pi_{ij}(\bq_1)\ep_{ik}^{(s_2)}(-\vbq_2)\ep_{kj}^{(s_3)}(-\vbq_3), \qquad
&&  \theta^{(s_2s_3)}(\bq_i) = \ep^{(s_2)}_{ij}(-\vbq_2)\ep^{(s_3)}_{ij}(-\vbq_3), \nn\\
&\Theta^{(s_1s_2s_3)}(\bq_i) = \ep^{(s_1)}_{ij}(-\vbq_1)\ep^{(s_2)}_{jk}(-\vbq_2)\ep^{(s_3)}_{ki}(-\vbq_3),
\end{align}
where the projection operator $\pi_{ij}$ is given in \eqref{projection_operators}.

We may explicitly evaluate these contractions in terms of the magnitudes $q_i$ of the momenta and the helicities $s_i$
by introducing a basis for the helicity tensors.
To do so, we first observe that the momenta $\vbq_i$ lie in a single plane due to momentum conservation.
Taking this plane to be the $(x,z)$ plane, we may then write
\[
\label{momconvs}
\vbq_i = \bq_i (\sin\theta_i, \,0,\,\cos\theta_i)
\]
where the magnitudes $\bq_i\ge 0$, and without loss of generality we may choose $\theta_1=0$, $0\le\theta_2\le\pi$ and $\pi\le\theta_3\le2\pi$ so that
\[
\label{trig_expressions}
 \cos\theta_2 = \frac{(\bq_3^2-\bq_1^2-\bq_2^2)}{2\bq_1\bq_2}, \quad \sin\theta_2 = \frac{\blambda}{2\bq_1\bq_2}, \quad 
\cos\theta_3 = \frac{(\bq_2^2-\bq_1^2-\bq_3^2)}{2\bq_1\bq_3}, \quad \sin\theta_3 = -\frac{\blambda}{2\bq_1\bq_3},
\]
with $\blambda$ as given in \eqref{blambda_def}.
The required helicity tensors then follow by rotation in the $(x,z)$ plane:
\vspace{0.3cm}
\begin{align}
\label{helicity_basis}
 \ep^{(s_i)}(\vbq_i) = \frac{1}{\sqrt{2}}\left(\begin{array}{ccc} \cos^2\theta_i & is_i\cos\theta_i & -\sin\theta_i\cos\theta_i \\ is_i\cos\theta_i & -1 & -is_i\sin\theta_i \\ -\sin\theta_i\cos\theta_i & -is_i\sin\theta_i & \sin^2\theta_i \end{array}\right). \end{align}
The contractions of helicity tensors used in this paper are then
\begin{align}
\label{Theta_results}
&\Theta_1^{(\pm)}(\bq_i) = -\frac{\blambda^2}{4\sqrt{2} \b_{13}^2},
&&\hspace{-4.5cm}\Theta_2^{(\pm)}(\bq_i) = -\frac{\blambda^2}{4\sqrt{2} \b_{23}^2},\nn\\[1ex]
&\Theta^{(+++)}(\bq_i) = -\frac{\blambda^2\a_{123}^2}{16\sqrt{2}\c_{123}^2},
&&\hspace{-4.5cm}\Theta^{(++-)}(\bq_i) = -\frac{\blambda^2}{16\sqrt{2}\c_{123}^2}(\bq_3-\a_{12})^2, \nn\\[1ex]
&\theta^{(++)}(\bq_i) = \frac{\a_{123}^2(\a_{23}-\bq_1)^2}{8\b_{23}^2},
&&\hspace{-4.5cm}\theta^{(+-)}(\bq_i) = \frac{(\a_{13}-\bq_2)^2(\a_{12}-\bq_3)^2}{8\b_{23}^2}, \nn\\[1ex]
&\Theta^{(++)}(\bq_i)  =
\frac{\a_{123}(\a_{23}-\bq_1)}{16\c_{123}^2}\,\big[2\bq_1^2\a_{123}(\a_{23}-\bq_1)-\blambda^2\big],&&  \nn\\[1ex]
&\Theta^{(+-)}(\bq_i) =
 \frac{(\a_{13}-\bq_2)(\a_{12}-\bq_3)}{16\c_{123}^2}\,\big[2\bq_1^2(\a_{13}-\bq_2)(\a_{12}-\bq_3)+\blambda^2\big]. &&
\end{align}

Let us now discuss the spinor helicity formalism introduced in \cite{Maldacena:2011nz}.
For any three-dimensional vector $\vec{\bar{q}}$ we can consider a four-dimensional vector $\bar{q}^\mu = ( \bar{q}, \vec{\bar{q}} )$ which satisfies $\bar{q}^\mu \bar{q}_\mu = 0$. Therefore, $\vec{\bar{q}}$ can be represented by spinors $\lambda_a$ as $\bar{q}^\mu = \sigma^{\mu}_{\dot{a} a} \lambda^a \bar{\lambda}^{\dot{a}}$,
where we use the same conventions as in \cite{Maldacena:2011nz}, namely
\begin{eqnarray}
\epsilon^{ab} & = & \epsilon^{\dot{a} \dot{b}} = \epsilon^{a \dot{a}} = \left( \begin{array}{cc} 0 & 1 \\ -1 & 0 \end{array} \right), \qquad
\sigma^{\mu a}_{\ \ \: b} =  ( - \delta^a_{\ b}, \vec{\sigma}^{\mu a}_{\ \ \: b} ), \nn \\[1ex]
\bar{\lambda}^{\dot{a}} & = & - \epsilon_{\dot{a} b} ( \lambda^b )^{\ast}, \qquad \qquad
\bar{q}_1^\mu \bar{q}_{2 \mu} = -2 \langle 1 2 \rangle \langle \bar{1} \bar{2} \rangle, \nn\\[1ex]
\langle 1 2 \rangle & = & \lambda_{1a} \lambda_2^{a} = \epsilon_{a b} \lambda_1^{a} \lambda_2^{b}, \qquad \langle \bar{1} \bar{2} \rangle = \bar{\lambda}_{1\dot{a}} \bar{\lambda}_2^{\dot{a}} = \epsilon_{\dot{a} \dot{b}} \bar{\lambda}_1^{\dot{a}} \bar{\lambda}_2^{\dot{b}}.
\end{eqnarray}
Here, $\vec{\sigma}^{\mu a}_{\ \ \: b}$ is a vector of Pauli matrices.  The spinors $\lambda_{1a}$ and $\lambda_{2a}$ corresponding to the two momenta $\vec{\bar{q}}_1$ and $\vec{\bar{q}}_2$
are denoted by $| 1 \rangle$ and $| 2 \rangle$, respectively. Spinor indices are raised and lowered by means of $\epsilon^{a b}$ and its inverse.

To compare our results with those of \cite{Maldacena:2011nz}, we need an explicit expression of the inner products $\langle 1 2 \rangle$, {\it etc.}, in terms of
momenta. A possible solution for a spinor\footnote{We corrected signs in (B.1) in v2 of \cite{Maldacena:2011nz} so that $\bar{q}^\mu = \sigma^{\mu}_{\dot{a} a} \lambda^a \bar{\lambda}^{\dot{a}}$.} is
\begin{equation}
\lambda^a = \left( \begin{array}{c}
\sqrt{ \frac{\bar{q} - \bar{q}_3}{2} } \\
\frac{-\bar{q}_1-i \bar{q}_2}{\sqrt{2 (\bar{q} - \bar{q}_3)}}
\end{array} \right)
=
\sqrt{\bar{q}} \left( \begin{array}{c}
\sin \left( \frac{1}{2} \theta \right) \\
- \cos \left( \frac{1}{2} \theta \right) e^{i \phi}
\end{array} \right),
\end{equation}
where the second expression makes use of the spherical coordinates
\begin{equation}
\vec{\bar{q}} = \bar{q} ( \sin \theta \cos \phi, \sin \theta \sin \phi, \cos \theta ).
\end{equation}
Choosing momenta $\vbq_1$ and $\vbq_2$ as in \eqref{momconvs}, and making use of \eqref{trig_expressions}, we find 
\[
\langle 1 2 \rangle =  - \sqrt{\bar{q}_1 \bar{q}_2} \sin \left( \frac{1}{2} \theta_2 \right)
 =  - \frac{1}{2} \sqrt{(\bar{q}_1 + \bar{q}_2)^2 - \bar{q}_3^2}.
\]
In general the sign depends on the orientation of $(\vec{\bar{q}}_1, \vec{\bar{q}}_2)$. Since we choose $0\le\theta_2\le\pi$, the orientation is assumed to be positive.
Note in particular that $\langle 2 1 \rangle = - \langle 1 2 \rangle$. Similarly, we find
\[
\langle 1 \bar{2} \rangle  =   \frac{1}{2} \sqrt{\bar{q}_3^2 - (\bar{q}_1 - \bar{q}_2)^2}\, , \qquad
\langle \bar{1} \bar{2} \rangle  =  - \frac{1}{2} \sqrt{(\bar{q}_1 + \bar{q}_2)^2 - \bar{q}_3^2} = \langle 1 2 \rangle\, ,
\]
in agreement with (B.6) of \cite{Maldacena:2011nz}. Combining these results we find
\begin{eqnarray}
\left[ \langle \bar{1} \bar{2} \rangle \langle \bar{2} \bar{3} \rangle \langle \bar{3} \bar{1} \rangle \right]^2 & = & \frac{\blambda^2}{64} \a_{123}^2 
= - \frac{1}{2 \sqrt{2}} \bar{c}^2_{123} \Theta^{(+++)}(\bar{q}_i), \nonumber \\
\left[ \langle \bar{1} \bar{2} \rangle \langle \bar{2} 3 \rangle \langle 3 \bar{1} \rangle \right]^2 & = & \frac{\blambda^2}{64}
(\a_{12}-\bq_3)^2
= - \frac{1}{2 \sqrt{2}} \bar{c}_{123}^2 \Theta^{(++-)}(\bar{q}_i), \label{Mus}
\end{eqnarray}
which we have made use of in the main text.

In four dimensions, the complexified symmetry group is locally isomorphic to $SL(2, \mathbb{C}) \times SL(2, \mathbb{C})$, in which case dotted and undotted indices transform independently. In our case, however, the symmetry group is that of rotations of three-dimensional space, which corresponds to $SL(2, \mathbb{C}) \hookrightarrow SL(2, \mathbb{C}) \times SL(2, \mathbb{C})$ embedded diagonally. An additional invariant tensor therefore exists, which we may choose to be $\epsilon^{a \dot{a}}$: this means that we are now allowed to contract dotted with undotted indices. In particular,
\begin{equation}
\langle \lambda \bar{\lambda} \rangle = \lambda^a \epsilon_{\dot{a} a} \bar{\lambda}^{\dot{a}}  = - \lambda^a ( \lambda^a )^{\ast} = - \bar{\lambda}^{\dot{a}} ( \bar{\lambda}^{\dot{a}} )^{\ast} = - \bar{q},
\end{equation}
motivating the following definition for complex conjugates
\begin{equation}
\lambda_{\dot{a}} = ( \bar{\lambda}^{\dot{a}} )^\ast = - \epsilon_{\dot{a} a} \lambda^a, \qquad \bar{\lambda}^a = (\lambda_a)^\ast = - \epsilon^{a \dot{a}} \bar{\lambda}_{\dot{a}}.
\end{equation}

The helicity tensors used in \cite{Maldacena:2011nz} may now be defined as
\begin{equation}
\epsilon_{\text{MP}}^{(s) a b \dot{a} \dot{b}} = \xi_{\text{MP}}^{(s) a \dot{a}} \xi_{\text{MP}}^{(s) b \dot{b}},
\end{equation}
where
\begin{equation}
\xi_{\text{MP}}^{(+) a \dot{a}} = \frac{\bar{\lambda}^a \bar{\lambda}^{\dot{a}}}{\langle \bar{\lambda} \lambda \rangle}, \qquad
\xi_{\text{MP}}^{(-) a \dot{a}} = \frac{\lambda^a \lambda^{\dot{a}}}{\langle \lambda \bar{\lambda} \rangle}.
\end{equation}
Contracting with Pauli matrices, we then find
\begin{equation}
\label{MPdef}
\xi_{\text{MP}}^{(+) \mu} = e^{- i \phi} \left( \begin{array}{c}
0 \\
\cos \theta \cos \phi - i \sin \phi \\
\cos \theta \sin \phi + i \cos \phi \\
- \sin \theta
\end{array} \right), \qquad
\xi_{\text{MP}}^{(-) \mu} = e^{i \phi} \left( \begin{array}{c}
0 \\
\cos \theta \cos \phi + i \sin \phi \\
\cos \theta \sin \phi - i \cos \phi \\
- \sin \theta
\end{array} \right).
\end{equation}
Since the time components vanish, we may regard the $\xi_{\text{MP}}^{(s) \mu}$ as three-dimensional vectors, and if $\bar{q}_\mu \xi_{\text{MP}}^{(s) \mu} = 0$ in four dimensions, then clearly $\bar{q}_i \xi_{\text{MP}}^{(s) i} = 0$ in three dimensions as well.

Let us now compare the vectors $\xi_{\text{MP}}^{(s) \mu}$ with those implicit in our own convention \eqref{helicity_basis} for the helicity tensors.   In our case, we started with
\begin{equation}
\xi^{(s) \mu}(\vec{\bar{q}}_1) = (0, 1, i s, 0) \quad \text{for} \quad \bar{q}_1^\mu = (\bar{q}_1,0,0,\bar{q}_1),
\end{equation}
and then obtained all other $\xi^{(s) \mu}(\vec{\bar{q}})$ by rotation in the $(xz)$ plane. In this way, we find that
\begin{equation}
\xi^{(+) \mu}_{\text{MP}}(\vec{\bar{q}}) = e^{- i \phi} \xi^{(+) \mu}(- \vec{\bar{q}}), \qquad \xi^{(-) \mu}_{\text{MP}}(\vec{\bar{q}}) = e^{i \phi} \xi^{(-) \mu}(- \vec{\bar{q}}).
\end{equation}
Our normalisation of helicity tensors is then such that
\begin{equation}
\epsilon^{(s)}_{ij} = \frac{1}{\sqrt{2}} \xi^{(s)}_i \xi^{(s)}_j,
\end{equation}
thus we find
\begin{equation}
\epsilon^{(s)}_{ij}(\vec{\bar{q}}) = \frac{1}{\sqrt{2}} \epsilon^{(s)}_{\text{MP}\: ij}(- \vec{\bar{q}}).
\end{equation}
Finally, we chose to
define $T^{(s)}(\vec{\bar{q}}) = \half \ep^{(s)}_{ij}(-\vec{\bar{q}}) T_{ij}(\vec{\bar{q}})$ (so that $T_{ij}(\vec{\bar{q}}) = T^{(s)}(\vec{\bar{q}}) \ep^{(s)}_{ij}(\vec{\bar{q}})$),
which leads to
\begin{equation}
T^{(s)}(\vec{\bar{q}}) = \frac{1}{2 \sqrt{2}} T^{(s)}_{\text{MP}}(\vec{\bar{q}}),
\end{equation}
where \cite{Maldacena:2011nz} defines instead $T^{(s)}_{\text{MP}}(\vec{\bar{q}}) = \epsilon_{\text{MP}}^{(s) ij}(\vec{\bar{q}}) T_{ij}(\vec{\bar{q}})$.

\section{Evaluation of integrals}
\label{methods_app}

To evaluate the holographic formulae \eqref{holo_zzg}-\eqref{holo_ggg} we must compute specific helicity projections of the stress tensor 3-point function.  One option, discussed in Section \ref{Method1} below, is to project into a helicity basis at the very outset of the calculation, leaving only relatively straightforward scalar integrals to evaluate.
A second option, discussed in Sections \ref{Method2} and \ref{Method3}, is to directly evaluate the tensor integrals for the full 3-point function, projecting into a helicity basis only as the final step.
While the evaluation of tensor integrals is more demanding, the Ward identities permit a useful consistency check of the results.
The required evaluation of tensor integrals may be accomplished using either a method due to Davydychev \cite{Davydychev:1991va, Davydychev:1992xr}, or else via a Feynman parametrisation approach, as discussed in Section \ref{Method2} and \ref{Method3} respectively.
In practice, we computed integrals using all three methods and cross-checked the results for each method against those of the others.

\subsection{Helicity projection to scalar integrals}
\label{Method1}

To illustrate the steps involved, let us consider the following integral derived from the result \eqref{min_scalars_int} for minimal scalars
\[
\label{sample_int}
\<\!\<T_\phi(\bq_1)T_\phi(\bq_2)T_\phi^{(s_3)}(\bq_3)\>\!\> = \mathcal{N}_\phi \bar{N}^2 \int[\d\bq]\frac{\bq\!\cdot\!(\bq+\bq_1) \,(\bq+\bq_1)\!\cdot\!(\bq-\bq_3)\,\ep_{ij}^{(s_3)}(-\vbq_3)\bq_i\bq_j}{\bq^2(\bq+\bq_1)^2(\bq-\bq_3)^2}.
\]
Making use of the explicit basis \eqref{helicity_basis}, we find
\[
\sqrt{2}\ep_{ij}^{(s_3)}(-\vbq_3)\bq_i\bq_j
=\bq_x^2\cos^2\theta_3+\bq_z^2\sin^2\theta_3-\bq_y^2-2\bq_x\bq_z\sin\theta_3\cos\theta_3+2is_3\bq_y\bq_z\sin\theta_3-2is_3\bq_x\bq_y\cos\theta_3.
\]
Since the external vectors $\vbq_i$ all lie in the $(x,z)$ plane and thus have no $y$-component,
the imaginary part of the integral \eqref{sample_int} is odd under $\bq_y\tto -\bq_y$ and therefore vanishes.
To deal with the remainder, it is then convenient to replace $\bq_y^2 = \bq^2-\bq_x^2-\bq_z^2$ and to substitute for $\bq_z$ and $\bq_x$ according to
\begin{align}
\bq_z &= \frac{\bq\cdot\bq_1}{\bq_1}, \qquad
\bq_x =\frac{1}{\bq_3 \sin\theta_3}\,\bq\cdot\bq_3-\frac{\cot\theta_3 }{\bq_1}\,\bq\cdot\bq_1.
\end{align}
Here, trigonometric expressions involving $\theta_3$ are equivalent to specific combinations of external momenta according to \eqref{trig_expressions}.
Finally, using the standard replacements $2\bq\cdot\bq_1 = (\bq+\bq_1)^2-\bq^2-\bq_1^2$, {\it etc.}, the integral \eqref{sample_int} may be reduced to a sum of elementary 2-point integrals and a single 3-point integral,
\[
\int [\d\bq]\frac{1}{\bq^2(\bq+\bq_1)^2(\bq-\bq_3)^2} = \frac{1}{8\bq_1\bq_2\bq_3}.
\]
(Note that this latter integral reduces to a standard 2-point integral upon substituting $\vbq\,' = \vbq/\bq^2$ and $\vbq\,'_i = \vbq_i/\bq_i^2$).

The evaluation of all remaining helicity-projected 3-point integrals proceeds in a similar fashion, the only complexity arising from  the need to keep track of moderately large expressions.

\subsection{Tensor integrals via Davydychev recursion}
\label{Method2}

An elegant general method for evaluating tensor Feynman integrals corresponding to arbitrary 1-loop $N$-point diagrams was proposed by Davydychev in \cite{Davydychev:1991va,  Davydychev:1992xr}.  
Here, we review its application to the tensor integrals appearing in our calculations of the stress tensor 3-point function.

\begin{figure}[t]
\begin{center}
\includegraphics[width=7cm]{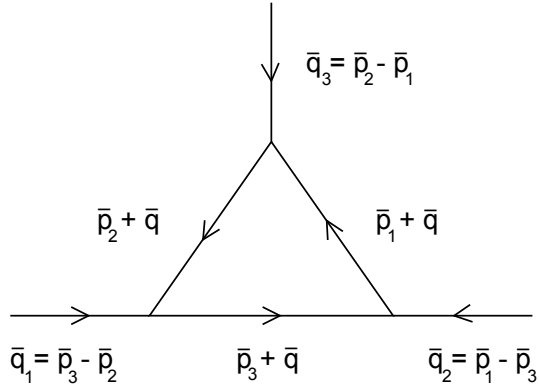}
\caption{\label{Tri}
Labelling of momenta}
\end{center}
\end{figure}

Our goal will therefore be to evaluate massless 1-loop 3-point integrals of the general form
\[
\label{tensorint1}
J_{\mu_1\ldots \mu_M}(n;\{\nu_i\}) \equiv \int \d^n \bq \frac{\bq_{\mu_1}\ldots \bq_{\mu_M}}{(\bq+\bp_1)^{2\nu_1}(\bq+\bp_2)^{2\nu_2}(\bq+\bp_3)^{2\nu_3}},
\]
where, for reasons that will be apparent shortly, we have kept the spacetime dimension $n$, as well as the powers $\nu_i$ (where $i=1\ldots 3$) appearing in the denominator, arbitrary.  We will temporarily denote spacetime indices with Greek letters
to avoid confusion with the index $i$.
Note also that our choice of Euclidean signature will result in a few minor changes
with respect to the corresponding formulae reported in \cite{Davydychev:1991va, Davydychev:1992xr}\footnote{Alternatively, one could use the Lorentzian formulae quoted in \cite{Davydychev:1991va, Davydychev:1992xr} and continue to Euclidean signature after completing all computations.}.
The symmetric form of the momenta in the denominator is convenient; to recover the form of the momenta in the denominator used in the main text one simply shifts $\bq_\mu \tto (\bq-\bp_3)_\mu$ (see Fig.~\ref{Tri}).

In \cite{Davydychev:1991va}, a general formula was derived allowing the tensor integral \eqref{tensorint1} to be expressed as a sum of symmetric tensors constructed from the spacetime metric and the external momenta, multiplied by coefficients given in terms of scalar integrals of the form
\[
\label{scalarint1}
J(n;\{\nu_i\})\equiv\int\d^n \bq\frac{1}{(\bq+\bp_1)^{2\nu_1}(\bq+\bp_2)^{2\nu_2}(\bq+\bp_3)^{2\nu_3}}.
\]
Explicitly, in Euclidean signature, this formula reads
\begin{align}
\label{gen_result}
J_{\mu_1\ldots\mu_M}(n;\{\nu_i\}) &= \sum_{\genfrac{}{}{0pt}{}{\lambda, \kappa_i}{2\lambda+\sum \kappa_i=M}}(-\frac{1}{2})^\lambda (-\frac{1}{\pi})^{M-\lambda} (\nu_1)_{\kappa_1}(\nu_2)_{\kappa_2}(\nu_3)_{\kappa_3}   \nn \\&\qquad \quad \times
\{[g]^\lambda[\bp_1]^{\kappa_1}[\bp_2]^{\kappa_2}[\bp_3]^{\kappa_3}\}_{\mu_1\ldots\mu_M}J(n+2(M-\lambda);\{\nu_i+\kappa_i\}),
\end{align}
where $(\nu)_\kappa \equiv \Gamma(\nu+\kappa)/\Gamma(\nu)$ is the Pochhammer symbol and  $\{[g]^\lambda[\bp_1]^{\kappa_1}[\bp_2]^{\kappa_2}[\bp_3]^{\kappa_3}\}_{\mu_1\ldots\mu_M}$ denotes the symmetric tensor constructed out of $\lambda$ copies of the metric tensor and $\kappa_i$ copies of each momenta $\bp_i$.
(Thus, for example,
$
\{g\bp_1\}_{\mu_1\mu_2\mu_3} = g_{\mu_1\mu_2}\bp_{1\mu_3}+g_{\mu_1\mu_3}\bp_{1\mu_2}+g_{\mu_2\mu_3}\bp_{1\mu_1},
$
where for present purposes the metric tensor $g_{\mu\nu}=\delta_{\mu\nu}$.)
In the formula \eqref{gen_result}, the sum runs over all possible non-negative values of $\lambda$ and $\kappa_i$, such that the total rank $2\lambda+\sum_i\kappa_i$ equals $M$.
Note in particular that the values of $n$ and $\nu_i$ appearing in the scalar coefficient integrals differ from those appearing in the original tensor integral \eqref{tensorint1}.

Equipped with the general formula \eqref{gen_result}, we may therefore reduce tensor integrals of the form \eqref{tensorint1} to scalar integrals of the form \eqref{scalarint1}.
The evaluation scheme is then completed
by a set of recursion relations enabling the scalar integrals \eqref{scalarint1} to be reduced to elementary integrals.
In \cite{Davydychev:1992xr}, it was shown that
\begin{align}
\label{recursion1}
&J(n;\{\nu_1,\nu_2,\nu_3+1\})  = \frac{1}{2\nu_3 \bq_2^2 \bq_1^2} \Big[\big(
(2\nu_1+\nu_2+\nu_3-n) \bq_1^2 \nn\\ & \qquad\qquad
+(2\nu_2+\nu_1+\nu_3-n)\bq_2^2 -(2\nu_3+\nu_1+\nu_2-n)\bq_3^2\big)J(n;\{\nu_1,\nu_2,\nu_3\}) \nn\\ &\qquad\qquad
+\nu_2\bq_1^2J(n;\{\nu_1-1,\nu_2+1,\nu_3\})+\nu_1\bq_2^2J(n;\{\nu_1+1,\nu_2-1,\nu_3\}) \nn\\&\qquad\qquad
+\nu_3\bq_1^2J(n;\{\nu_1-1,\nu_2,\nu_3+1\})-\nu_1\bq_3^2J(n;\{\nu_1+1,\nu_2,\nu_3-1\}) \nn\\&\qquad\qquad
+\nu_3\bq_2^2J(n;\{\nu_1,\nu_2-1,\nu_3+1\})-\nu_2\bq_3^2J(n;\{\nu_1,\nu_2+1,\nu_3-1\})\Big],
\end{align}
with similar formulae for $J(n;\{\nu_1+1,\nu_2,\nu_3\})$ and $J(n;\{\nu_1,\nu_2+1,\nu_3\})$ following by permutation of indices.
If we regard the indices $(\nu_1,\nu_2,\nu_3)$ as coordinates on an integer lattice, these recursion
relations allow us to construct three integrals in the plane $\sum_i \nu_i = \sigma+1$ in terms of six contiguous integrals in the plane $\sum_i\nu_i = \sigma$.
Now, in general, we are interested in the region $\nu_i\ge 0$.
Any integrals on the boundary of this region may be evaluated trivially: if more than one of the $\nu_i$ vanish the integral is zero in dimensional regularisation, and if only one of the $\nu_i$ vanishes, the integral reduces to the standard 2-point integral
\[
\label{bdy_int}
J(n;\{\nu_1,\nu_2,0\}) = \frac{\Gamma(\nu_1+\nu_2-n/2)\Gamma(n/2-\nu_1)\Gamma(n/2-\nu_2)}{\Gamma(\nu_1)\Gamma(\nu_2)\Gamma(n-\nu_1-\nu_2)}\,\pi^{n/2} (\bq_3^2)^{n/2-\nu_1-\nu_2}.
\]
The first non-trivial integral for which all the $\nu_i>0$ is therefore $J(n;\{1,1,1\})$, which sits in the plane $\sigma=3$.  From this integral, plus the appropriate `boundary' integrals, we may then use the recursion relations \eqref{recursion1} to construct all the non-trivial integrals in the plane $\sigma=4$, namely $J(n;\{1,1,2\})$ and its permutations.  Through repeated application of the recursion relations, we may proceed to evaluate any integral with positive integer $\{\nu_i\}$ in terms of the initial integral $J(n;\{1,1,1\})$ and boundary integrals of the form \eqref{bdy_int}.

Examining the form of \eqref{gen_result}, we see that to evaluate a tensor integral of rank $M$ in three dimensions, we need to evaluate the corresponding scalar integrals $J(n;\{\nu_i\})$, and hence initial integrals $J(n;\{1,1,1\})$, in all odd dimensions $3\le n \le 3+2M$.
As noted in the previous subsection, the initial integral in three dimensions may be evaluated by inverting all momenta yielding
\[
\label{special1}
J(3;\{1,1,1\}) = \frac{\pi^3}{\bq_1\bq_2\bq_3}. 
\]
The higher odd-dimensional initial integrals may then be obtained using the additional recursion relation
\begin{align}
J(n+2;\{1,1,1\})=\frac{2\pi}{(n-2)(2H_4-H_2^2)}\Big[ & \frac{\pi^{n/2+1}\Gamma(n/2)}{\Gamma(n-1)}\mathrm{cosec}\(\frac{n\pi}{2}\)(H_{n-2}H_2-2H_n) \nn\\&
+(\bq_1\bq_2\bq_3)^2 J(n;\{1,1,1\})\Big],
\end{align}
where
\[
H_n \equiv \bq_1^n+\bq_2^n+\bq_3^n,
\]
which may be derived from \eqref{recursion1} and the Euclidean analogue of equation (6) in \cite{Davydychev:1991va}.

Armed with the above analysis, the evaluation of tensor integrals of the form \eqref{tensorint1} is now straightforward.
For the computation of stress tensor 3-point functions in the main text, we need to evaluate the six tensor integrals with $\nu_1=\nu_2=\nu_3=1$, $n=3$ and ranks $M=1\ldots 6$.
For the lower ranks the calculation may easily be executed by hand, yielding for example
\begin{align}
\int [\d\bq] \frac{\bq_\mu}{\bq^2 (\bq-\bq_1)^2(\bq+\bq_2)^2} &= \frac{1}{8\a_{123}\c_{123}}(\bq_2\bq_{1\mu}-\bq_1\bq_{2\mu}), \nn\\[1ex]
\int [\d\bq] \frac{\bq_\mu\bq_\nu}{\bq^2 (\bq-\bq_1)^2(\bq+\bq_2)^2} &=
\frac{1}{16 \a_{123}^2\c_{123}}\big(\a_{123}\c_{123}\delta_{\mu\nu}+\bq_2(\a_{13}+2\bq_2)\bq_{1\mu}\bq_{2\nu} \nn\\&\qquad\qquad\qquad
+\bq_1(\a_{23}+2\bq_1)\bq_{2\mu}\bq_{2\nu}-\b_{12}\bq_{1\mu}\bq_{2\nu}\big),
\end{align}
but for the higher ranks it is convenient to automate the process.
In the highest rank case $M=6$, we see from \eqref{gen_result} we need to evaluate scalar integrals in planes up to $\sigma=9$ (thus requiring up to six iterations of the recursion relation \eqref{recursion1}), for odd spacetime dimensions up to $n=15$.
Having explicitly computed all tensor integrals up to rank six, any 3-point function $\<\!\<T_{ij}(\bq_1)T_{kl}(\bq_2)T_{mn}(\bq_3)\>\!\>$ may now be directly evaluated.  After checking against the Ward identities for consistency, the result may then be projected into the helicity basis.

\subsection{Tensor integrals via Feynman parametrisation}
\label{Method3}

In this third method, the correlation functions were again calculated directly in the tensor representation $\lla T_{i_1 j_1} T_{i_2 j_2} T_{i_3 j_3} \rra$ and then projected into the helicity basis or traced. All 3-point functions we consider may be expressed as a sum of the integrals of the form
\begin{equation}
\int [ \d \bar{q} ] \frac{t_{i_1 j_1 i_2 j_2 i_3 j_3}}{\bar{q}^2 (\bar{q} - \bar{q}_1)^2 (\bar{q} + \bar{q}_2)^2},
\end{equation}
where $t_{i_1 j_1 i_2 j_2 i_3 j_3}$ is a tensor build up with $\bar{q}, \bar{q}_1, \bar{q}_2$ and a metric $\delta$. In order to calculate this integral, Feynman parameters $x_1, x_2, x_3$, such that $x_1 + x_2 + x_3 = 1$, may be introduced. This leads to the substitution $\bar{q} = \bar{l} + x_2 \bar{q}_1 - x_1 \bar{q}_2$ and the integral takes the form
\begin{equation} \label{e:myint1}
\int [ \d \bar{q} ] \frac{t_{i_1 j_1 i_2 j_2 i_3 j_3}}{\bar{q}^2 (\bar{q} - \bar{q}_1 )^2 (\bar{q} + \bar{q}_2)^2} = 2 \int_{[0,1]^3} \d X \int [ \d \bar{l} ] \frac{t_{i_1 j_1 i_2 j_2 i_3 j_3}}{(\bar{l}^2 + \Delta)^3},
\end{equation}
where
\begin{eqnarray}
\d X & = & \d x_1 \d x_2 \d x_3 \delta(x_1 + x_2 + x_3 - 1), \nn\\
\Delta & = & \bar{q}_1^2 x_2 (1 - x_2) + \bar{q}_2^2 x_1 (1 - x_1) + 2 (\bar{q}_1 \cdot \bar{q}_2) x_1 x_2 \nn\\
& = & \bar{q}_1^2 x_2 x_3 + \bar{q}_2^2 x_1 x_3 + \bar{q}_3^2 x_1 x_2,
\end{eqnarray}
and the integration is over the cube $(x_1, x_2, x_3) \in [0,1]^3$. Finally, we decompose the integral \eqref{e:myint1} into a linear combination of integrals of the form
\begin{equation} \label{e:fullI}
2 \int \d X P(x_1, x_2, x_3) \int [ \d \bar{l} ] \frac{\bar{l}_{i_1} \bar{l}_{j_1} \ldots \bar{l}_{i_m} \bar{l}_{j_m}}{(\bar{l}^2 + \Delta)^3},
\end{equation}
where $P(x_1, x_2, x_3)$ is some polynomial in Feynman parameters. The integral over momenta may be evaluated by means of the formula
\begin{equation} \label{e:regI}
2 \int [ \d \bar{l} ] \frac{\bar{l}_{i_1} \bar{l}_{j_1} \ldots \bar{l}_{i_m} \bar{l}_{j_m}}{(\bar{l}^2 + \Delta)^3} = \frac{\Gamma(3/2 - m)}{(4 \pi)^{3/2}} \frac{S_{i_1 j_1 \ldots i_m j_m}}{2^m} \Delta^{m - 3/2},
\end{equation}
where $S_{i_1 j_1 \ldots i_m j_m}$ is a completely symmetric tensor constructed from metric tensors with all coefficients equal to one. Due to the $\bar{l} \mapsto -\bar{l}$ symmetry the integrals with an odd number of momenta $\bar{l}$ vanish.

The remaining task is to evaluate the integrals
\begin{equation} \label{e:intF}
\int \d X P(x_1, x_2, x_3) \Delta^{m - 3/2}
\end{equation}
over the Feynman parameters. For $d = 3$, the r.h.s.~of \eqref{e:regI} is a well-defined expression for any integer $m$, and \eqref{e:intF} exists for any polynomial $P$ and any non-negative $m$. It turns out that in order to find all the integrals we need of the form \eqref{e:intF}, it is enough to evaluate only one integral. This integral, coming from six $l$'s in the numerator of \eqref{e:fullI}, is
\begin{equation}
\int \d X \Delta^{3/2} = \frac{\pi}{640 \: \bar{a}_{123}^3} \left[ 3 \bar{a}_{123}^6 - 9 \bar{a}_{123}^4 \bar{b}_{123} + 3 \bar{a}_{123}^2 \bar{b}_{123}^2 + 3 \bar{a}_{123}^3 \bar{c}_{123} + 3 \bar{a}_{123} \bar{b}_{123} \bar{c}_{123} + 2 \bar{c}_{123}^2 \right].
\end{equation}
The remaining integrals we need may now be evaluated by the following tricks:
\begin{itemize}
\item Differentiating an integral with respect to $\bq_3$ introduces Feynman parameters $x_1 x_2$, {\it e.g.}
\begin{equation}
\int \d X x_1 x_2 \Delta^{1/2} = \frac{1}{3 \bar{q}_3} \cdot \frac{\partial}{\partial \bar{q}_3} \int \d X \Delta^{3/2}.
\end{equation}
Notice that this operation decreases the power of $\Delta$ by $1$.

\item Integrals such as $\int \d X x_1^2 \Delta^{1/2}$ cannot be obtained by the above method. In this case, we may use the following formulae
\begin{align}
& k (\bar{q}_3^2 - \bar{q}_2^2) \int \d X x_1^{n+1} \Delta^{k-1} 
\nn \\& \qquad \qquad
= ( \bar{q}_3^{2k} - \bar{q}_2^{2k} ) B(n+k+1, k+1) - \: k \bar{q}_1^2 \int \d X x_1^n (x_3 - x_2) \Delta^{k-1}, \nn\\[1ex]
& k (\bar{q}_3^2 - \bar{q}_2^2) \int \d X x_1^{n+1} x_2^m \Delta^{k-1} \nn \\
& \qquad\qquad = \bar{q}_3^{2k} B(n+k+1, m+k+1) - k \bar{q}_1^2 \int \d X x_1^n x_2^m (x_3 - x_2) \Delta^{k-1} \nn\\
& \qquad \qquad\qquad - m \int \d X x_1^n x_2^{m-1} \Delta^k, \qquad \text{for } m > 0,
\end{align}
with numbers $k$, $m$ and $n$ such that these expressions exist, and where $B$ is  Euler's beta function. For example, taking $k=3/2$, $m=0$, $n=1$ we find
\begin{equation}
\int \d X x_1^2 \Delta^{1/2} = \frac{\bar{q}_1^2}{\bar{q}_3^2 - \bar{q}_2^2} \left[ \int \d X x_1 x_2 \Delta^{1/2} - \int \d X x_1 x_3 \Delta^{1/2} \right] + \frac{\pi}{128} \frac{\bar{q}_2^2 + \bar{q}_2 \bar{q}_3 + \bar{q}_3^2}{\bar{q}_2 + \bar{q}_3}.
\end{equation}

\item Integrals with odd numbers of Feynman parameters may be obtained from the integrals with even numbers of Feynman parameters by utilising the fact that $x_1 + x_2 + x_3 = 1$. For example,
\begin{equation}
\int \d X x_1 \Delta^{1/2} = \int \d X x_1^2 \Delta^{1/2} + \int \d X x_1 x_2 \Delta^{1/2} + \int \d X x_1 x_3 \Delta^{1/2},
\end{equation}
where the integrals on the r.h.s.~may be found in previous points.

\item Iterating the trick described above we may find integrals with different powers of $\Delta$, {\it e.g.}
\begin{align}
& \int \d X \Delta^{1/2} = \frac{\pi}{24 \: \bar{a}_{123}^2} \left[ \bar{a}_{123}^3 - \bar{a}_{123} \bar{b}_{123} - \bar{c}_{123} \right], \nn\\
& \int \d X \Delta^{-1/2} = \frac{\pi}{\bar{a}_{123}}, \qquad \qquad \int \d X \Delta^{-3/2} = \frac{2 \pi}{\bar{c}_{123}}.
\end{align}

\end{itemize}

This method allows the exact tensor representation of any 3-point function we consider in this paper to be calculated. However, since we are interested in the helicity representation, it can be significantly simplified. Any rank-six tensor $t_{i_1 j_1 i_2 j_2 i_3 j_3}$ built out of the metric and two independent momenta may be represented as a sum of $499$ simple tensors. In general, the independent momenta may be different for different tensor indices: we choose to use momenta $\bar{q}_1$ and $\bar{q}_2$ for $i_1$ and $j_1$ indices, $\bar{q}_2$ and $\bar{q}_3$ for $i_2$ and $j_2$ and $\bar{q}_3, \bar{q}_1$ for $i_3, j_3$. Due to various symmetries and Ward identities on $t$, however, the number of independent tensors is usually much smaller. If we consider a 3-point function $\lla T_{a_1 b_1} T_{a_2 b_2} T_{a_3 b_3} \rra$ projected onto the transverse-traceless part, we find only five independent coefficients, {\it i.e.},
\begin{eqnarray}
&& \Pi_{i_1 j_1 a_1 b_1} \Pi_{i_2 j_2 a_2 b_2} \Pi_{i_3 j_3 a_3 b_3} \lla T_{a_1 b_1} T_{a_2 b_2} T_{a_3 b_3} \rra \label{e:ttbasis}\\
&& \qquad = \Pi_{i_1 j_1 a_1 b_1} \Pi_{i_2 j_2 a_2 b_2} \Pi_{i_3 j_3 a_3 b_3} \left[
A_1(\bar{q}_1, \bar{q}_2, \bar{q}_3) \delta^{a_1 b_2} \delta^{a_2 b_3} \delta^{a_3 b_1} \right. \nn \\
&& \qquad \qquad + \: A_2(\bar{q}_1, \bar{q}_2, \bar{q}_3) \delta^{a_1 b_2} \delta^{a_2 b_3} \bar{q}_1^{a_3} \bar{q}_2^{b_1} + ( 1 \leftrightarrow 3 ) + ( 2 \leftrightarrow 3 ) \nn \\
&& \qquad \qquad + \: A_3(\bar{q}_1, \bar{q}_2, \bar{q}_3) \delta^{a_1 a_2} \delta^{b_1 b_2} \bar{q}_1^{a_3} \bar{q}_1^{b_3} + ( 1 \leftrightarrow 3 ) + ( 2 \leftrightarrow 3 ) \nn \\
&& \qquad \qquad + \: A_4(\bar{q}_1, \bar{q}_2, \bar{q}_3) \delta^{a_1 a_2} \bar{q}_2^{b_1} \bar{q}_3^{b_2} \bar{q}_1^{a_3} \bar{q}_1^{b_3} + ( 1 \leftrightarrow 3 ) + ( 2 \leftrightarrow 3 ) \nn \\
&& \left. \qquad \qquad + \: A_5(\bar{q}_1, \bar{q}_2, \bar{q}_3) \bar{q}_2^{a_1} \bar{q}_2^{b_1} \bar{q}_3^{a_2} \bar{q}_3^{b_2} \bar{q}_1^{a_3} \bar{q}_1^{b_3} \right]. \nn
\end{eqnarray}
The coefficients $A_j$ may be easily expressed in terms of the coefficients of various tensors appearing in $\lla T_{i_1 j_1} T_{i_2 j_2} T_{i_3 j_3} \rra$.  Specifically, we see that
\begin{eqnarray*}
A_1(\bar{q}_1, \bar{q}_2, \bar{q}_3) & = & 8 \cdot \text{ coefficient of } \delta^{i_1 j_2} \delta^{i_2 j_3} \delta^{i_3 j_1}, \\
A_2(\bar{q}_1, \bar{q}_2, \bar{q}_3) & = & 8 \cdot \text{ coefficient of } \delta^{i_1 j_2} \delta^{i_2 j_3} \bar{q}_1^{i_3} \bar{q}_2^{j_1}, \\
A_3(\bar{q}_1, \bar{q}_2, \bar{q}_3) & = & 2 \cdot \text{ coefficient of } \delta^{i_1 i_2} \delta^{j_1 j_2} \bar{q}_1^{i_3} \bar{q}_1^{j_3}, \\
A_4(\bar{q}_1, \bar{q}_2, \bar{q}_3) & = & 4 \cdot \text{ coefficient of } \delta^{i_1 i_2} \bar{q}_2^{j_1} \bar{q}_3^{j_2} \bar{q}_1^{i_3} \bar{q}_1^{j_3}, \\
A_5(\bar{q}_1, \bar{q}_2, \bar{q}_3) & = & \text{coefficient of } \bar{q}_2^{i_1} \bar{q}_2^{j_1} \bar{q}_3^{i_2} \bar{q}_3^{j_2} \bar{q}_1^{i_3} \bar{q}_1^{j_3}.
\end{eqnarray*}
In other words, it is enough to calculate only {\it five scalar integrals} in order to evaluate the five independent coefficients $A_j$.

Finally, to obtain the result in the helicity basis we may contract \eqref{e:ttbasis} with helicity tensors. Using the identities \eqref{PiTTdecomp} and \eqref{Theta_def} one finds five contractions of helicity tensors corresponding to the independent transverse-traceless tensors.
These results were checked by a simple computer algebra program which carried out a brute force calculation of all $499$ coefficients in $\lla T_{i_1 j_1} T_{i_2 j_2} T_{i_3 j_3} \rra$ before projecting the result into transverse-traceless and helicity bases. (This procedure also enables checking against the Ward identities.)

Note this method also works if some indices are traced. In this case, the situation is analogous to that for tensors of ranks two and four. Transverse-traceless tensors of rank four have three independent coefficients, while those of rank two have only one. The coefficient can be evaluated by the same method as described above.

\section{Contribution of ghosts and gauge-fixing terms}

\label{App_ghosts}

To evaluate the gauge field contribution to 3-point
functions we must gauge-fix and introduce ghost fields. This procedure
generates a new contribution to the stress tensor
that depends on the gauge-fixing part of the Lagrangian.
Here we show that this part does not contribute to the
3-point functions. The general argument is based on the fact that the full
Lagrangian for the gauge field is
\begin{equation}
S_{\text{YM}} = \frac{1}{g_{\text{YM}}^2} \int \d^3 x \: \tr \Big[ \frac{1}{4} F_{ij}^I F_{ij}^I + \delta_B \mathcal{O} \Big],
\end{equation}
where $\mathcal{O}$ is a gauge-fixing part containing ghosts and
$\delta_B$ is an infinitesimal BRST transformation. The full stress tensor is therefore
\begin{equation}
T^{\text{YM}}_{ij} = T^A_{ij} + T^{\text{gf}}_{ij},
\end{equation}
where $T^{\text{gf}}_{ij}$ is a BRST-exact operator. Since physical states correspond to the cohomology of the BRST transformation, $T^{\text{gf}}_{ij}$ vanishes when acting on such states. Therefore, inside any vacuum correlation function, $T^{\text{YM}}_{ij}$ can be replaced by $T^A_{ij}$.

As this is a formal argument, we will also present now an explicit perturbative
proof that the gauge-fixing part does not contribute to any correlation functions. We 
work in the $R_\xi$ gauge and to first order in $g^2_{\mathrm{YM}}$. The
ghost part and gauge-fixing part of the action may be written as
\begin{equation}
S_{\xi} = - \frac{1}{g_{\mathrm{YM}}^2} \int \d^3 x \tr \Big[ \frac{\xi}{2} (B^I)^2 + A_i^I \partial_i B^I \Big], \qquad S_{\text{gh}} = \frac{1}{g_{\mathrm{YM}}^2} \int \d^3 x \: \tr \big[\partial_i \bar{c}^I \partial_i c^I \big].
\end{equation}
where $\bar{c}^I$ and $c^I$ are the antighost and the ghost fields,
and $B^I$ is the BRST auxiliary field. All fields are in the adjoint representation and are regarded as traceless hermitian matrices. The full Yang-Mills theory is given by the action
\begin{equation} \label{e:fullSYM}
S_{\text{YM}} = \frac{1}{g_{\text{YM}}^2} \int \d^3 x \: \tr \Big[ \frac{1}{4} F_{ij}^I F_{ij}^I \Big] + S_{\xi} + S_{\text{gh}}.
\end{equation}
This leads to the following propagators
\begin{equation} \label{e:BBandBF}
\lla B^I(\bar{q}) B^{J}(-\bar{q}) \rra = 0, \qquad \qquad
\lla B^I(\bar{q}) F^{J}_{ij}(-\bar{q}) \rra = 0,
\end{equation}
and
\begin{equation} \label{e:BAandcc}
-\lla A_i^{I a}(\bar{q}) B^{J b}(- \bar{q}) \rra =
\lla (\partial_i \bar{c}^{I a})(\bar{q}) \: c^{J b}(-\bar{q}) \rra = \delta^{ab} \delta^{I J} \frac{i g_{\text{YM}}^2 \bar{q}_i}{\bar{q}^2}.\end{equation}
Here, by $(\partial_i \bar{c}^{I a})(\bar{q})$, we denote the Fourier transform of $\partial_i \bar{c}^{I a}(x)$.

The stress tensor and  the  $\Upsilon$ tensor defined in \eqref{Upsilon_def}
corresponding
to each component of the action is given by
\begin{eqnarray}
T_{ij}^A & = & \frac{1}{g_{\mathrm{YM}}^2} \tr [ F^I_{ik} F^I_{jk} - \delta_{ij} \frac{1}{4} F^I_{kl} F^I_{kl} ], \nn\\
T_{ij}^\xi & = & \frac{1}{g_{\mathrm{YM}}^2} \tr [ - P_{ijkl} A_k^I \partial_l B^I + \delta_{ij} \frac{\xi}{2} (B^I)^2 ], \nn\\
T_{ij}^\text{gh} & = & \frac{1}{g_{\mathrm{YM}}^2} \tr [ P_{ijkl} \partial_k \bar{c} \partial_l c ], \nn\\
\Upsilon^A_{ijkl} & = & -\frac{1}{2} \left[ \delta_{ij} T^A_{kl} + P_{ijkl} T^A  + Q_{ijklmn} T^A_{mn} \right] \delta(x - y), \nn\\
\Upsilon_{ijkl}^\xi & = & \frac{1}{g_{\mathrm{YM}}^2} \tr [ - \delta_{i(k} \delta_{l)j} A_m^I \partial_m B^I + \delta_{ij} A_{(k}^I \partial_{l)} B^I - \delta_{i(k} \delta_{l)j} \frac{\xi}{2} (B^I)^2 ] \delta(x - y), \nn\\
\Upsilon_{ijkl}^\text{gh} & = & \frac{1}{g_{\mathrm{YM}}^2} \tr [ \delta_{i(k} \delta_{l)j} \partial_m \bar{c}^I \partial_m c^I - \delta_{ij} \partial_{(k} \bar{c}^I \partial_{l)} c^I ] \delta(x - y).
\end{eqnarray}
where $Q_{ijklmn}$ is defined in (\ref{Odef}).
The full stress tensor and $\Upsilon$ tensor is a sum
\begin{equation}
T^{\text{YM}}_{ij} = T^A_{ij} + T^\xi_{ij} + T^{\text{gh}}_{ij}, \qquad \qquad \Upsilon^{\text{YM}}_{ijkl} = \Upsilon^A_{ijkl} + \Upsilon^\xi_{ijkl} + \Upsilon^{\text{gh}}_{ijkl}.
\end{equation}

The mechanism for cancellation of ghost and gauge-fixing terms is very general. Let us consider a set of general gauge-invariant operators $\mathcal{F}^{(\alpha)}$ of arbitrary tensor structure, indexed by $\alpha$, quadratic in field strengths $F^I$.  Consider moreover gauge dependent terms $\mathcal{B}^{(\alpha)}$ and ghost terms $\mathcal{C}^{(\alpha)}$ of the schematic form
\begin{eqnarray}
\mathcal{B}^{(\alpha)} & = & \frac{1}{g_{\text{YM}}^2} \tr \Big[ A^I_i \hat{O}^{A, (\alpha)}_i[B^I] + \hat{O}^{B, (\alpha)}[(B^I)^2] \Big], \\
\mathcal{C}^{(\alpha)} & = & \frac{1}{g_{\text{YM}}^2} \tr \Big[ \partial_i \bar{c}^I \hat{O}^{C, (\alpha)}_i [ c^I ] \Big],
\end{eqnarray}
where $\hat{O}^{A, (\alpha)}_i$ is linear in $B^I$, $\hat{O}^{B, (\alpha)}$ is
quadratic in $B^I$ and
$\hat{O}^{C, (\alpha)}_i$ is linear in $c^I$,
but are otherwise operators of arbitrary tensor
structure which \textit{may} contain derivatives but no other fields. We
consider operators $\mathcal{O}^{(\alpha)} = \mathcal{F}^{(\alpha)} +
\mathcal{B}^{(\alpha)} + \mathcal{C}^{(\alpha)}$ and their $n$-point
function in the Yang-Mills theory with the action
\eqref{e:fullSYM}. The stress tensor and the $\Upsilon$ tensor are
of this form. We find
\begin{eqnarray}
\langle \mathcal{O}^{(1)}  \mathcal{O}^{(2)} \ldots \mathcal{O}^{(n)} \rangle & = & \langle \mathcal{F}^{(1)} \mathcal{F}^{(2)} \ldots \mathcal{F}^{(n)} \rangle + \nn\\
&& + \: \langle \mathcal{B}^{(1)} \mathcal{F}^{(2)} \ldots \mathcal{F}^{(n)} \rangle + \langle \mathcal{F}^{(1)} \mathcal{B}^{(2)} \ldots \mathcal{F}^{(n)} \rangle + \ldots + \langle \mathcal{F}^{(1)} \mathcal{F}^{(2)} \ldots \mathcal{B}^{(n)} \rangle \nn\\
&& + \: \langle \mathcal{B}^{(1)} \mathcal{B}^{(2)} \mathcal{F}^{(3)} \ldots \mathcal{F}^{(n)} \rangle + \text{ perms} \nn\\
&& + \: \ldots \nn\\
&& + \: \langle \mathcal{B}^{(1)} \mathcal{B}^{(2)} \ldots \mathcal{B}^{(n)} \rangle + \langle \mathcal{C}^{(1)} \mathcal{C}^{(2)} \ldots \mathcal{C}^{(n)} \rangle
\end{eqnarray}
since there is no interaction between ghosts and any other fields at leading order in $g^2_{\text{YM}}$. We will now show that all terms but the first one cancel.

To begin, we observe that all terms containing at least one $\mathcal{F}$ and at least one $\mathcal{B}$ vanish. Indeed, when Wick's theorem is applied, there must be at least one contraction between $F$ and $B$ fields, or between $B$ and another $B$ field, which gives zero by \eqref{e:BBandBF}.

Now consider the term with $\mathcal{B}$ operators only. When expanded, it has $2^n$ terms, but every term containing $(B^I)^2$ must evaluate to zero as there must be at least one $B$-$B$ contraction. Therefore, only one term survives, namely
\begin{equation} \label{e:Bterm}
\langle \mathcal{B}^{(1)} \ldots \mathcal{B}^{(n)} \rangle
= \frac{1}{g_{\text{YM}}^{2n}} \langle \tr
\left( A^{I_1}_{j_1} \hat{O}^{A, (1)}_{j_1} [B^{I_1}]  \right) \cdot \ldots \cdot \tr \left(A^{I_n}_{j_n} \hat{O}^{A, (n)}_{j_n}[B^{I_n}] \right) \rangle.
\end{equation}
The only non-vanishing way of contracting fields is to contract auxiliary fields with gauge fields. This gives precisely the same possible contractions as in the ghost part, which is
\begin{equation} \label{e:Cterm}
\langle \mathcal{C}^{(1)} \ldots \mathcal{C}^{(n)} \rangle
= \frac{1}{g_{\text{YM}}^{2n}} \langle \tr \left( \partial_{j_1} \bar{c}^{I_1} \hat{O}^{C, (1)}_{j_1} [c^{I_1}] \right) \cdot \ldots \cdot \tr \left( \partial_{j_n} \bar{c}^{I_n} \hat{O}^{C, (n)}_{j_n} [c^{I_n}] \right) \rangle.
\end{equation}
It follows that if $- \hat{O}^{A, (\alpha)}_i
= \hat{O}^{C, (\alpha)}_i$ for all $\alpha$, then \eqref{e:Bterm} and
\eqref{e:Cterm} cancel each other out, due to \eqref{e:BAandcc} and
the anti-commuting nature of ghost fields.

In our case, and to this order in $g^2_{\mathrm{YM}}$, there are no gauge boson interactions and so the gauge group $G$ is effectively $U(1)^{\dim G}$. For $\langle T^{\text{YM}}_{i_1 j_1} \: T^{\text{YM}}_{i_2 j_2} \: T^{\text{YM}}_{i_3 j_3} \rangle$ we therefore find
\begin{equation}
\mathcal{F}^{\alpha} = T^A_{i_\alpha j_\alpha}, \qquad \mathcal{B}^{\alpha} = T^{\xi}_{i_\alpha j_\alpha}, \qquad \mathcal{C}^{\alpha} = T^{\text{gh}}_{i_\alpha j_\alpha},
\end{equation}
and
\begin{equation}
- \hat{O}^{A, (\alpha)}_{i_\alpha j_\alpha, k} = \hat{O}^{C, (\alpha)}_{i_\alpha j_\alpha, k} = P_{i_\alpha j_\alpha k l} \partial_l
\end{equation}
for $\alpha = 1,2,3$. For $\langle \Upsilon^{\text{YM}}_{ijkl} \: T^{\text{YM}}_{mn} \rangle$, we have
\begin{eqnarray}
\mathcal{F}^{(1)} = \Upsilon^A_{ijkl}, \qquad & \mathcal{B}^{(1)} = \Upsilon^{\xi}_{ijkl}, & \qquad \mathcal{C}^{(1)} = \Upsilon^{\text{gh}}_{ijkl}, \nn\\
\mathcal{F}^{(2)} = T^A_{mn}, \qquad & \mathcal{B}^{(2)} = T^{\xi}_{mn}, & \qquad \mathcal{C}^{(2)} = T^{\text{gh}}_{mn}
\end{eqnarray}
and
\begin{eqnarray}
- \hat{O}^{A, (1)}_{ijkl, m} = \hat{O}^{C, (1)}_{ijkl, m} & = & \delta_{i(k} \delta_{l)j} \partial_m - \delta_{ij} \delta_{m(k} \partial_{l)}, \nn\\
- \hat{O}^{A, (2)}_{mn, i} = \hat{O}^{C, (2)}_{mn, i} & = & P_{mn ik} \partial_k.
\end{eqnarray}
It follows that the contribution due to the gauge-fixing part of the action
indeed cancels out.

\section{Further Ward identities} \label{app_WI}

\subsection{Conformal Ward identities}
\label{app_CWI}

The conformal Ward identities are given by
\begin{align}
0 & =  \left[ \bar{q}_i^{2 \Delta_i - d - 1} \frac{\partial}{\partial \bar{q}_i} \left( \frac{1}{\bar{q}_i^{2 \Delta_i - d - 1}} \frac{\partial}{\partial \bar{q}_i}
 \right) - (i \leftrightarrow j)\right]
\lla \mathcal{O}_1(\bar{q}_1) \mathcal{O}_2(\bar{q}_2) \mathcal{O}_3(\bar{q}_3) \rra, \quad (i,j{=}1,2,3) \\
0 & =  \left[ 2 (\Delta_2 - d) \partial_{2\mu} + \sum_{j=1}^2
\left( - 2 \bar{q}_j^\nu  \partial_{j\nu} \partial_{j\mu} + \bar{q}_{j\mu} \partial_j^2 \right) \right] \lla T_{i_1 j_1}(\bar{q}_1) \mathcal{O}_2(\bar{q}_2) \mathcal{O}_3(\bar{q}_3) \rra \\
& \qquad + \: 2 \left[ (\delta_{i_1\mu} \partial_1^{a_1}
- \delta_{\mu}^{a_1} \partial_{1 i_1}
) \delta_{j_1}^{b_1}
+ (a_1 \leftrightarrow  b_1, i_1 \leftrightarrow  j_1)
\right]
\lla T_{a_1 b_1}(\bar{q}_1) \mathcal{O}_2(\bar{q}_2) \mathcal{O}_3(\bar{q}_3) \rra, \nonumber \\
0 & =  \left[ \sum_{j=1}^2 \left( - 2 \bar{q}_j^\nu \partial_{j\nu}
\partial_{j\mu} + \bar{q}_{j\mu} \partial_j^2 \right) \right] \lla T_{i_1 j_1}(\bar{q}_1) T_{i_2 j_2}(\bar{q}_2) \mathcal{O}(\bar{q}_3) \rra \\
& \qquad + \: 2 \left[ (\delta_{i_1 \mu} \partial_1^{a_1}
- \delta_{\mu}^{a_1} \partial_{1 i_1}
)\delta_{j_1}^{b_1}
+ (a_1 \leftrightarrow  b_1, i_1 \leftrightarrow  j_1)
 \right]
\lla T_{a_1 b_1}(\bar{q}_1) T_{i_2 j_2}(\bar{q}_2) \mathcal{O}(\bar{q}_3) \rra
\nonumber \\
&\qquad + \: 2 \left[ (\delta_{i_2\mu} \partial_2^{a_2}
- \delta_{\mu}^{a_2} \partial_{2 i_2}
) \delta_{j_2}^{b_2}
+ (a_2 \leftrightarrow  b_2, i_2 \leftrightarrow  j_2)
 \right]
\lla T_{i_1 j_1}(\bar{q}_1) T_{a_2 b_2}(\bar{q}_2) \mathcal{O}(\bar{q}_3) \rra,
\nonumber \\
0 & =  \left[ \sum_{j=1}^2 \left( - 2 \bar{q}_j^\nu \partial_{j\nu}
\partial_{j\mu} + \bar{q}_{j\mu} \partial_j^2 \right) \right] \lla T_{i_1 j_1}(\bar{q}_1) T_{i_2 j_2}(\bar{q}_2)  T_{i_3 j_3}(\bar{q}_3) \rra \\
& \qquad + \: 2 \left[ (\delta_{i_1 \mu} \partial_1^{a_1}
- \delta_{\mu}^{a_1} \partial_{1 i_1}
) \delta_{j_1}^{b_1}
+ (a_1 \leftrightarrow  b_1, i_1 \leftrightarrow  j_1)
 \right]
\lla T_{a_1 b_1}(\bar{q}_1) T_{i_2 j_2}(\bar{q}_2)  T_{i_3 j_3}(\bar{q}_3) \rra
\nonumber \\
&\qquad + \: 2 \left[ (\delta_{i_2 \mu} \partial_2^{a_2}
- \delta_{\mu}^{a_2} \partial_{2 i_2}
) \delta_{j_2}^{b_2}
+ (a_2 \leftrightarrow  b_2, i_2 \leftrightarrow  j_2)
 \right]
\lla T_{i_1 j_1}(\bar{q}_1) T_{a_2 b_2}(\bar{q}_2)  T_{i_3 j_3}(\bar{q}_3) \rra,
\nonumber
\end{align}
where $\mathcal{O}_i$ is taken to have dimension $\Delta_i$ and
$\partial_{j\mu} \equiv \partial/\partial \bar{q}_j^\mu$.

\subsection{Diffeomorphism Ward identity}
\label{app_Ward}

The diffeomorphism Ward identity for 3-point functions may be evaluated by functionally differentiating $\nabla^i\<T_{ij}(x)\>_s=0$ twice with respect to the metric \cite{Osborn:1993cr}, yielding
\begin{align}
 0 &= \bq_{1i}\<\!\<T_{ij}(\bq_1)T_{kl}(\bq_2)T_{mn}(\bq_3)\>\!\>
-2\bq_{1i}\<\!\<\Upsilon_{ijmn}(\bq_1,\bq_3)T_{kl}(\bq_2)\>\!\>
-2\bq_{1i}\<\!\<\Upsilon_{ijkl}(\bq_1,\bq_2)T_{mn}(\bq_3)\>\!\> \nn\\&\quad
-2\bq_{1(k}\<\!\<T_{l)j}(\bq_3)T_{mn}(-\bq_3)\>\!\>
-2\bq_{1(m}\<\!\<T_{n)j}(\bq_2)T_{kl}(-\bq_2)\>\!\>
-\delta_{kl}\bq_{2 p}\<\!\<T_{pj}(\bq_3)T_{mn}(-\bq_3)\>\!\>\nn\\&\quad
-\delta_{mn}\bq_{3 p}\<\!\<T_{pj}(\bq_2)T_{kl}(-\bq_2)\>\!\>
+\bq_{2j}\<\!\<T_{kl}(\bq_3)T_{mn}(-\bq_3)\>\!\>
+\bq_{3j}\<\!\<T_{mn}(\bq_2)T_{kl}(-\bq_2)\>\!\>.
\end{align}
We explicit checked that all our 3-point functions satisfy this identity. 
Note that our result differs from that quoted in \cite{Osborn:1993cr} due to a difference in the definition of the 3-point function: here, we define the 3-point function by the insertion of three copies of the operator $T_{ij}$, whereas in \cite{Osborn:1993cr}, the 3-point function is defined via functionally differentiating the generating functional three times.  These two definitions differ from each other by semi-local terms (see the discussion around (\ref{osb})) .

\bibliography{QFT9}
\bibliographystyle{h-ieeetr}

\end{document}